\documentclass[aps,pra,twocolumn,showpacs,groupedaddress,notitlepage,superscriptaddress,floatfix,longbibliography,nofootinbib]{revtex4-2}
\usepackage{amsmath,amsthm,amssymb,amsfonts,enumitem,float,graphicx,hyperref}
\usepackage{color}
\usepackage[dvipsnames]{xcolor}
\usepackage[normalem]{ulem}
\usepackage{subfigure}
\hypersetup{
    colorlinks=true,       
    linkcolor=cyan,          
    citecolor=magenta,        
    filecolor=magenta,      
    urlcolor=cyan,           
    runcolor=cyan
}

\usepackage{algorithm}
\usepackage{algpseudocode}
\usepackage{setspace}
\usepackage{subfigure}
\usepackage{mathtools}
\usepackage{placeins}
\usepackage{ulem}

\usepackage{braket}

\def\>{\rangle}
\def\<{\langle}

\newcommand{\ketb}[2]{|{#1}\>\!\<#2|}

\newcommand{\pqty}[1]{\left(#1\right)}
\newcommand{\Bqty}[1]{\left\{#1\right\}}

\newcommand{\bes} {\begin{subequations}}
\newcommand{\ees} {\end{subequations}}

\begin{document}

\title{
Breakdown of the weak-coupling limit in quantum annealing}
\author{Yuki Bando}
\thanks{Present address: Arithmer Inc., R\&D Headquarters, Terashimahonchonishi, Tokushima-shi, Tokushima 770-0831, Japan}
\affiliation{Institute of Innovative Research, Tokyo Institute of Technology, Yokohama, Kanagawa 226-8503, Japan}

\author{Ka-Wa Yip}
\affiliation{Center for Quantum Information Science \& Technology, University of Southern California, Los Angeles, California 90089, USA}
\affiliation{Department of Physics \& Astronomy, University of Southern California, Los Angeles, California 90089, USA}

\author{Huo Chen}
\affiliation{Center for Quantum Information Science \& Technology, University of Southern California, Los Angeles, California 90089, USA}
\affiliation{Department of Electrical \& Computer Engineering, University of Southern California, Los Angeles, California 90089, USA}

\author{Daniel A. Lidar}%
\affiliation{Center for Quantum Information Science \& Technology, University of Southern California, Los Angeles, California 90089, USA}
\affiliation{Department of Physics \& Astronomy, University of Southern California, Los Angeles, California 90089, USA}
\affiliation{Department of Electrical \& Computer Engineering, University of Southern California, Los Angeles, California 90089, USA}
\affiliation{Department of Chemistry, University of Southern California, Los Angeles, California 90089, USA}

\author{Hidetoshi Nishimori}
\affiliation{International Research Frontiers Initiative, Tokyo Institute of Technology, Yokohama, Kanagawa 226-8503, Japan}
\affiliation{Graduate School of Information Sciences, Tohoku University, Sendai, Miyagi 980-8579, Japan}
\affiliation{RIKEN Interdisciplinary Theoretical and Mathematical Sciences (iTHEMS), Wako, Saitama 351-0198, Japan}

\begin{abstract}
Reverse annealing is a variant of quantum annealing, in which the system is prepared in a classical state, reverse-annealed to an inversion point, and then forward-annealed. 
We report on reverse annealing experiments using the D-Wave 2000Q device, with a focus on the $p=2$ $p$-spin problem, which undergoes a second order quantum phase transition with a gap that closes polynomially in the number of spins. 
We concentrate on the total and partial success probabilities, the latter being the probabilities of finding each of two degenerate ground states of all spins up or all spins down, the former being their sum. The empirical partial success probabilities exhibit a strong asymmetry between the two degenerate ground states, depending on the initial state of the reverse anneal. To explain these results, we perform open-system simulations using master equations in the limits of weak and strong coupling to the bath. The former, known as the adiabatic master equation (AME), with decoherence in the instantaneous energy eigenbasis, predicts perfect symmetry between the  two degenerate ground states, thus failing to agree with the experiment. In contrast, the latter, known as the polaron transformed Redfield equation (PTRE), is in close agreement
with experiment. 
Thus our results present a challenge to the sufficiency of the weak system-bath coupling limit in describing the dynamics of current experimental quantum annealers, at least for reverse annealing on timescales of a $\mu$sec or longer.
\end{abstract}

 \maketitle

\section{Introduction}
\label{sec:INTRO} 

Quantum annealing is a quantum metaheuristic originally conceived as a method to obtain the global solution of an optimization problem by using quantum fluctuations to escape from local minima~\cite{Kadowaki1998,Farhi2001,Santoro2002} (see Refs.~\cite{Das2008, Morita2008,Albash2018, Hauke2020,crosson2020prospects} for reviews).

Commercial devices developed by D-Wave Systems that realize programmable quantum annealing in hardware have now been available for more than a decade~\cite{Dwave}. Not only can these devices be used to solve optimization problems by physically realizing quantum annealing, but they are also able to perform physics experiments, for example, spin-glass phase transitions~\cite{Harris2018}, the Kosterlitz-Thouless phase transition~\cite{King2018, King:2019aa}, alternating-sector ferromagnetic chains~\cite{Mishra:2018}, the Kibble-Zurek mechanism~\cite{Gardas2018, Weinberg2020, Bando2020}, the Griffiths-McCoy singularity~\cite{Nishimura2020}, spin ice~\cite{King2020}, the Shastry-Sutherland model~\cite{Kairys:2020vv},
field theory~\cite{Abel2021}, and spin liquids~\cite{Zhou2021}.

In this work we use a D-Wave 2000Q quantum annealer as a simulator of a simple spin system: the $p$-spin model~\cite{Derrida1981,Gross1984,Bapst2012} with $p=2$. But rather than using traditional ``forward'' annealing, wherein the Hamiltonian of a system initialized in the ground state of a transverse field evolves to a longitudinal target Hamiltonian~\cite{Kadowaki1998}, here we focus on \emph{reverse annealing}, which is defined as the process of choosing an appropriate classical state as the initial state, starting from the target Hamiltonian and mixing it with the transverse field Hamiltonian to return to the original target Hamiltonian~\cite{Perdomo-Ortiz2011}. Reverse annealing is conceptually richer than forward annealing, since it introduces at least two additional parameters: the inversion point of the anneal and the pause duration. These parameters can be chosen to make forward annealing a special case
of (the forward direction of) reverse annealing. Moreover, one can also iterate the process, leading to a strategy known as iterated reverse annealing~\cite{Yamashiro:2019aa,King:2019aa}. We note that there is some evidence that reverse annealing can outperform forward annealing in solving optimization problems~\cite{Chancellor2017,Ohkuwa:2018aa,Ottaviani2018,Yamashiro:2019aa,Venturelli2019,Marshall2019,9259948,Rocutto:2021wt}. However, our focus in this work is not on algorithmic performance, but rather on using the rich playground provided by the $p=2$ $p$-spin model under reverse annealing to answer the question of which of a variety of models best describes the results obtained from the D-Wave annealer. This question has been addressed before many times~\cite{q-sig,q108,Smolin,Boixo:2014yu,Albash:2014if,SSSV,albash2015consistency,PhysRevX.6.031015,Mishra:2018,King2018, King:2019aa,Bando2020}, but as we shall show, the $p=2$ $p$-spin model under reverse annealing allows us to rather clearly reveal a deficiency of one of the most popular and successful models, the weak-coupling adiabatic master equation (AME)~\cite{Albash2012}. 
A relatively recent model,
with strong system-bath coupling,
the polaron-transformed Redfield master equation (PTRE)~\cite{xu_non-canonical_2016,chen_hoqst_2020},
provides a closer match to the empirical data we present. 

We remark that the $p$-spin model has been studied before in the context of reverse annealing, for $p=3$~\cite{Passarelli2019}. The choice of $p=3$ 
was motivated by the fact that this model has a first order phase transition in the thermodynamic limit~\cite{Derrida1981,Gross1984,Bapst2012} (gap closing exponentially in the system size) and thus is a hard problem for conventional (forward) quantum annealing. However, this model does not have a direct physical embedding in current quantum annealing hardware, since it involves three-body interactions. While gadgets can be used to embed such a model in physical hardware supporting only two-body interactions, this comes at the cost of using up qubits and also introduces various errors~\cite{Babbush:2013aa}. Hence the study~\cite{Passarelli2019} was purely numerical. The $p$-spin model with $p=2$, which undergoes a second-order phase transition (polynomially closing gap), can be directly represented in the hardware graph of the D-Wave devices, and we do this here. The direct embedding allows us to avoid gadgetization errors and also gives us access to larger system sizes than the $p=3$ case:
we experiment with up to $20$ fully connected (logical) qubits.
Another reason for our choice of $p=2$ is the existence of ground-state degeneracy, which leads to additional information and the crucial ability to distinguish the AME from the other models, as will be described below. 

The structure of this paper is as follows.  We first define the $p$-spin model with $p=2$ in Sec.~\ref{sec:problem} and describe the reverse annealing protocol we employed in this study. In Sec.~\ref{sec:exp} we present and discuss the empirical results from the D-Wave 2000Q device, for a variety of different parameter settings, including initial conditions, annealing time, pause duration, problem size, and the effect of iteration. This is followed in Sec.~\ref{sec:simulations} by an examination of numerical results based first on the closed-system Schr\"{o}dinger equation, followed by quantum adiabatic master equation with both independent and collective system-bath coupling, and the PTRE.
In all cases we compare and contrast the predictions of the various models to the empirical D-Wave data. Conclusions are presented in Sec.~\ref{section:Conc}, and  appendixes provide further technical details, as well as additional results.

\section{Problem definition and reverse annealing protocol}
\label{sec:problem}

In this section we describe the problem Hamiltonian and the reverse annealing protocol we used both in our experiments on the D-Wave device and in numerical simulations.

\subsection{The $p$-spin model with $p=2$}
\label{sec:p-spin}

\begin{figure}[htb]
\includegraphics[width=\columnwidth]{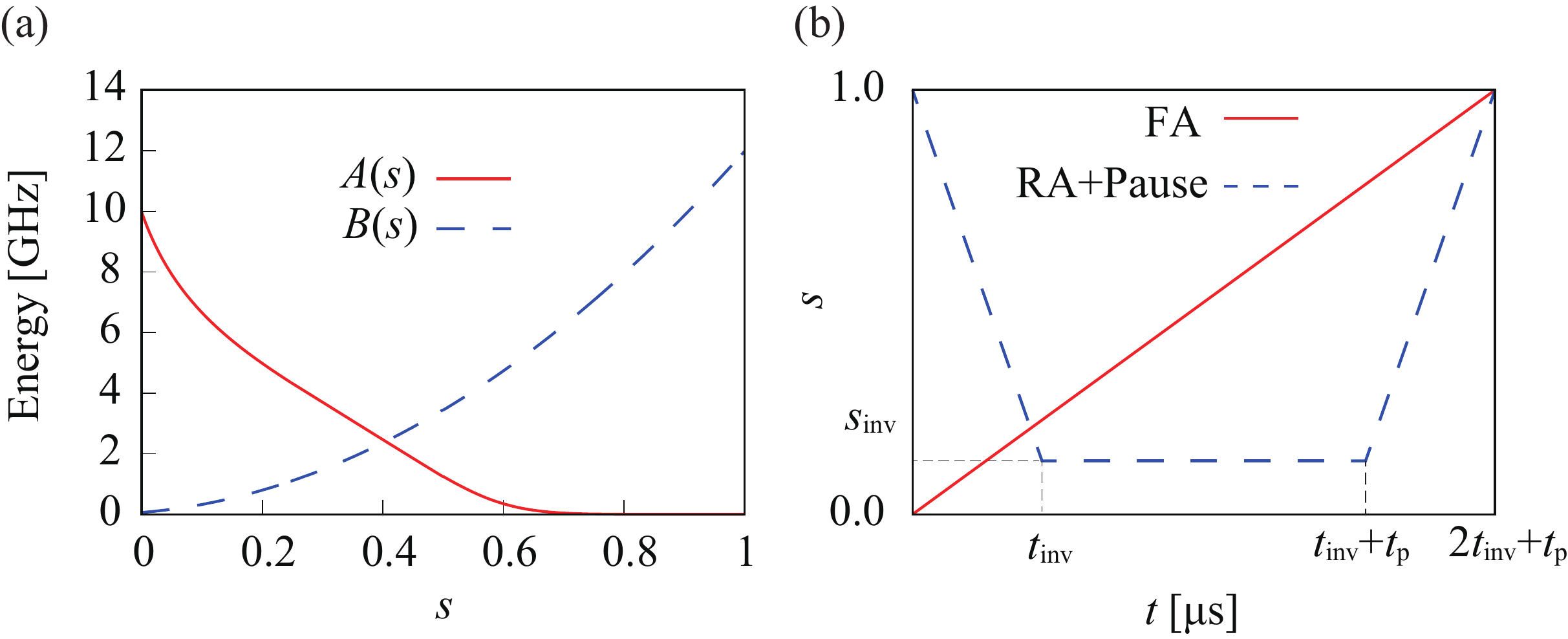}%
\caption{(a) Annealing schedule of D-Wave 2000Q with the ``DW\_2000Q\_6'' solver. (b) Forward (FA) and reverse annealing (RA) protocols as defined in Eqs.~\eqref{eq:FA} and~\eqref{eq:RA}, respectively. The latter incorporates a pause of duration $t_p$. Here $t_a=1.2~\mathrm{\mu s}$ for forward annealing, and $\tau=1~\mathrm{\mu s}$ and $s_\mathrm{{inv}}=0.2$ for reverse annealing.}
\label{fig:schedule}
\end{figure}

We consider a quantum annealing Hamiltonian comprising a driver Hamiltonian $H_{D}$ and a target Hamiltonian $H_T$, the latter encoding the combinatorial optimization problem represented as an Ising model, as a function of dimensionless time $0\le s(t)\le 1$:
\begin{equation}
\label{eq:H}
H(s) = \frac{A(s)}{2} H_{D} + \frac{B(s)}{2}H_T\ .
\end{equation}
Here, $A(s)$ and $B(s)$ are device-dependent annealing schedules. The D-Wave device used in the present experiment has $A(s)$ and $B(s)$ as shown in Fig.~\ref{fig:schedule}(a). We work in units where $\hbar=1$ and $k_\mathrm{B}=1$ except in Figs.~\ref{fig:schedule}(a), \ref{fig:gap}, \ref{fig:myspectrum}, and \ref{fig:surface}, where we opted to use units where $h=1$ in accordance with the conventions of the D-Wave device documentation~\cite{DWave_unit}.

The final values of the schedule functions are $A(s=1)\approx0$ and $B(s=1)>0$ so that ideally the ground state of $H_T$ is realized as the final state. The time dependence of $s(t)$, in particular forward or reverse annealing as depicted in Fig.~\ref{fig:schedule}(b), corresponds to different variants of the general quantum annealing algorithm.
The driver Hamiltonian $H_{D}$ is usually chosen as
\begin{equation}\label{eq:QA}
H_{D}=-\sum_{i=1}^{N}\sigma_{i}^{x}
\end{equation}
where $N$ is the number of qubits and $\sigma_{i}^{x}$ is the $x$ component of the Pauli matrix acting on the $i$th qubit. 

We study the $p$-spin model,
\begin{equation}
H_T=-N\left(\frac{1}{N}\sum_{i=1}^{N}\sigma_{i}^{z}\right)^p
\end{equation}
 with $p=2$.
This Hamiltonian couples every qubit to every other qubit, which exceeds the ``Chimera" graph connectivity of the D-Wave 2000Q device. Thus the qubits in $H_T$ should be viewed as logical qubits, to be represented in the actual device by physical qubits. These physical qubits are coupled in ferromagnetic chains to form logical qubits, a procedure known as minor embedding~\cite{Choi2008}
\footnote{
Embedding a fully-connected graph was implemented in the standard way; see, e.g., Fig.~1 of Ref.~ \cite{albash2016}.}.
We chose the ferromagnetic interaction between physical qubits in a logical qubit to be $J_{\text{F}}=-1.0$, as discussed in App.~\ref{append:optJF}.

\subsection{Reverse annealing protocol}

The traditional protocol of forward  annealing has
\begin{equation}
\label{eq:FA}
s(t)=\frac{t}{t_{a}}\ \ \ t\in[0,t_{a}],
\end{equation}
where $t_a$ is the total annealing time; see the solid red line in Fig.~\ref{fig:schedule}(b). The initial state of forward annealing is the ground state of the driver Hamiltonian $H_D$.

In reverse annealing, we start from $s=1$ and decrease $s$ to an intermediate value $s_{\mathrm{inv}}$, pause, then increase $s$ to finish the process at $s=1$. The explicit time dependence $s(t)$ realized on the D-Wave device is
\begin{equation}
\label{eq:RA}
s(t)=\begin{cases}
1-\,\displaystyle\frac{\,t}{\tau} & 0 \leq t \leq t_{\rm inv} \\
1-\displaystyle\frac{t_{\rm inv}}{\tau} = s_{\mathrm{inv}} & t_{\rm inv}\le t \leq t_{\rm inv}+t_p \\
2s_{\mathrm{inv}} - 1 
-\displaystyle\frac{t_p}{\tau}
+ \displaystyle\frac{t}{\tau} & t_{\rm inv} +t_p< t \leq 2t_{\rm inv}+t_p,
\end{cases}
\end{equation} 
where $1/\tau$ is the annealing rate,  $t_{\rm inv}$ is the inversion time defined as $t_{\rm inv}=\tau(1-s_{\mathrm{inv}})$, and $t_p$ is the duration of the intermediate pause. Our reverse annealing begins from $s(t=0)=1$ and ends at $s(t=t_a)=1$, where
\begin{equation}
t_a = 2t_{\rm{inv}}+t_p = 2 \tau(1-s_{\mathrm{inv}}) + t_p  ,
\label{eq:t_a}
\end{equation}
passing through the inversion point $s_{\mathrm{inv}}$; see the blue dashed line in Fig.~\ref{fig:schedule}(b). To distinguish between $\tau$ and $t_a$, we henceforth refer to the former as the ``annealing time" and the latter as the ``total annealing time". The initial state of reverse annealing is a classical state, usually a candidate solution to the combinatorial optimization problem, i.e, a state that is supposed to be close to the solution.

We can iteratively repeat the process of reverse annealing with the final state of a cycle as the initial condition of the next cycle. This protocol is called iterated reverse annealing~\cite{Yamashiro:2019aa} and we denote the iteration number by $r$. Additional details are provided in App.~\ref{app:RA}.

\section{Empirical results}
\label{sec:exp}

In this section we report the results of performing reverse annealing experiments for the $p=2$ $p$-spin model on the D-Wave 2000Q device. 

\begin{figure}[t]
\includegraphics[width=0.7\columnwidth]{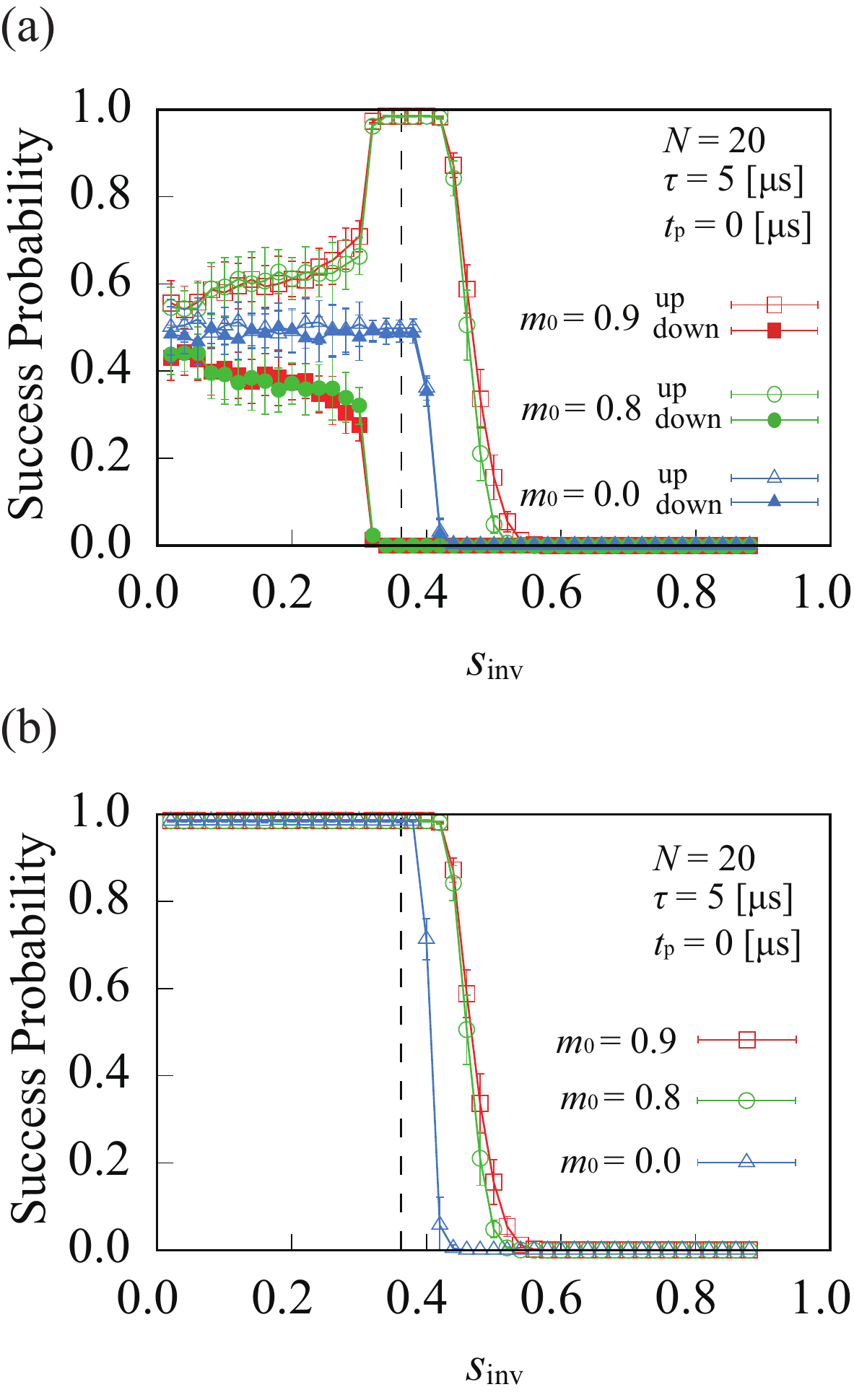}%
\caption{Empirical success probabilities for different initial conditions $m_0$ (magnetization) in reverse annealing on the D-Wave 2000Q device as a function of the inversion point $s_{\mathrm {inv}}$. The dashed line is the minimum-gap point at $s_{\Delta}\approx 0.36$ for $N=20$. When $s_{\mathrm {inv}} < s_{\Delta}$ the reverse direction of the anneal goes through and past the minimum gap point, and then crosses it again during the forward anneal. When $s_{\mathrm {inv}} > s_{\Delta}$ there is no crossing of the minimum gap point.
(a) Partial success probabilities for
$m_0=0, 0.8$, and $0.9$. (b) Total success probabilities for the same set of $m_0$ as in (a). Here and in all subsequent figures the labels ``up" and ``down" mean the populations of the all-up state and the all-down state, respectively; $s_{\mathrm {inv}}$ was incremented by steps of $0.02$, and error bars denote one standard deviation; see App.~\ref{app:RA} for details on the calculation of error bars.}
\label{fig:RA_20_initial}
\end{figure}

\subsection{Dependence on initial conditions}
\label{sec:initcond}

Figure~\ref{fig:RA_20_initial} shows the empirical success probability, i.e., the probability that the final state observed is one of the two degenerate ground states. This is shown for $N=20$ and no pause. The initial condition is specified by the initial value of the normalized total magnetization,
\begin{align}
\label{eq:mo}
    m_0=\frac{1}{N}\sum_{i=1}^N \langle \psi_0 |\sigma_i^z |\psi_0\rangle, 
\end{align}
where $|\psi_0\rangle$ denotes the initial wave function (a classical state). Since the doubly degenerate ground state of the problem Hamiltonian $H_T$ has $\pm 1$ as the normalized magnetization, a value of $m_0$ closer to 1 (or $-1$) represents an initial condition exhibiting higher overlap with the ground state. In Fig.~\ref{fig:RA_20_initial} we present results for a completely unbiased initial condition ($m_0=0$) and initial conditions strongly biased toward the all-up state $\sigma_i^z=1~\forall i$ ($m_0=0.8,0.9$). The state with $m_0=0$ is the highest excited state, and the state with $m_0=0.9$ is the first excited state.

\subsubsection{$m_0=0$}

Figure~\ref{fig:RA_20_initial}(a) shows the probability that the system reaches the all-up state (denoted ``up") and the all-down state ($\<\sigma_i^z\>=-1,~\forall i$, denoted ``down"). We call these the partial success probabilities.

When the initial condition is $m_0=0$, the up and down probabilities are equal to within the error bars for all $s_{\rm {inv}}$, as expected because the initial state is unbiased. Note that both success probabilities are close to $0.5$ for $s_{\mathrm{inv}}<0.4$ whereas they are zero for $s_{\mathrm{inv}}>0.5$. The vertical dashed line at $s_{\mathrm{inv}}\approx 0.36$ indicates the minimum-gap point, i.e., the point where the energy gap between the ground and the second excited states becomes minimum, which we denote by $s_{\Delta}$ (the first excited state becomes degenerate with the ground state at the end of the anneal, hence it is the second excited state that is relevant). It is likely that the system remains close to its initial state for $s_{\mathrm{inv}}>0.5$, where the transverse field and the associated quantum fluctuations are small, and the true final ground state with $m_0=\pm 1$ is hard to reach, since the initial state with $m_0=0$ has small overlap with the latter. The situation is different when $s_{\mathrm{inv}}<0.4$, i.e., when the system traverses the minimum-gap region. In this case the system enters the paramagnetic (quantum-disordered) phase dominated by the driver $H_D$, so that the initial condition is effectively erased. This renders the process similar to conventional forward quantum annealing, by which the two true ground states are reached with nearly equal probability $0.5$. Recall that the phase transition around $s_{\Delta}$ is of second order in the present problem, and therefore the system finds (one of) the true final ground states relatively easily by forward annealing because of the mild, polynomial, closing of the energy gap, as we discuss in more detail later [see Fig.~\ref{fig:gap}].

\subsubsection{$m_0=0.8,0.9$: up/down symmetry breaking for $0.3 \lesssim s_{\mathrm{inv}} \lesssim 0.5$}

For the initial states with $m_0=0.8$ and $m_0=0.9$, the experimental results shown in Fig.~\ref{fig:RA_20_initial}(a) reveal significant differences between the probabilities of the final all-up and all-down states. Of course, the initial conditions $m_0=0.8,0.9$ have much larger overlap with the all-up state, and this fact alone suggests a mechanism for breaking the symmetry between the probabilities of the final state being the all-up or all-down state. However, it is remarkable that the success probability for the all-up state is almost 1 in a region $0.3 \lesssim s_{\mathrm{inv}}\lesssim 0.5 $ both to the left and to the right of $s_{\Delta}\approx 0.36$ (we discuss the additional asymmetry for $s_{\mathrm{inv}} \lesssim 0.3$ below). As we explain in Sec.~\ref{sec:AME}, this behavior is inconsistent with a model of an open quantum system that is weakly coupled to its environment and is thus described by the adiabatic master equation~\cite{Albash2012}. However, it is consistent with both a quantum model of a system that is strongly coupled to its environment, as explained in Sec.~\ref{sec:PTRE}, and with a simple semiclassical model captured by the spin-vector Monte Carlo algorithm~\cite{SSSV,albash2020comparing}, as explained in App.~\ref{sec:classical}.

\subsubsection{$s_{\mathrm{inv}} \gtrsim 0.5 > s_{\Delta}$: freezing}
\label{sec:freezeout}

Note that for all three initial conditions the success probability eventually vanishes for $s_{\mathrm{inv}} \gtrsim 0.5 > s_{\Delta}$. The reason is that in all cases the system is initialized in an excited state, and remains in an excited state at the end of the anneal, since there is no mechanism for thermal relaxation to a lower energy state when $s_{\mathrm{inv}}$ is large. This is a manifestation of the phenomenon of freezing~\cite{Albash2012,Amin2015,Boixo:2014yu}, i.e., the extreme slowdown of relaxation due to the fact that the system-bath interaction (nearly) commutes with the system Hamiltonian when the transverse field is very small. In addition, the annealing timescale is manifestly too slow for downward diabatic transitions. This is true even with the discontinuity in the derivative of the schedule depicted in Fig.~\ref{fig:schedule}(b).
Despite this, the reversal of the anneal direction is apparently too slow in practice to have a non-adiabatic effect, or if diabatic transitions do occur, then they exclusively populate higher excited states.

\subsubsection{$s_{\mathrm{inv}} \lesssim 0.3 < s_{\Delta}$: spin-bath polarization}
\label{sec:spinbathpol}

It is also noteworthy from Fig.~\ref{fig:RA_20_initial}(a) that the initial condition results in different probabilities for the all-up and all-down states even in the paramagnetic region $s_{\mathrm{inv}} \lesssim 0.3 < s_{\Delta}$, where quantum fluctuations are large and the all-up and all-down states are expected to have the same probability in equilibrium. The system ``remembers" the initial condition to a certain extent, an anomaly that may be attributable to spin-bath polarization~\cite{spin-bath-polarization,lanting2020probing}. Namely, the persistent current flowing in the qubit body during the anneal produces a magnetic field that can partially align or polarize an ensemble of environmental spins local to the qubit wiring, with a much slower relaxation time than the anneal duration. Given the polarized initial condition $m_0=0.8$ or $0.9$, this polarized spin-bath will be aligned with the all-up state even after the system crosses into the paramagnetic phase, thus preventing the system from equilibrating and explaining the observed memory effect. Spin-bath polarization is expected to be particularly pronounced under reverse annealing, since the strong polarization of the initial state will act to polarize the spin bath, more so than in forward annealing, where the initial state is unpolarized.

\begin{figure}[t]
\includegraphics[width=0.7\columnwidth]{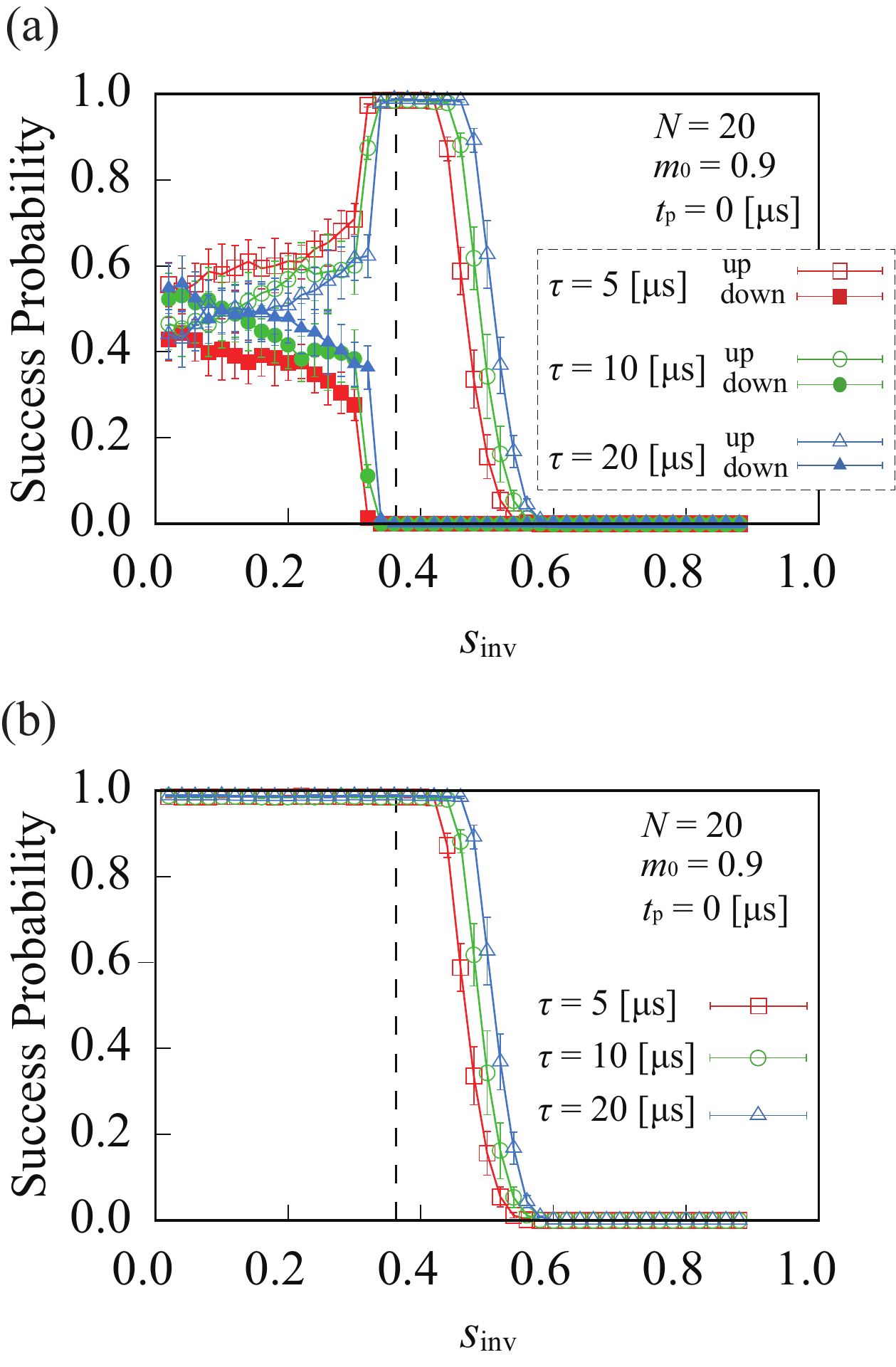}%
\caption{Empirical success probabilities for different annealing times $\tau$ in reverse annealing on the D-Wave 2000Q device as a function of $s_{\mathrm{inv}}$.
(a) Partial success probabilities for $\tau=5, 10$, and $20\mu s$. (b) Total success probabilities for the same set of $\tau$ as in (a). The dashed line is the minimum-gap point at $N=20$, $s_{\Delta}\approx 0.36$.}
\label{fig:RA_20_diff_tau}
\end{figure}

\begin{figure}[t]
\includegraphics[width=0.72\columnwidth]{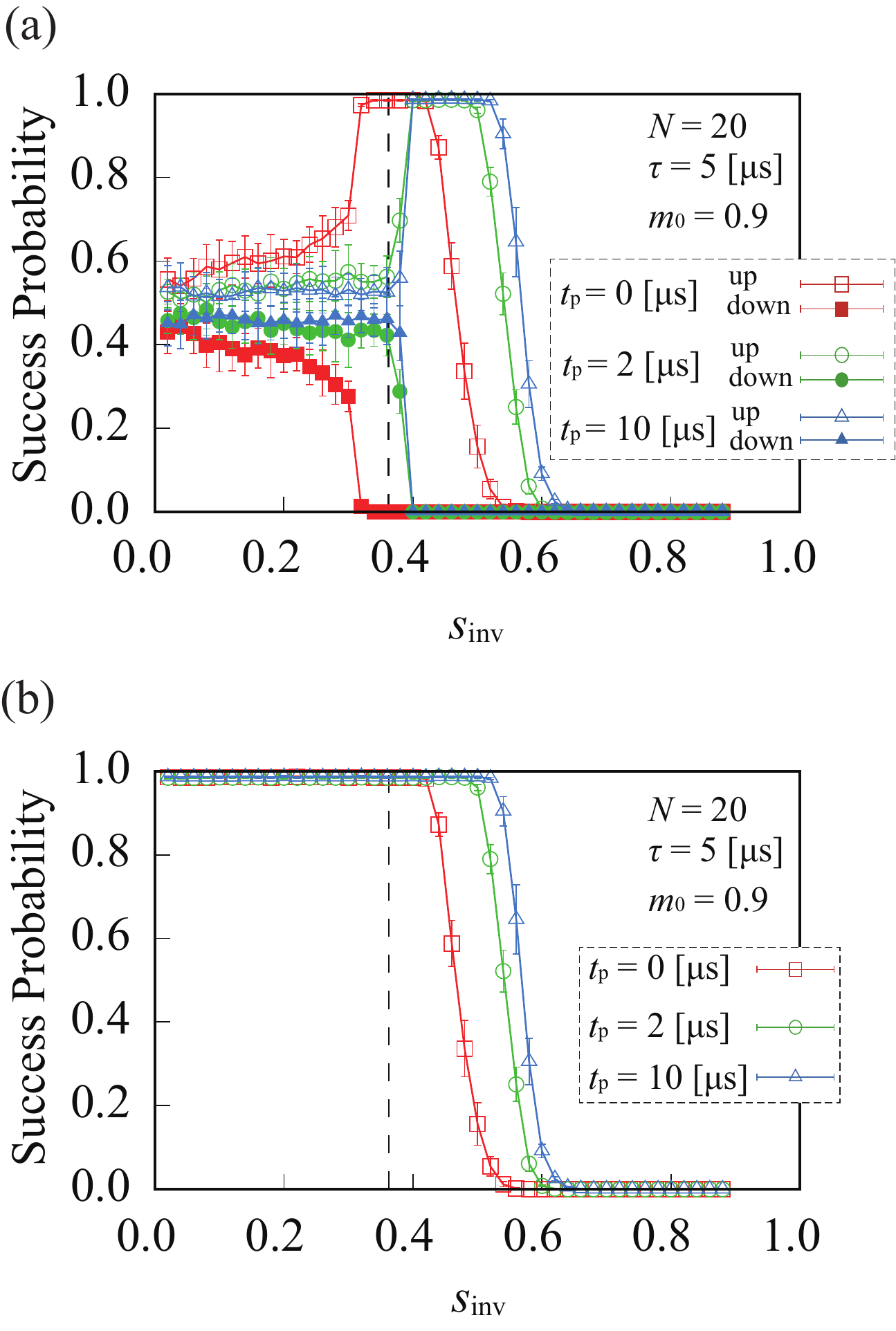}%
\caption{Empirical success probabilities for different pausing times $t_p$ in reverse annealing as a function of $s_{\mathrm{inv}}$. 
(a) Partial success probabilities for pause $t_p=0, 2$, and $10~\mu s$. (b) Total success probabilities for the same set of $t_p$ as in (a). The dashed line is the minimum-gap point at $N=20$, $s_{\Delta}\approx 0.36$.}
\label{fig:RA_20_pause}
\end{figure}

\begin{figure}[t]
\subfigure{\includegraphics[width=0.7\columnwidth]{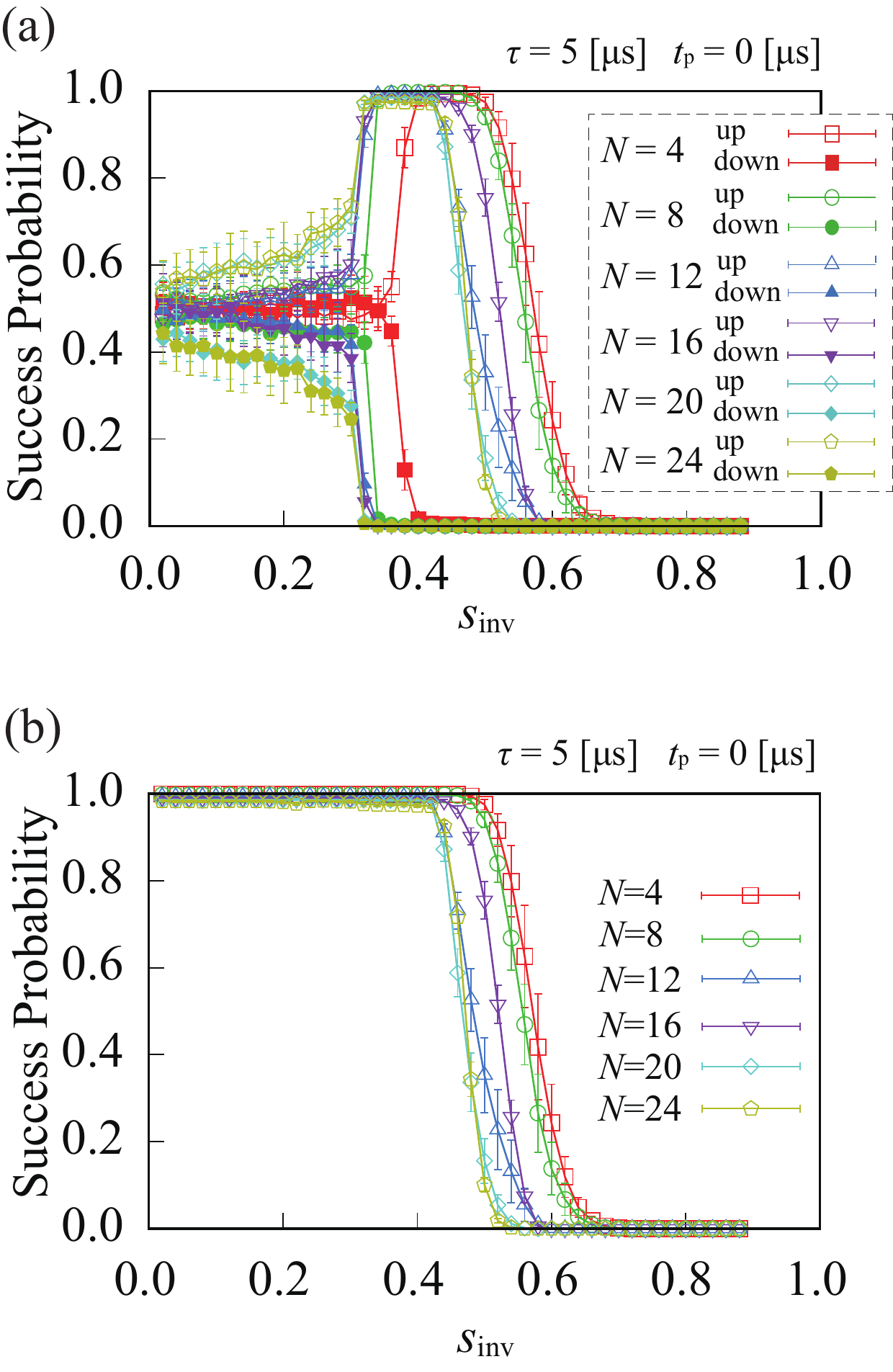}}
\caption{Empirical success probabilities for different system sizes $N$ in reverse annealing  as a function of $s_{\mathrm{inv}}$. The initial state for each $N$ is a state with one spin flipped.
(a) Partial success probabilities for system size $N\in\{4, 8, 12, 16, 20, 24\}$. (b) Total success probabilities for the same set of $N$ as in (a). } 
\label{fig:RA_20_size}
\end{figure}

\begin{figure}[t]
\subfigure{\includegraphics[width=0.75\linewidth]{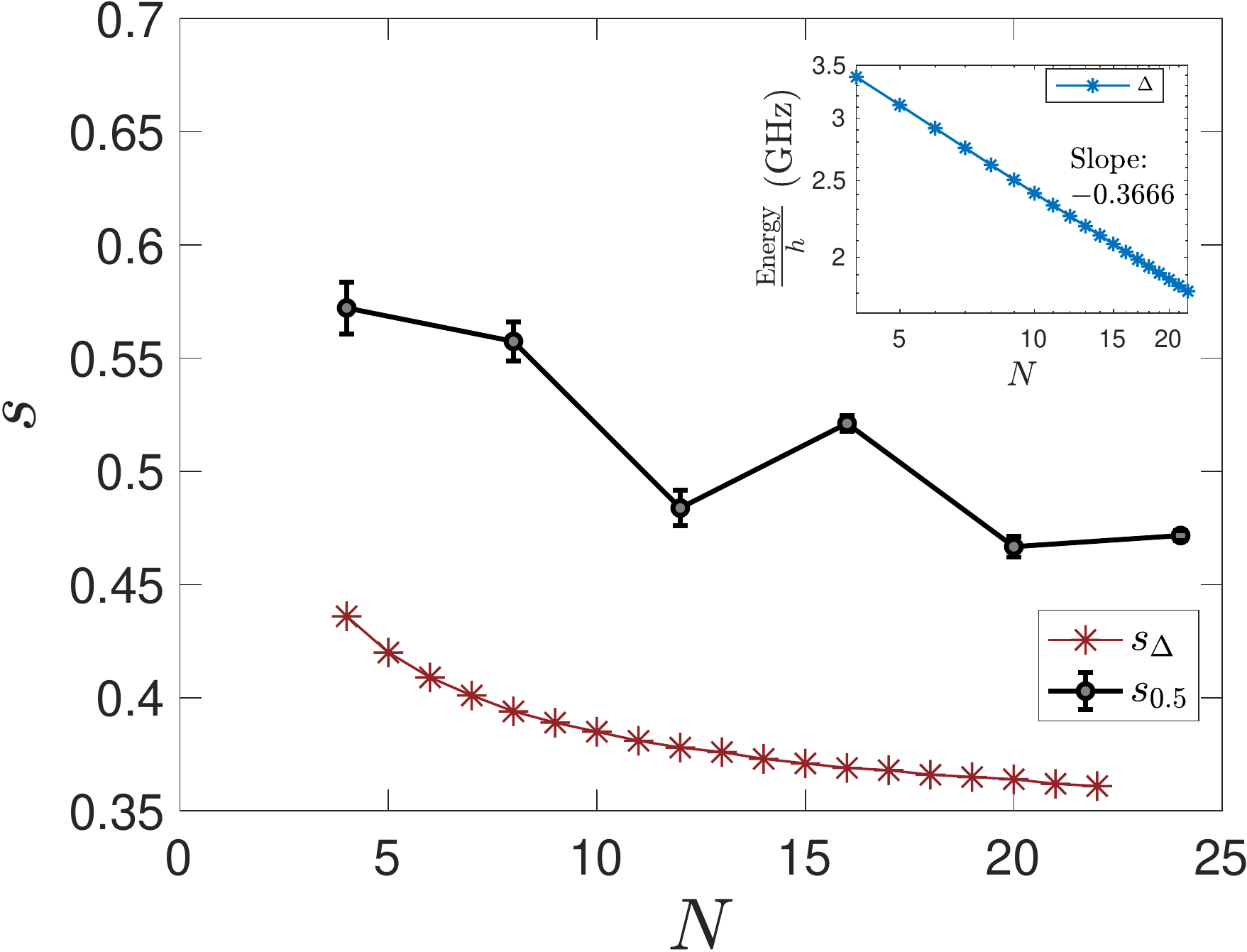}}
\caption{The position $s_\Delta$ of the minimum gap $\Delta$ as a function of system size $N$, along with the value $s_{0.5}$ of $s_{\mathrm{inv}}$ at which the total success probability $=0.5$ for each $N$. The jump at $N=16$ is due to the non-monotonicity seen for the up probabilities in Fig.~\ref{fig:RA_20_size}(a). 
Inset: the minimum gap for each $N$. The gap obeys a power law scaling $N^{-11/30}$.}
\label{fig:gap}
\end{figure}

\subsubsection{Total success probability}

Figure~\ref{fig:RA_20_initial}(b) shows the total success probability, i.e., the sum of the final up and down probabilities. The resulting curve stays almost flat and close to $1$ for $s_{\mathrm{inv}}<0.4$.  This constant $1$ is a mixture of the two effects manifest in Fig.~\ref{fig:RA_20_initial}(a), the genuine effects of reverse annealing around $s_{\mathrm{inv}}\approx 0.4$ for $m_0=0.9$ and $0.8$, and the effectively-forward-annealing-like behavior for $m_0=0$. Indeed, the case with $s_{\mathrm{inv}}=0$ is essentially equivalent to standard, forward annealing: the inversion point is set at $s=0$, so that the anneal restarts from a Hamiltonian that is completely dominated by the transverse field. That the empirical total success probability is $1$ in this case shows that the $p=2$ problem is easy also for forward annealing, in contrast to the $p=3$ case studied in Ref.~\cite{Passarelli2019}, which numerically found very small values of success probability near $s_{\mathrm{inv}}=0$. This may be explained by the very fast forward annealing in this parameter region in Ref.~\cite{Passarelli2019}, which keeps the system almost unchanged from the quantum-disordered state at $s=s_{\mathrm{inv}}$.
Aside from this subtlety, our experimental data are consistent with the numerical results of Ref.~\cite{Passarelli2019}, supporting the latter's expectation that the system-environment interaction (through spin-boson coupling) prompts relaxation of the system toward the ground state around the minimum-gap region. 

We shall see in Sec.~\ref{sec:simulations-init} that all the salient features seen in Fig.~\ref{fig:RA_20_initial}(b), such as the shift to the left with decreasing $m_0$, are captured by our open system simulations.

\subsection{Dependence on annealing time}
As seen from Eq.~\eqref{eq:t_a}, the total annealing time $t_a$ is linearly dependent on the annealing time $\tau$, and proportional to $\tau$ when $t_p=0$.
Figure~\ref{fig:RA_20_diff_tau} shows the success probability for different $\tau$ with $N=20$, $m_0=0.9$, and no pausing.  The overall trend is similar to Fig.~\ref{fig:RA_20_initial}.

As shown in Fig.~\ref{fig:RA_20_diff_tau}(a), an increase of $\tau$ leads to slightly higher all-up success probabilities for $s_{\mathrm{inv}}> s_{\Delta}$ but the other way around for $s_{\mathrm{inv}} < s_{\Delta}$. This can be explained in terms  of an increased relaxation to the ferromagnetic ground state near the minimum energy gap for larger $\tau$ and an enhanced relaxation to the paramagnetic ground state for $s_{\mathrm{inv}}<s_{\Delta}$.
As seen in Fig.~\ref{fig:RA_20_diff_tau}(b), the total success probability stays close to $1$ for $s_{\mathrm{inv}}<s_{\Delta}$ and benefits slightly from increasing $\tau$ for $s_{\mathrm{inv}}>s_{\Delta}$.

\subsection{Effects of pausing}

Figure~\ref{fig:RA_20_pause} shows success probabilities for pausing time $t_p\in\{0, 2, 10\}~\mu s$ as a function of $s_{\mathrm{inv}}$  with $N=20$, $m_0=0.9$ and $\tau=5~\mu s$. 

The overall trend is similar to Fig.~\ref{fig:RA_20_diff_tau}, i.e., a longer pause leads to an increased relaxation, but the effects are more significant in the present case. Pausing at $0.5<s_{\mathrm{inv}}<0.6$ greatly improves the success probability from nearly 0 ($t_p= 0~\mu s$) to nearly 1 ($t_p= 2~\mu s$ and $10~\mu s$), implying that pausing in a relatively narrow region slightly past the minimum gap  point is most effective. This is in line with previous findings~\cite{Marshall2019,chen2020pausing,albash2020comparing}. 

Another significant difference between pausing and not pausing is that the partial success probabilities shown in Fig.~\ref{fig:RA_20_pause}(a) are nearly piecewise flat for $t_p>0$, 
showing that the effect responsible for the anomaly disappears under pausing. This is consistent with the spin-bath polarization effect discussed in Sec.~\ref{sec:initcond}, in that the pause provides the time needed for this polarization to relax to equilibrium, on a timescale of a few $\mu s$.

\subsection{Size dependence}
\label{sec:N}

Figures~\ref{fig:RA_20_size}(a) and~\ref{fig:RA_20_size}(b), respectively, show the partial and total success probability for different system sizes $N$ as a function of $s_{\mathrm{inv}}$ with $\tau=5~\mu s$. The initial state for each $N$ is the first excited state with $m_0<1.0$ (closest to the ground state for which $m_0=1.0$), i.e., a state with one spin flipped. We observe a statistically significant slight non-monotonicity with $N$ in the region $s_{\mathrm{inv}}\geq s_{\Delta}$ of the up success probabilities [Fig.~\ref{fig:RA_20_size}(a)] and the total success probabilities [Fig.~\ref{fig:RA_20_size}(b)], namely, the ordering of the shift to the left with $N$ is: $\{4,8,16,12,20,24\}$; we do not have an explanation for this effect, though it could plausibly be an anomaly related to the minor embedding procedure. However, the down success probabilities are monotonic in $N$, and the overall trend is clear, namely, the drop-off to zero success probability occurs at smaller $s_{\mathrm{inv}}$ as $N$ is increased.\footnote{
We have confirmed that this tendency persists for larger systems from preliminary data for up to $N=48$.  We did not collect data systematically for larger systems because of technical difficulties.
}
This is consistent with the reduction in the position of the minimum gap, $s_\Delta$, as a function of $N$, shown in Fig.~\ref{fig:gap}, which is tracked by the value of $s_{\mathrm{inv}}$ at which the total success probability equals $0.5$. This suggests that the success probabilities (both partial and total) are sensitive to the location of the minimum quantum gap. See App.~\ref{append:spectrum} for more details about the spectrum of the $p$-spin problem.

Additionally, the asymmetry between the all-up and all-down states for $s_{\mathrm{inv}}<s_\Delta$ is enhanced with increasing $N$. This is consistent with the formation of larger domains of spin-bath polarization, which would be expected to take longer to dissipate due to their size as $N$ increases.

Finally, the overlap of data for $N=20$ and $24$ 
suggests that large-$N$ effects have already converged at these sizes, i.e., that $N\sim20$ is  sufficiently large to infer the behavior at larger $N$. Also, even the smallest systems with $N=4$ and $8$ already share qualitative features with larger systems.

\subsection{Effects of iteration}

\begin{figure}
\includegraphics[width=1.03\columnwidth]{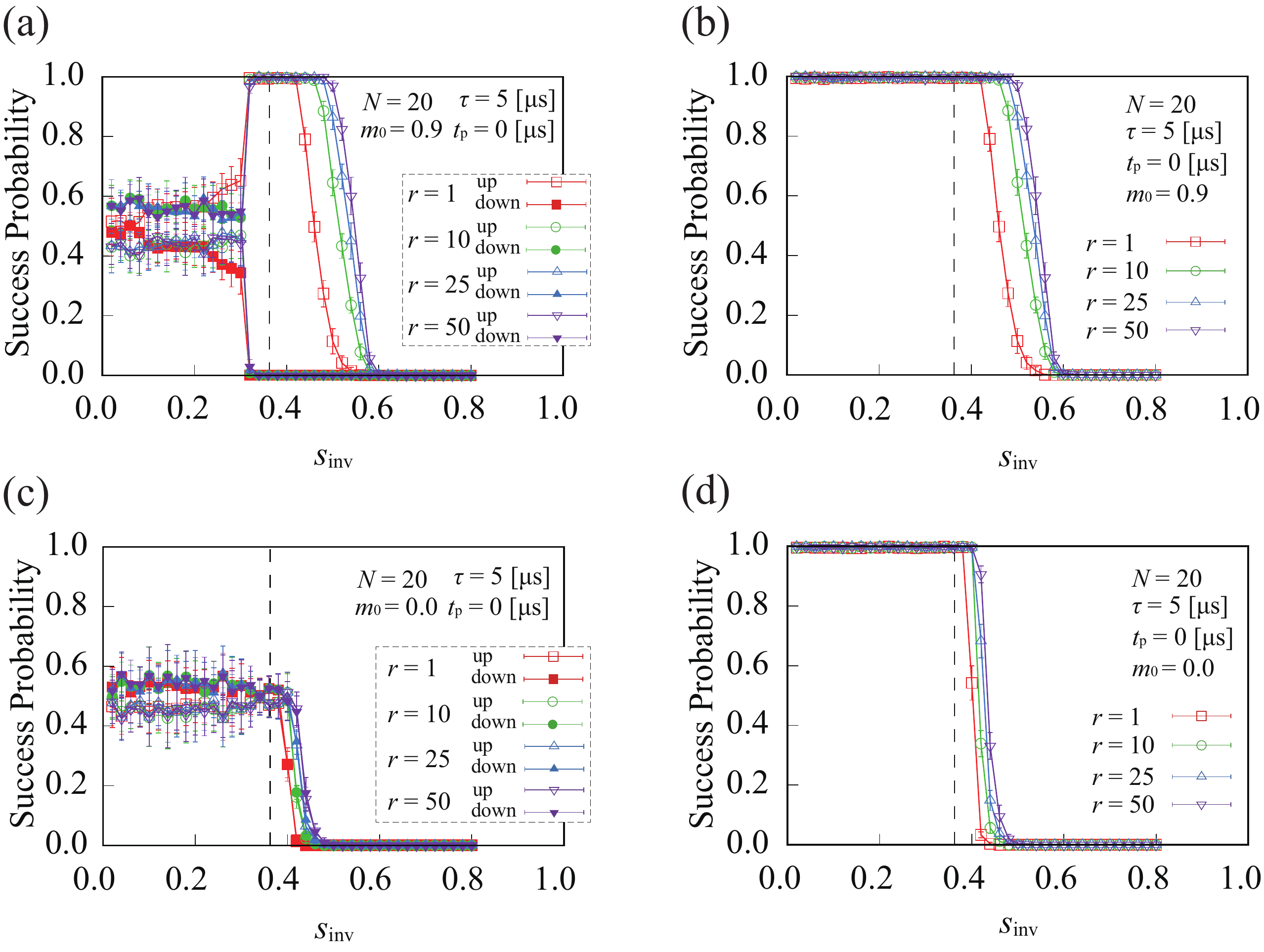}%
\caption{Empirical success probabilities for different number of iterations $r$ in iterated reverse annealing  as a function of $s_{\mathrm{inv}}$. Panels (a) and (b) are the partial and total success probabilities with the initial state $m_0=0.9$ (the first excited state), respectively. Panels (c) and (d) are the partial and total success probabilities with the initial state $m_0=0$ (the highest excited state). The dashed line is the minimum-gap point $s_{\Delta}\approx 0.36$ for $N=20$.}
\label{fig:IRA}
\end{figure}

We next study the effects of iteration on reverse annealing, i.e., how the success probability depends on the number of iterations $r$. For this purpose, we used the {\tt reinitialize\_state=False} setting of the D-Wave device, meaning that the output state of the previous iteration was the initial state of the next iteration. Figure~\ref{fig:IRA} shows the results for $r\in\{1, 10, 25, 50\}$ as a function of $s_{\mathrm{inv}}$, with $N=20$, $m_0=0.9$, $\tau=5~\mu s$, and no pausing.
The success probabilities improve in the region $s_{\mathrm{inv}}\geq s_{\rm \Delta}$ as the number of iterations $r$ increases, regardless of the initial state $m_0$. We expect the relaxation to the low-energy state due to coupling to the environment to be induced by successive iterations, and the occupation of the ground states to increase correspondingly. The results in Fig.~\ref{fig:IRA} confirm this expectation, in that the success probability is larger at given $s_{\mathrm{inv}} \ge s_\Delta$ as $r$ is increased.

Comparing Figs.~\ref{fig:IRA}(a) and (b) with $m_0=0.9$ and Figs.~\ref{fig:IRA}(c) and (d) with $m_0=0$, we can see that the initial state with $m_0=0.9$ has a larger improvement in success probability with fewer iterations $r$ in the region where $s_{\mathrm{inv}} \geq s_{\Delta}$. The initial state with $m_0 = 0$ deviates greatly from the ground states, and there are many excited states between it and the latter. Therefore, when the system transitions from the initial state of $m_0 = 0$ to other states by relaxation due to coupling to the environment, those excited states become populated. Many iterations are expected to be necessary to obtain a significant improvement in the success probability. On the other hand, there are only a few excited states between the initial state $m_0\lesssim 1.0$ and the all-up ground state. Therefore, it is expected that only a few iterations will be sufficient to make the transition to this ground state. Our results support this picture of improved performance under iterated reverse annealing.

\section{Numerical Results}
\label{sec:simulations}

We next present closed and open-systems simulations and compare the results with the data presented in the previous section. We choose relatively small system sizes $N=4$ and $8$ to facilitate numerical computations. 
Because our experimental data do not show a strong size dependence, as discussed in Sec.~\ref{sec:N}, the qualitative comparison our numerical results enable should suffice to draw relevant physical conclusions.

In all our numerical studies we simulate the logical problem directly, i.e., we do not simulate the embedded problem which replaces every logical spin by a ferromagnetic chain, as in our D-Wave experiments.

For simplicity, we focus on reverse annealing without pausing (i.e., $t_p=0$) in this section. We expect that in general pausing will lead to an overall success probability increase in open-system simulations.

\subsection{Closed system model}
An analysis of the closed system case, while being unrealistic due to the absence of thermal effects, is instrumental in isolating the effect and importance of diabatic transitions in explaining our experimental results.

The time evolution of reverse annealing in a closed system is described as follows.
The state after a single iteration (cycle) is
\begin{equation}
\ket{\psi(2t_{\text{inv}})} = U(2t_{\text{inv}},0)\ket{\psi(0)}\,,
\end{equation}
where $\ket{\psi(0)}$ is the initial state and
\begin{equation}
    U(2t_{\text{inv}}, 0) = {\cal T}\exp\left[-i\int_0^{2t_{\text{inv}}}H(t')dt'\right]
\end{equation}
is the unitary time-evolution operator and ${\cal T}$ denotes forward time ordering. 
At the end of $r$ cycles, the final state $\ket{\psi(2rt_{\text{inv}})}$ can be expressed as:
\begin{equation}
\ket{\psi(2rt_{\text{inv}})} = U(2rt_{\text{inv}}, 0)\ket{\psi(0)}\,,
\end{equation}
with
\begin{equation}
    U(2rt_{\text{inv}}, 0) = \prod_{i=0}^{r-1}U(2(i+1)t_{\text{inv}}, 2it_{\text{inv}}) \,.
\end{equation}
Here, the final state of the $r$th cycle is the initial state of the next cycle. This condition is shared by the experiments in the previous section. The solution states are doubly degenerate, and the total success probability at the end of $r$ cycles is
\begin{equation}
p(r)=|\braket{\psi(2rt_{\text{inv}})|\text{up}}|^2 + |\braket{\psi(2rt_{\text{inv}})|\text{down}}|^2 \,,
\end{equation}
where $\ket{\text{up}} = \ket{\uparrow}^{\otimes N}$ and $\ket{\text{down}} = \ket{\downarrow}^{\otimes N}$.
For higher computational efficiency without loss of accuracy, we rotate the state vector into the instantaneous energy eigenbasis representation at each time step in our numerical simulations. 

\subsubsection{Dependence on system size and annealing time}

We initialize the state to have a single spin down, i.e., as the computational basis state $\ket{0001}$ [$\ket{0} \equiv \ket{\uparrow}$, $\ket{1} \equiv \ket{\downarrow}, (m_0=0.5)$] for $N=4$ and $\ket{0000001}~(m_0=0.75)$ for $N=8$, respectively.
Note that in our simulations these are not exact eigenstates of $H(1)$, due to a very small residual transverse field at $s=1$, as in the D-Wave annealing schedule shown in Fig.~\ref{fig:schedule}.\footnote{$A(1) = 1.9\times 10^{-6}$\;GHz, $B(1) = 11.97718$\;GHz, $A(1)/B(1) = 1.58635004\times 10^{-8}$.}
When the initial state $\ket{\psi(0)}$ is a computational basis state, there is a simple upper bound on the success probability achievable in closed system reverse annealing: the population of the initial state in the maximum-spin sector (see App.~\ref{append:successbound}). This upper bound explains why the following closed system results have relatively low success probabilities.

\begin{figure}[t]
\subfigure[]{\includegraphics[width=0.49\columnwidth]{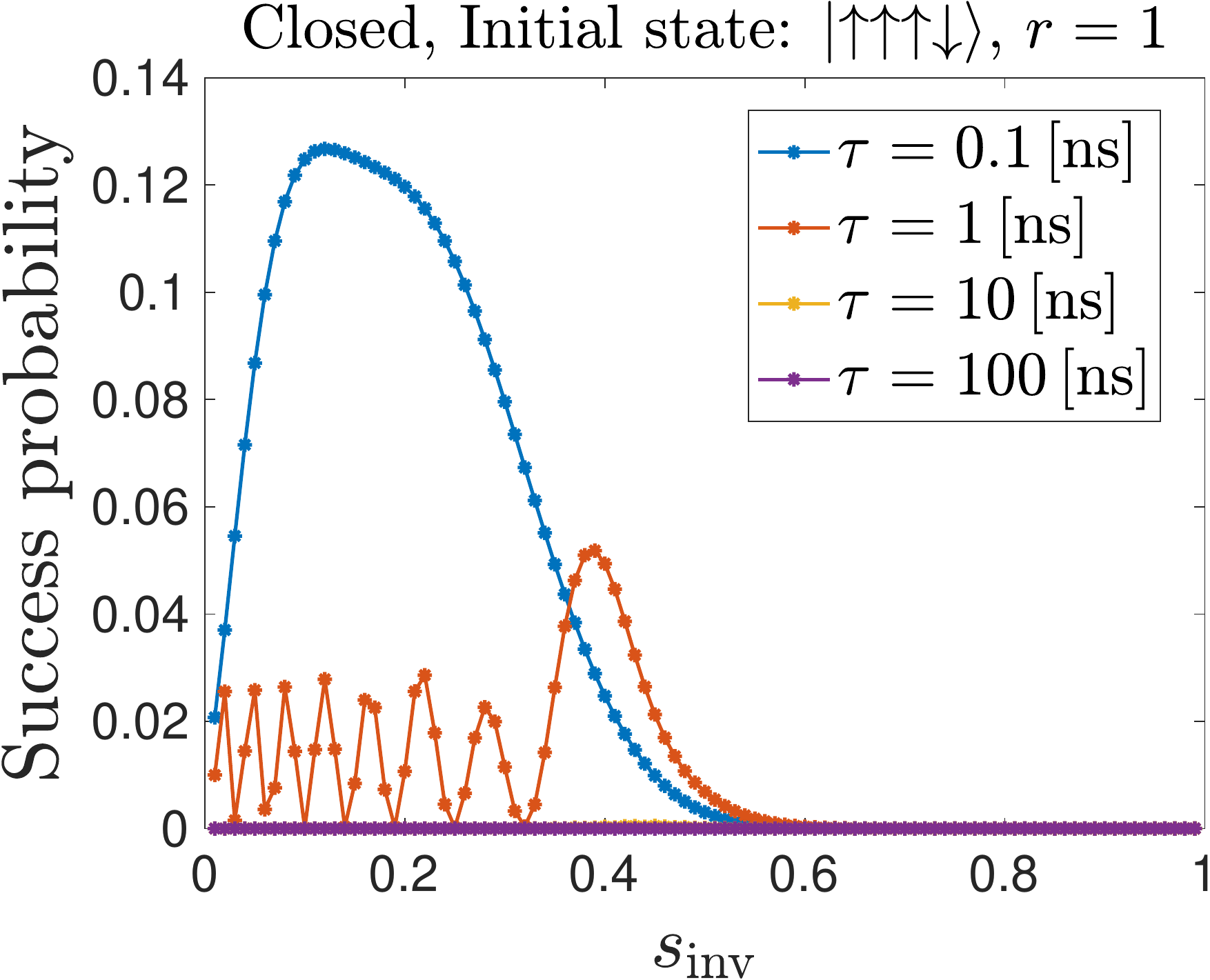}\label{fig:4qubitsunitary}}
\subfigure[]{\includegraphics[width=0.4939\columnwidth]{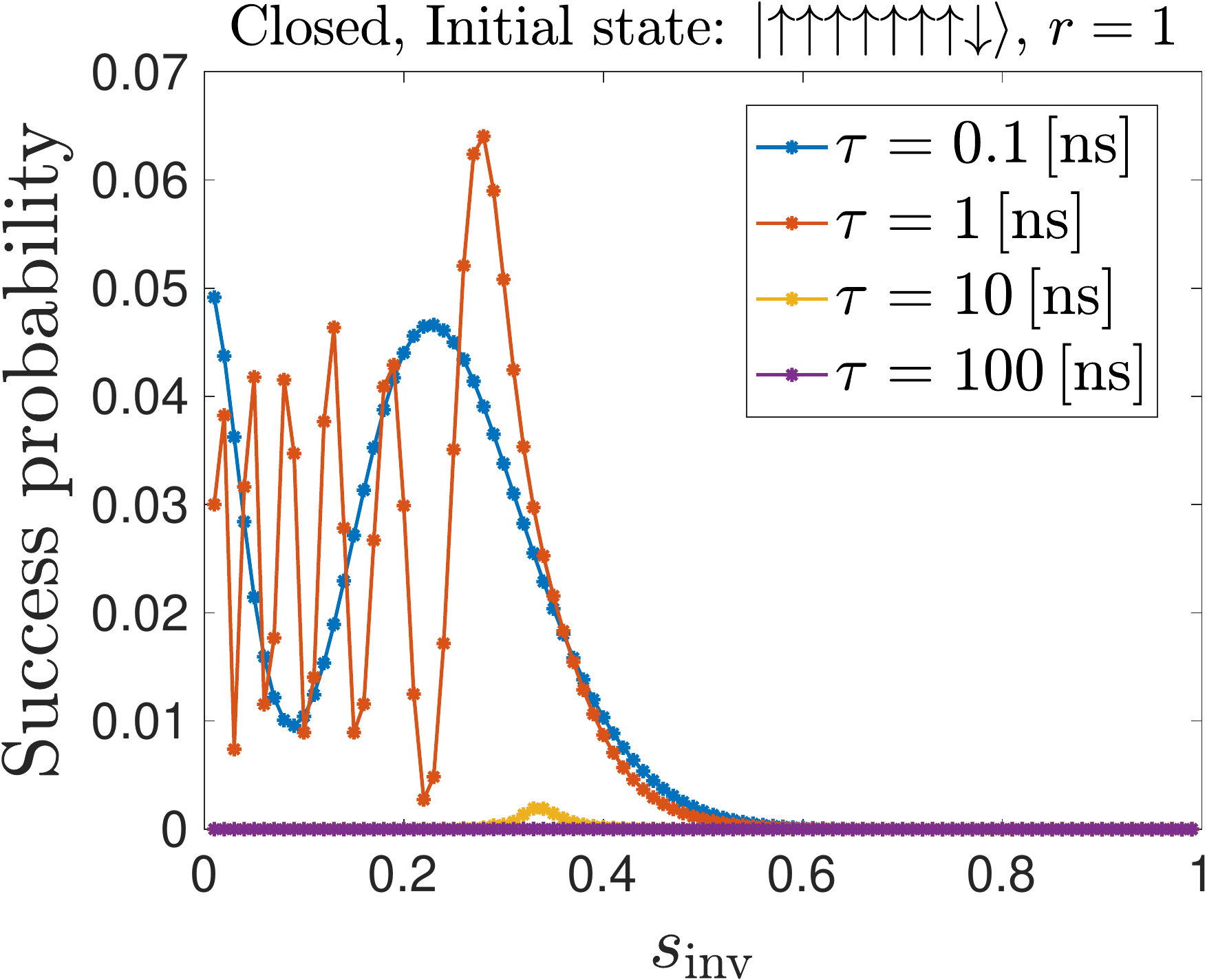}\label{fig:8qubitsunitary}}
\caption{Total success probabilities as computed by solution of the Schr\"odinger equation for a closed system with $r=1$ (single cycle) and a single-spin down as the initial state. (a) $N=4$, (b) $N=8$. Here and in the subsequent figures the nanoscecond timescale is set by the energy scale of the D-Wave annealing schedule shown in Fig.~\ref{fig:schedule}. Note the different scale of the vertical axes.}
\label{fig:close0001_r1}
\end{figure}

We plot in Fig.~\ref{fig:close0001_r1} the simulation results for the total success probability for various annealing rates, subject to the D-Wave annealing schedule shown in Fig.~\ref{fig:schedule}. We see that the success probability after a single cycle is non-negligible only when $\tau$ is small enough ($\tau < 1$\;ns), in which case diabatic transitions to states with lower energies take place when $s_{\mathrm{inv}}< s_\Delta$. However, the success probability remains small even in this case, and in any case much smaller than in our experimental results where thermal relaxation plays the dominant role.

\subsubsection{Dependence on initial state and number of iterations}
\label{sec:closed2}

We next choose the initial state $\ket{\psi(0)}$ with two spins down $\ket{0011}~(m_0=0)$ for $N=4$ and  $\ket{00000011}~(m_0=0.5)$ for $N=8$, respectively, i.e., the second excited states of $H_T$ for these respective system sizes.

\begin{figure}[t]
\subfigure[]{\includegraphics[width=0.4939\columnwidth]{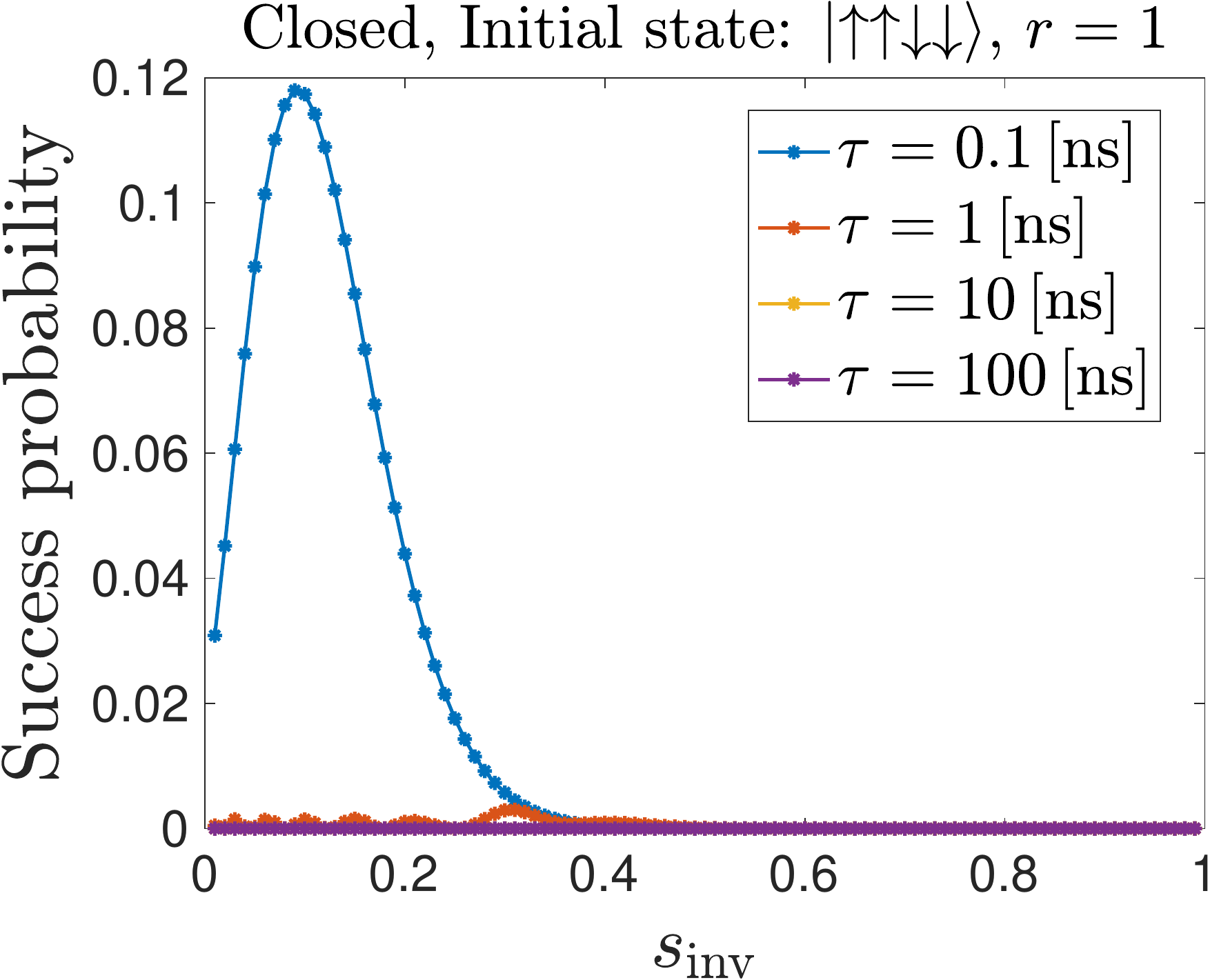}\label{fig:4qubitsunitary2}}
\subfigure[]{\includegraphics[width=0.4939\columnwidth]{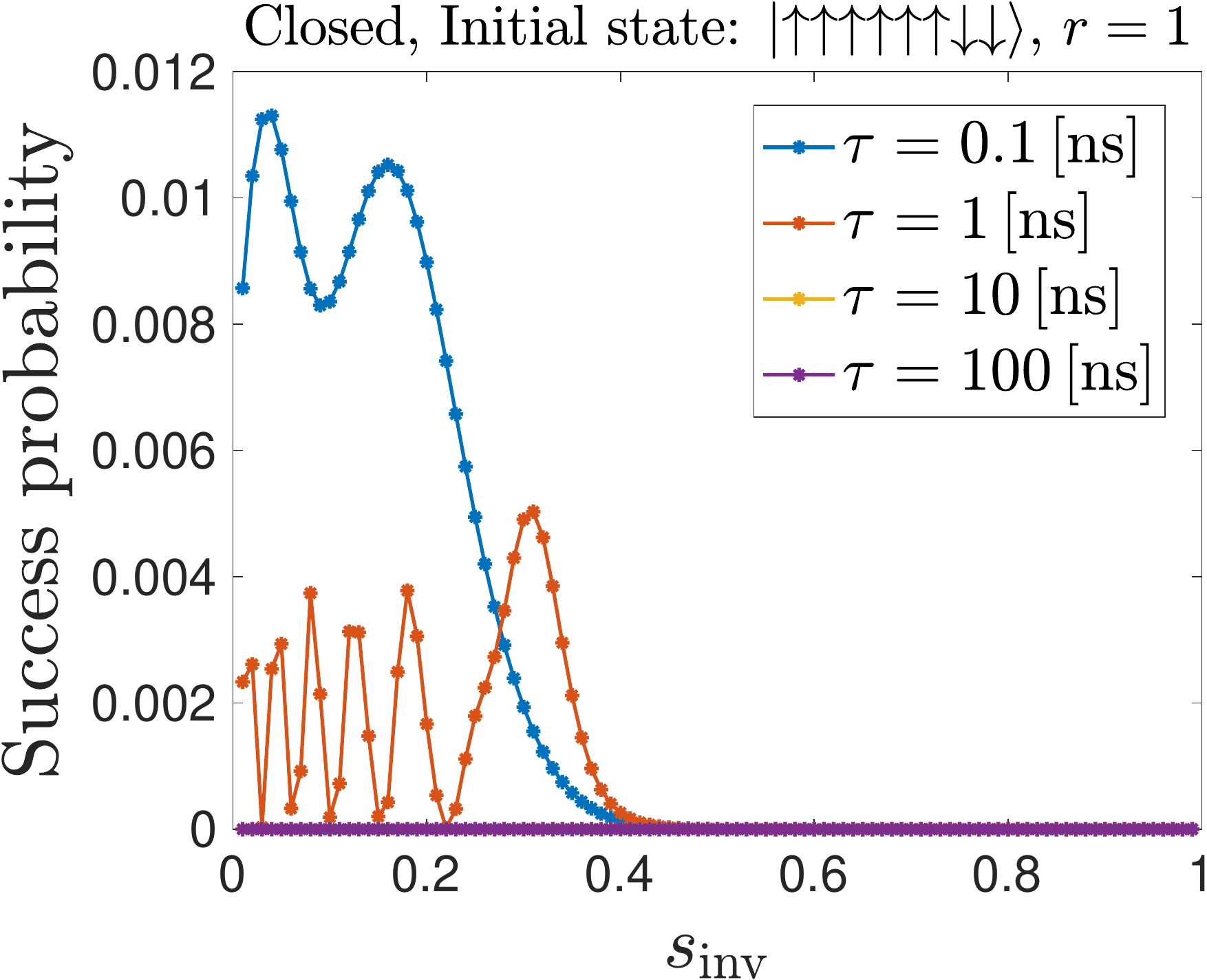}\label{fig:8qubitsunitary2}}
\subfigure[]{\includegraphics[width=0.4939\columnwidth]{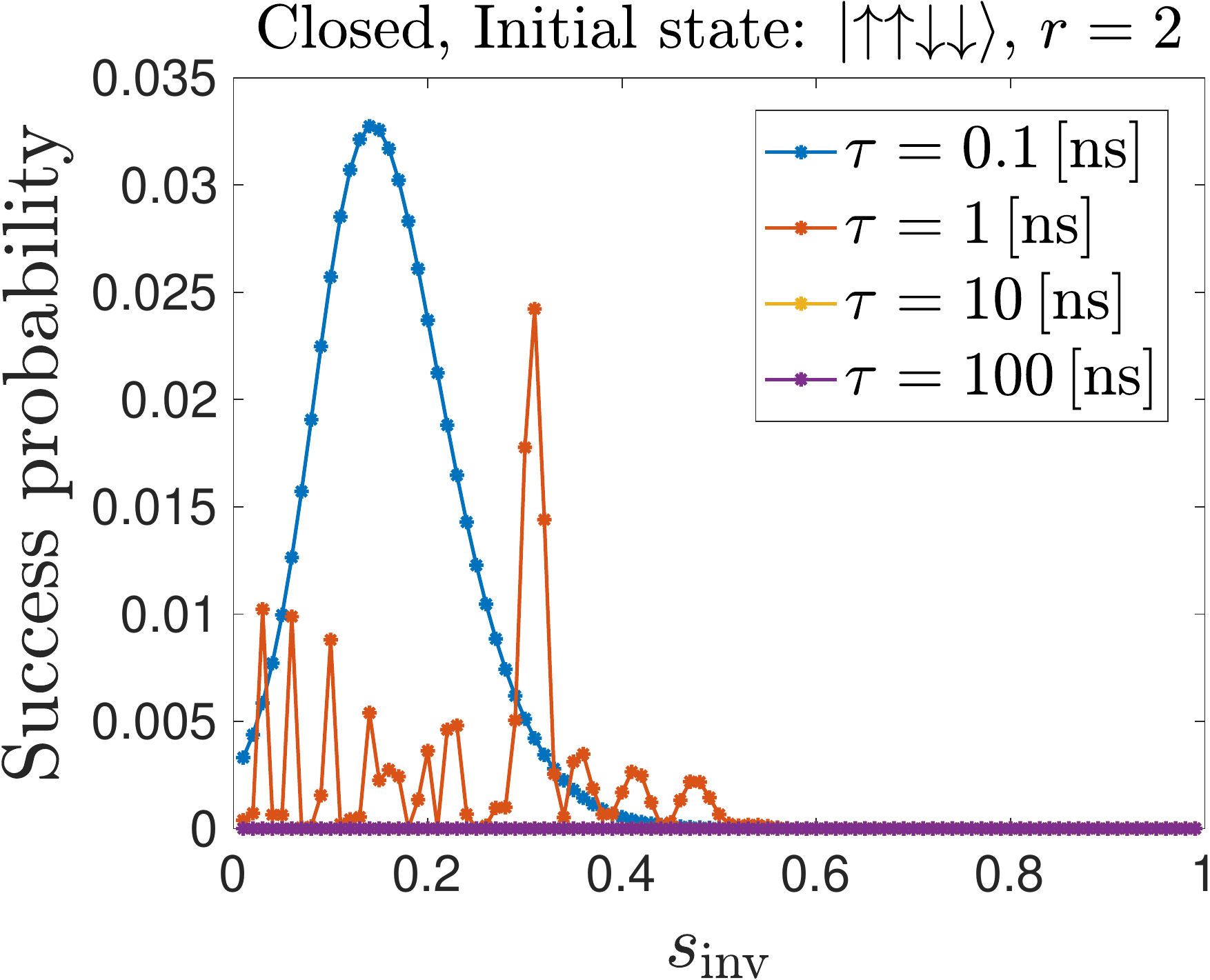}\label{fig:4qubitsunitary3}}
\subfigure[]{\includegraphics[width=0.4939\columnwidth]{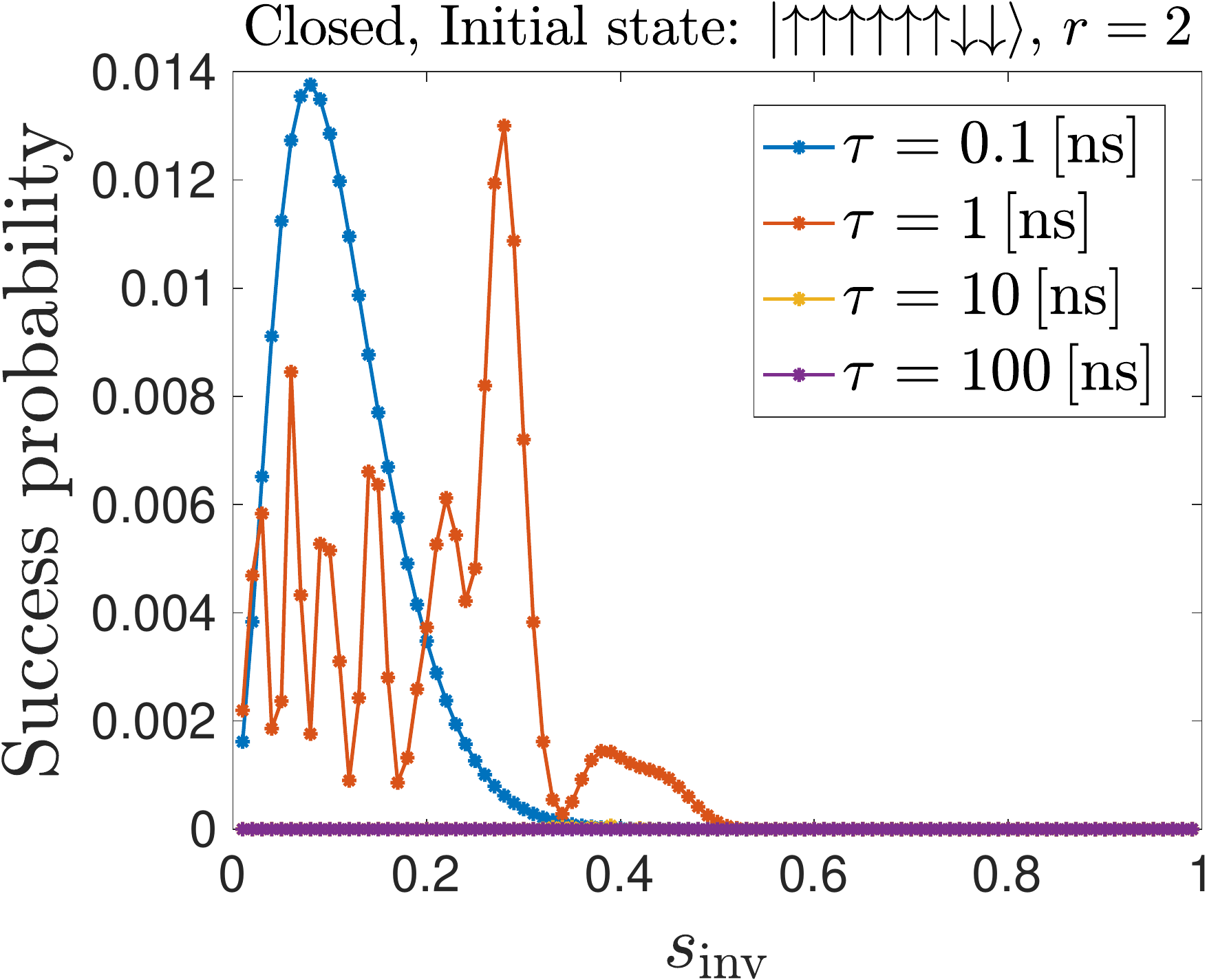}\label{fig:8qubitsunitary3}}
\caption{Total success probabilities as computed by solution of the Schr\"odinger equation for a closed system with two spins down as the initial state. (a) $N=4$ and $r=1$, (b) $N=8$ and $r=1$, (c) $N=4$ and $r=2$, (d) $N=8$ and $r=2$. Note the different scale of the vertical axes.}
\label{fig:close0011_r1}
\end{figure}

Figure~\ref{fig:close0011_r1} (top row) shows that as expected, for $N=4$ and $r=1$, the success probability is overall lower than the case when the initial state is the first excited state, Fig.~\ref{fig:close0001_r1}. Only for the highest annealing rate (smallest $\tau$, most diabatic) is the maximum success probability similar to the case of the first excited state. For $N=8$, the success probability is much smaller, no matter how diabatic the process is. This indirectly confirms the dominant role played by thermal effects in our experimental results.

We plot in Fig.~\ref{fig:close0011_r1} (bottom row) the results for $r=2$ cycles, where we see that the success probability decreases compared to the $r=1$ case. This is consistent with the conclusions of Ref.~\cite{Yamashiro:2019aa}. 

Recalling the experimental results reported in the previous section, we may therefore safely conclude that, as expected, the closed-system picture is far from the experimental reality of the D-Wave device.

\subsection{Open system model: setup}

For an open system, the state after the first cycle is
\begin{equation}
\rho(2t_{\text{inv}}) = V(2t_{\text{inv}},0)\rho(0) \,,
\end{equation}
where 
\begin{equation}
    V(t,0) = {\cal T}\exp\left[\int_0^{t}\mathcal{L}(t')dt'\right] \,.
\label{eq:vpropagator}    
\end{equation}
$\mathcal{L}(t)$ is the time-dependent Liouville superoperator, which generates a master equation of the form
\bes
\label{eq:15}
\begin{align}
	\frac{d\rho(t)}{dt} 
	&= \mathcal{L}(t)\rho(t)\\
	&= i \bigl[\rho(t), H(t) + H_\text{LS}(t)\bigr] + \mathcal{D}\bigl[\rho(t)\bigr] \,.
\end{align} 
\ees
Here, $ H_\text{LS}(t) $ is the Lamb shift term and $\mathcal{D}$ is the dissipator. Again, the time dependence of the integrand in Eq.~\eqref{eq:vpropagator} is incorporated into $s(t)$.

At the end of $r$ cyles, the final state $\rho(2rt_{\text{inv}})$ is expressed as:
\begin{equation}
\rho(2rt_{\text{inv}}) = V(2rt_{\text{inv}}, 0)\rho(0)\,,
\end{equation}
with
\begin{equation}
    V(2rt_{\text{inv}}, 0) = \prod_{i=0}^{r-1}V(2(i+1)t_{\text{inv}}, 2it_{\text{inv}}) \,.
\end{equation}

We next consider both the weak and the strong coupling cases, which give rise to different Liouville superoperators.

\subsection{Weak coupling: adiabatic master equation}
\label{sec:AME}

In this subsection we use the adiabatic master equation (AME), which holds under weak coupling to the environment~\cite{Albash2012}. In this limit $\mathcal{D}$ can be expressed in a diagonal form with Lindblad operators $L_{i,\omega}(t)$:
\begin{eqnarray}
\label{eq:dissdiag}
{\mathcal{D}}_{}  [\rho(t)]  & = &\sum_{i}\sum_{\omega}\gamma_i(\omega)\left(
L_{i,\omega}(t) \rho(t) L^\dagger_{i,\omega}(t)\phantom{\frac{1}{2}} \right. \notag \\
&& \hspace{-0.5cm} \left. \qquad\qquad - \frac{1}{2} \left\{ L_{i,\omega}^{\dagger}(t) L_{i,\omega}(t) , \rho(t) \right\} \right) \,,
\end{eqnarray}
where the summation runs over the qubit index $i$ and the Bohr frequencies $\omega$ [all possible differences of the time-dependent eigenvalues of $H(t)$]. This dissipator expresses decoherence in the energy eigenbasis~\cite{Albash:2015nx}: quantum jumps occur only between the eigenstates of $H(t)$~\cite{Yip2018}.

We consider two different models of system-bath coupling: independent and collective dephasing~\cite{LidarWhaley:03}. 
In the first case, we assume that the qubit system is coupled to independent, identical bosonic baths, with the bath and interaction Hamiltonians being 
\begin{subequations}
\begin{align}
\label{eq:SBm}
H_B &= \sum_{i=1}^N \sum_{k=1}^\infty \omega_k b_{k,i}^\dagger b_{k,i} \ ,  \\
H_{SB}^{\text{ind}} &= g \sum_{i=1}^N
\sigma_i^z \otimes \sum_k \left(b_{k,i}^\dagger + b_{k,i} \right) \ ,
\end{align}
\end{subequations}
where $b_{k,i}^\dagger$ and $b_{k,i}$ are, respectively, the raising and lowering operators for the $k$th oscillator mode with natural frequency $\omega_k$.  The rates appearing in Eq.~\eqref{eq:dissdiag} are given by
\begin{equation}
\gamma_{i}(\omega) = 2\pi \eta g^2\frac{\omega e^{-|\omega|/\omega_c}}{1-e^{-\beta\omega}}  \ ,
\label{eq:Ohmic}
\end{equation}
arising from an Ohmic spectral density, and satisfying the Kubo-Martin Schwinger (KMS) condition~\cite{kubo1957statistical, martin1959theory} $\gamma_{i}(- \omega) = e^{-\beta\omega} \gamma_{i}(\omega)$, with $\beta = 1/T$ the inverse temperature.  We use $\eta g^2 = 10^{-3}$, the cutoff frequency $\omega_c = 1$\;THz, and the D-Wave device operating temperature $T= 12.1 \text{mK} = 1.57$\;GHz. With the independent dephasing assumption, the Lindblad operators are:
\begin{equation}
   L_{i,\omega}(t) = \sum_{\epsilon_b-\epsilon_a = \omega} | \varepsilon_a(t) \rangle \langle \varepsilon_a(t) | \sigma^z_{i} | \varepsilon_b(t) \rangle \langle \varepsilon_b(t) | \,,
   \label{eq:21}
\end{equation}
corresponding to dephasing in the instantaneous eigenbasis $\{|\varepsilon_a(t) \rangle\}$ of $H(t)$. Similarly to the closed system case, we rotate the density matrix and Lindblad operators into the instantaneous energy eigenbasis at each time step in our numerical simulations. This keeps the matrices sparse, without loss of accuracy. For $N = 8$ we truncate the system size to the lowest $n=18$ levels out of $256$ [$18 = 2(1+8)$: the total number of degenerate ground and first excited states at $s=1$].

We also consider the collective dephasing model, where all the qubits are coupled to a collective bath with the same coupling strength $g$, which preserves the spin symmetry. In this case, the interaction Hamiltonian becomes
\begin{align}
    H_{SB}^{\text{col}} = g S^z \otimes B,
\end{align}
where
\begin{align}
    S^z = \sum_{i} \sigma_i^z, \quad B = \sum_k \left(b_{k}^\dagger + b_{k} \right).
\end{align}
With this assumption, we can group together the Lindblad operators corresponding to different qubits $i$ into a single one:
\begin{equation}
\label{eq:AME-Lind}
   L_{\omega}(t) = \sum_{\epsilon_b-\epsilon_a = \omega} | \varepsilon_a(t) \rangle \langle \varepsilon_a(t) | S^z | \varepsilon_b(t) \rangle \langle \varepsilon_b(t) | \,.
\end{equation}
The resulting number of Lindblad operators is a factor of $N$ smaller than that of the independent system-bath coupling model.

The total success probability at the end of the $r$ cycles is
\begin{equation}
p(r) = \bra{\text{up}}\rho(2rt_{\text{inv}})\ket{\text{up}} + \bra{\text{down}}\rho(2rt_{\text{inv}})\ket{\text{down}} \,.
\end{equation}

Any relaxation during the reverse annealing dynamics to the global instantaneous ground state or the instantaneous first excited state of $H(t)$ is beneficial, the latter since it becomes degenerate with the ground states $\{\ket{\text{up}},\ket{\text{down}}\}$ of $H_T$ at the end of the anneal (see App.~\ref{append:spectrum}).

\subsubsection{Dependence on the annealing time.}
\label{sec:AME-tau}

\begin{figure}[t]
\subfigure[]{\includegraphics[width=0.4939\columnwidth]{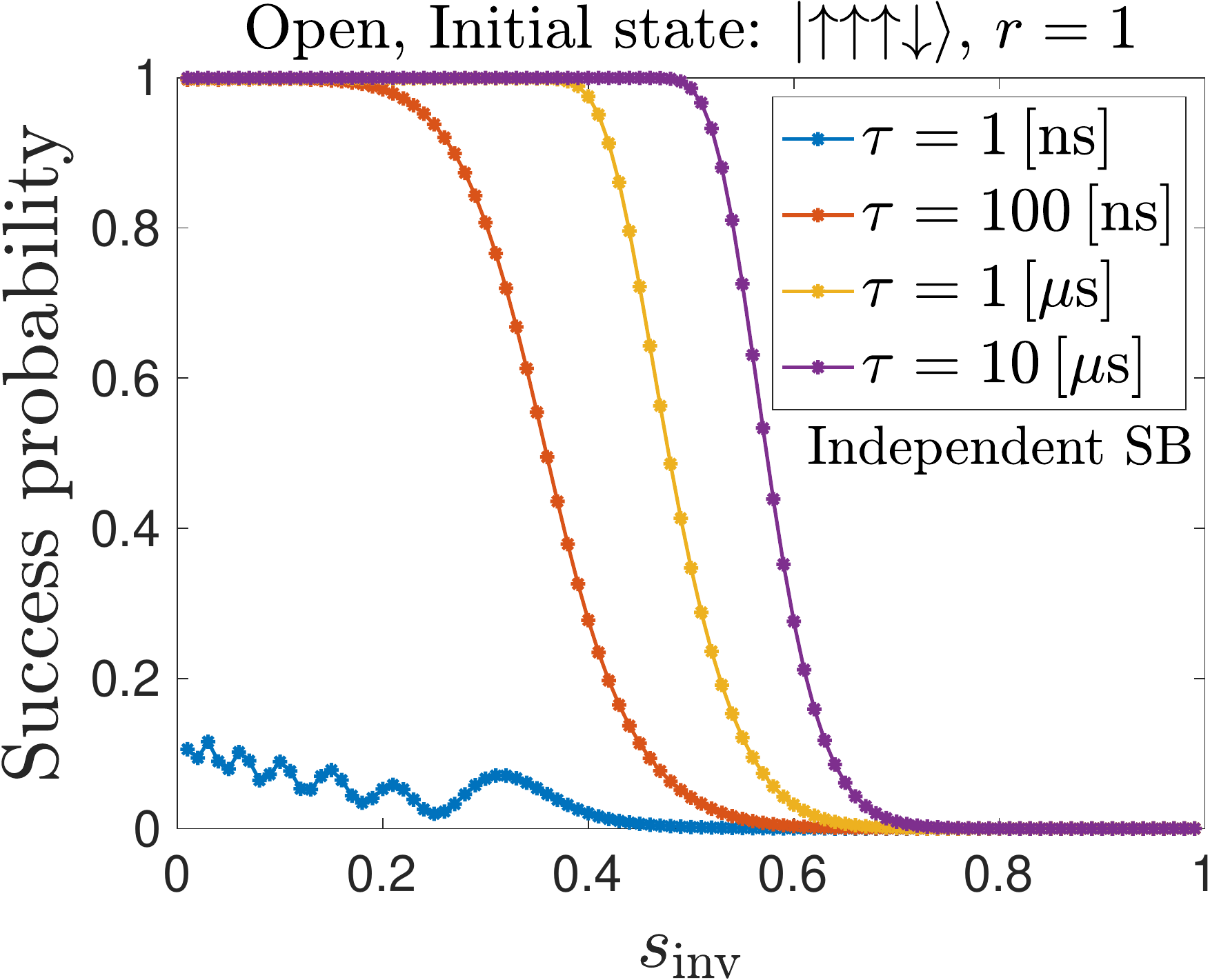}\label{fig:4qubitopena}}
\subfigure[]{\includegraphics[width=0.4939\columnwidth]{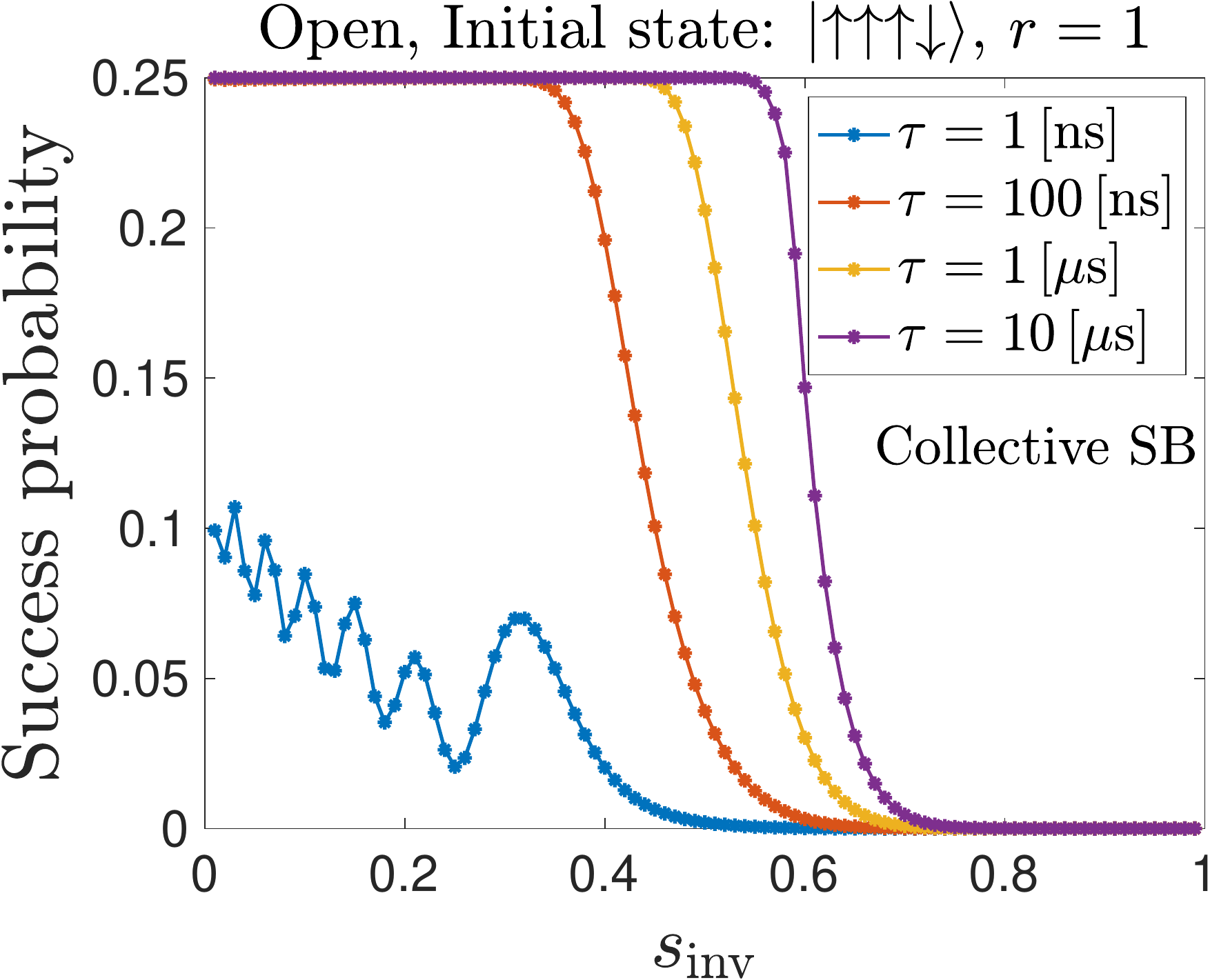}\label{fig:4qubitopenb}}
\subfigure[]{\includegraphics[width=0.4939\columnwidth]{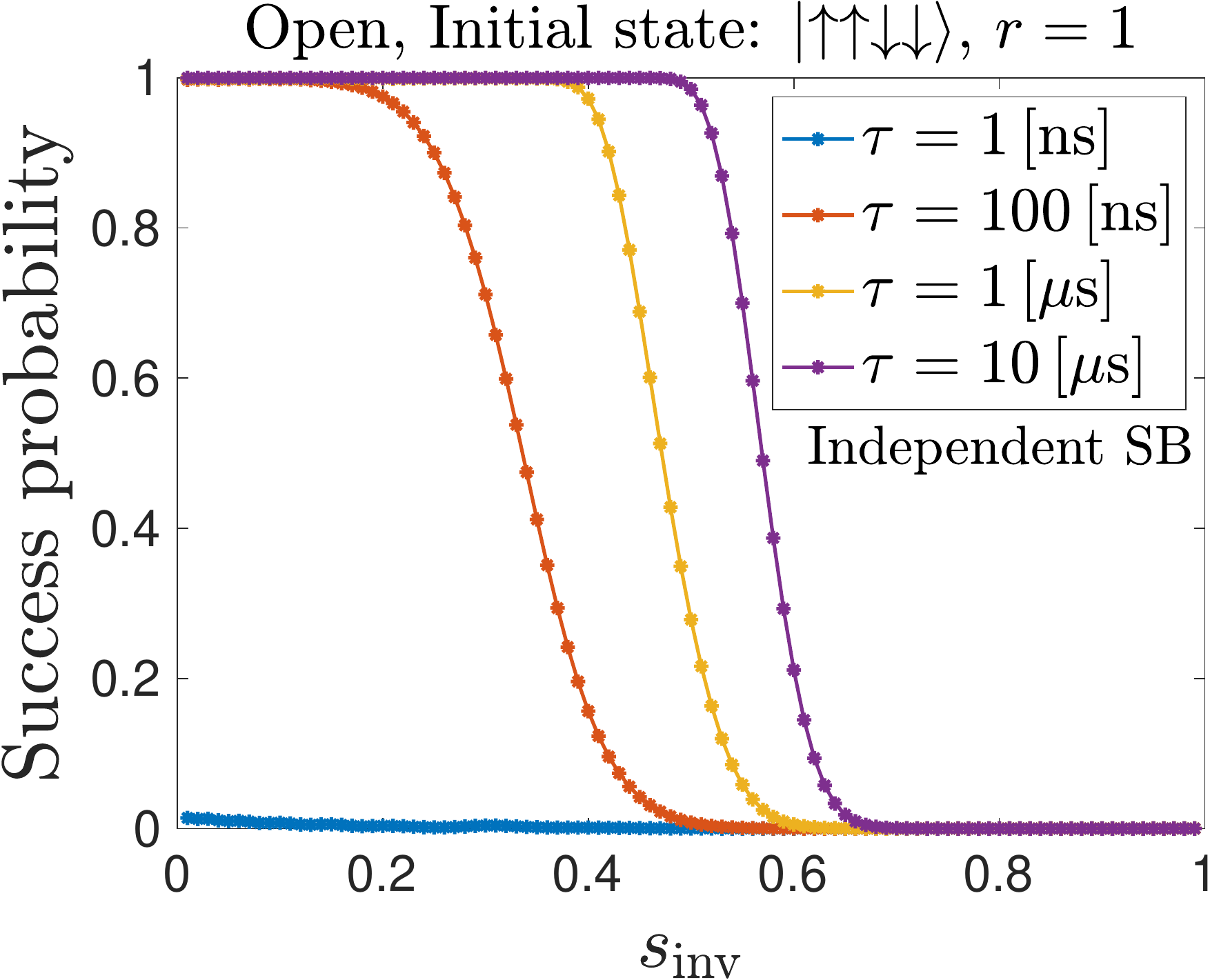}}
\subfigure[]{\includegraphics[width=0.4939\columnwidth]{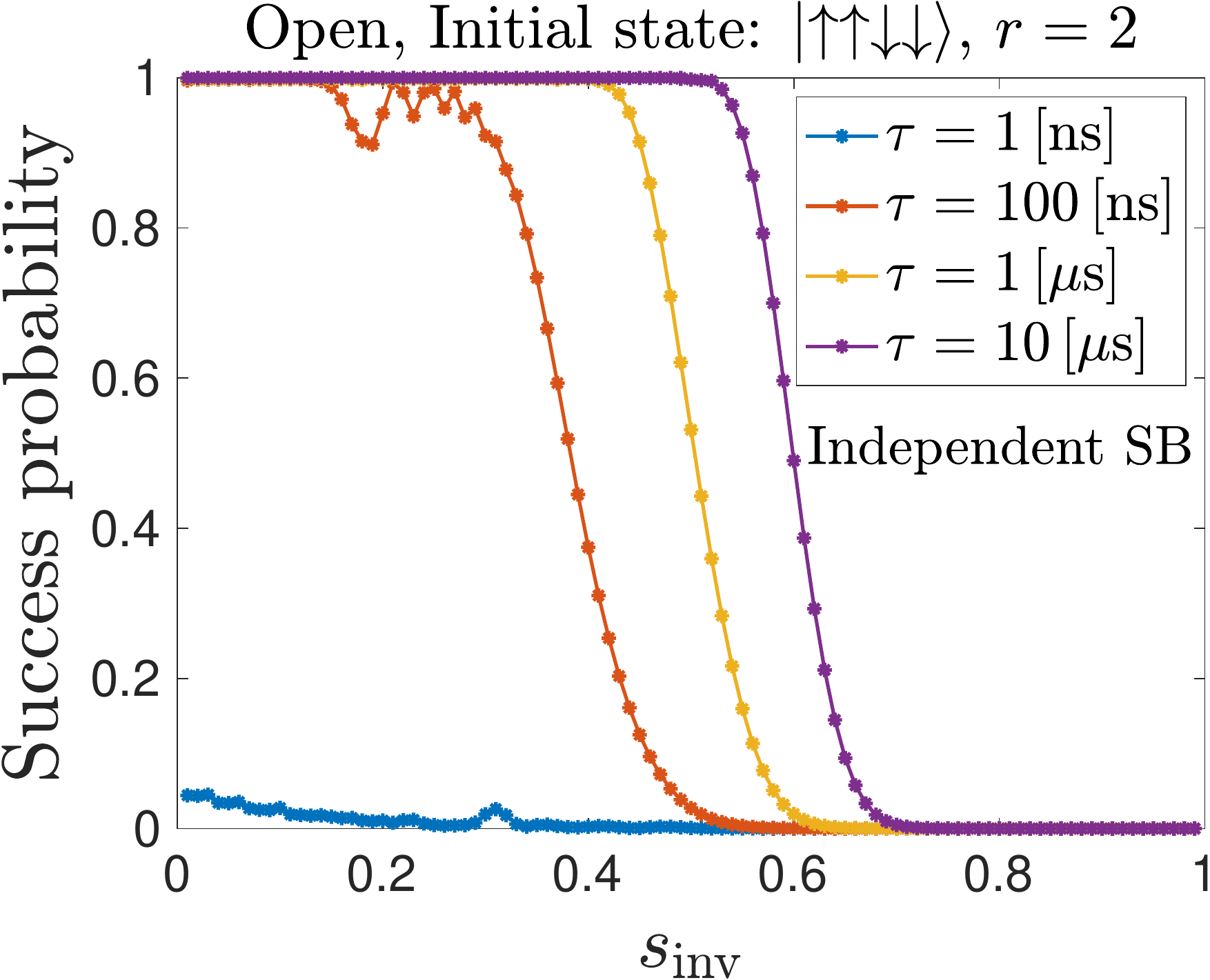}}
\caption{Top row: Total success probabilities as computed by the AME as a function of $s_{\mathrm{inv}}$ with (a) independent and (b) collective dephasing (SB denotes system-bath coupling). Initial state: $\ket{0001}~(m_0=0.5)$ and $r=1$, a single cycle. Bottom row: Total success probabilities as computed by the AME with different $\tau$. Initial state: $\ket{0011}~(m_0=0)$. 
(c)
$r=1$, 
(d)
$r=2$. Note the different scale of the vertical axis of panel (b).}
\label{fig:4open_0001_r1}
\end{figure}

Figure~\ref{fig:4open_0001_r1} (top row) shows the AME simulation results for the success probability as a function of $s_{\mathrm{inv}}$ with $N=4$, various $\tau$, and the initial state $\ket{0001}~(m_0=0.5)$, using the independent and collective dephasing models.

The independent dephasing model leads to a maximum success probability of $1$ for large $\tau$ and small $s_{\mathrm{inv}}$. However, the collective dephasing model leads to a maximum success probability of $1/4$. The reason is the same as in the closed system simulations. For the initial state $\ket{0001}$, only $1/4$ of its population is in the subspace of maximum quantum spin number. 
However, the global (instantaneous) ground state and the first excited state of the annealing Hamiltonian $H(s)$ both belong to the maximum-spin subspace. Since the collective dephasing model preserves the spin symmetry and the dynamics is restricted to each subspace, at most $1/4$ of the initial population can be relaxed to the instantaneous ground state and the first excited state, and reach the correct solutions at the end of the anneal (see App.~\ref{append:successbound} for more details). It is also noteworthy that coherent oscillations are visible for $\tau=1$\;ns [compare with Fig.~\ref{fig:close0001_r1}(a)], but not for the larger values of $\tau$ we have simulated. Recall that the experimental timescale in Sec.~\ref{sec:exp} is on the order of a $\mu$s.

Comparing the simulation results in Fig.~\ref{fig:4open_0001_r1} and the experimental results in Fig.~\ref{fig:RA_20_diff_tau}(b), we observe that they share the same main features. Namely, the success probability increases with $\tau$; and, as $\tau$ increases the maximum success probability can be achieved with a larger inversion point $s_{\text{inv}}$. Most notably, the total success probability also drops to zero for sufficiently large $s_{\text{inv}}$ in our simulations. This is the freeze-out effect that is well captured by the adiabatic master equation (as reported in Ref.~\cite{Albash2012}), since  thermal relaxation is suppressed when the transverse field magnitude becomes so small that the system and system-bath Hamiltonians effectively commute.

Note that the experimental results in Sec.~\ref{sec:exp} have a maximum success probability as high as $1$, while our closed system simulations and open-system simulations with the collective dephasing model have success probabilities upper bounded by some constants $<1$, as already discussed. The high total success probability observed in our experiments is evidence that, as expected, the dynamics in the D-Wave device do not preserve spin symmetry. Collective system-bath coupling cannot explain the experimental results, leaving independent system-bath coupling as the only candidate consistent with the experiments according to our simulations.

\subsubsection{Dependence on initial condition and the number of cycles.}
\label{sec:simulations-init}

For the initial state $\ket{0011}~(m_0=0)$ and with $r=1$, the independent dephasing model still gives a maximum success probability of $1$ as seen in Fig.~\ref{fig:4open_0001_r1}(c). The collective dephasing model (not shown) has a maximum success probability of $1/6$, since the initial state has $1/{{4}\choose{2}} = 1/6$ of its population in the maximum-spin subspace.

Comparing Fig.~\ref{fig:4open_0001_r1}(a) and (c), the dependence of the success probability on $s_{\text{inv}}$ is  similar to that of the initial state $\ket{0001}$. The $\tau = 1$\;ns coherent oscillations visible for the latter are more attenuated for $\ket{0011}$, but this was also the case for the closed system simulations [contrast Fig.~\ref{fig:close0001_r1}(a) and Fig.~\ref{fig:close0011_r1}(a) at $\tau = 1$\;ns]. 
The main feature distinguishing Fig.~\ref{fig:4open_0001_r1}(a) and (c) is the shift to the left of the $\tau \geq 100$\;ns curves for the initial state $\ket{0011}$, i.e., as $m_0$ is reduced from $0.5$ to $0$. This is consistent with our experimental results, as can be seen in Fig.~\ref{fig:RA_20_initial}(b).

In Fig.~\ref{fig:4open_0001_r1}(d), we consider the dependence on the number of cycles $r$, using the independent system-bath model. For $r=2$, we see that the results are similar to those of a single cycle. However, we do see a small improvement in the sense of a slight shift to the right of the curves with $\tau \geq 100$\;ns compared with $r=1$, which is consistent with the experimental result shown in Fig.~\ref{fig:IRA}(d). We note that this improvement was not observed in the closed system case, as can be seen by contrasting Figs.~\ref{fig:close0011_r1}(a) and (c). Interestingly, there is also a small signature of coherent oscillations for $\tau \leq 100$\;ns.

Finally, we note that compared to the closed systems case, the results depend much less on the initial condition.

\subsubsection{Size dependence and partial success probability.}

\begin{figure}[t]
\includegraphics[width=0.8\columnwidth]{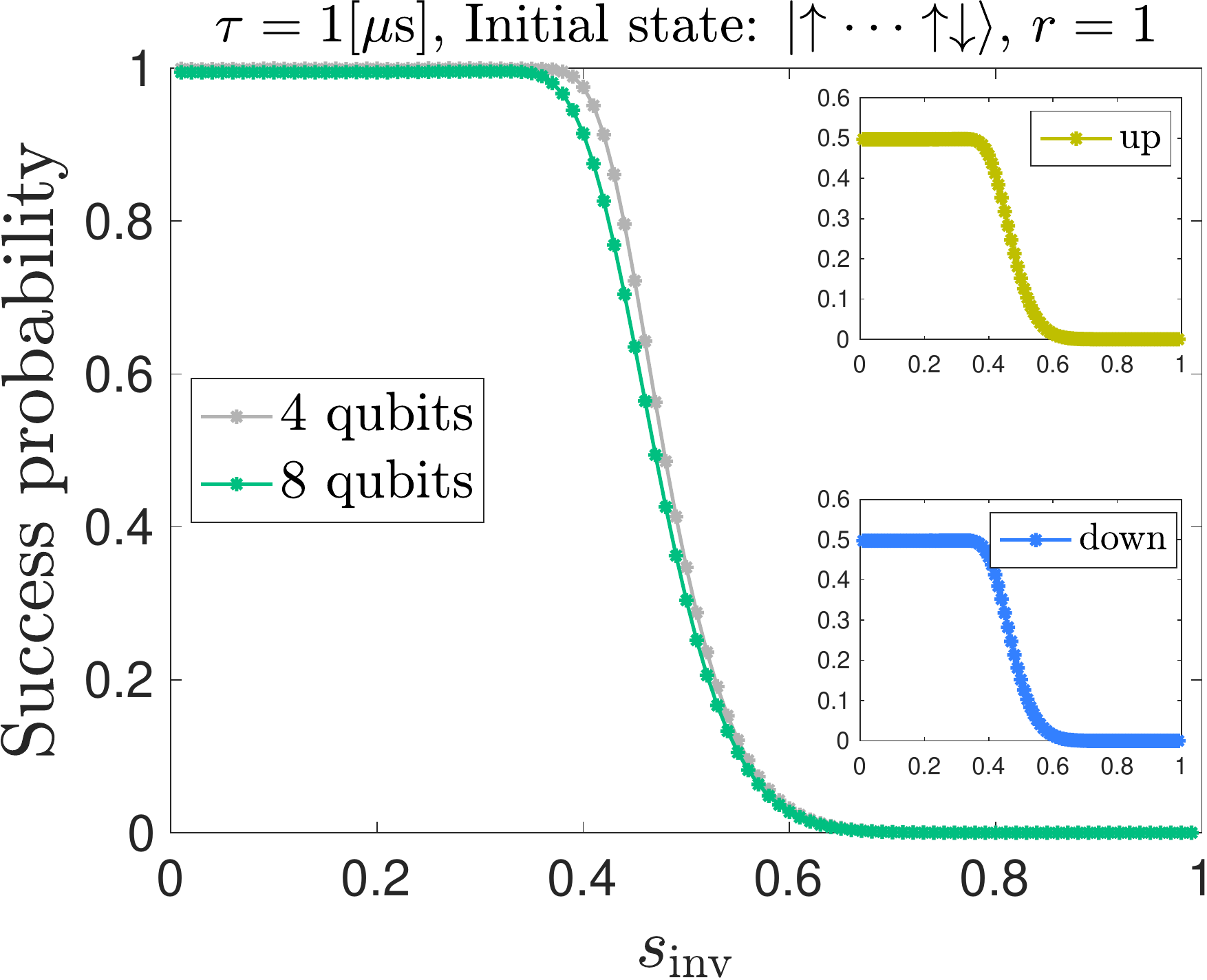}
\caption{Total and partial success probabilities as computed by the AME as a function of $s_{\mathrm{inv}}$ for independent dephasing. The initial state has one spin down and $\tau=1\;\mu$s. The main plot shows the total success probabilities for $N=4$ and $8$. The inset shows the partial success probabilities for $N=8$.}
\label{fig:4vs8open}
\end{figure}

In Fig.~\ref{fig:4vs8open}, we show the success probability for two different system sizes $N=4$ and $8$. The initial state has a single spin down. We observe that the results do not depend much on the system size, with the total success probabilities slightly larger (shifted to the right) for $N=4$. This is consistent with the experimental results for $N=4$ and $8$ shown in Fig.~\ref{fig:RA_20_size}(b).

While the total success probabilities of our open-system simulations with independent dephasing are generally consistent with the experimental total success probabilities, our open-system simulations always produce symmetric partial success probabilities as shown in Fig.~\ref{fig:4vs8open}, in stark contrast with the experiments, where the all-up state is strongly favored over the all-down state in the region around the minimum gap (see all the top panels in the figures in Sec.~\ref{sec:exp}). Clearly, this reflects a significant failure of the AME model.

One reason for this discrepancy is that our simulations do not include a mechanism such as the polarized spin-bath which we believe explains the experimentally observed asymmetry for small $s_{\mathrm{inv}}$ (see Sec.~\ref{sec:spinbathpol}). Moreover, our adiabatic master equation simulations result in a symmetric final all-up and all-down population 
since any relaxation event during the anneal (at any $s$) to either the global instantaneous ground state $\ket{\epsilon_0(s)}$ or the instantaneous first excited state $\ket{\epsilon_1(s)}$ of $H(s)$, which become degenerate at $s=1$, 
eventually contributes amplitude to $\ket{\epsilon_0(s=1)} = ({\ket{\uparrow}^{\otimes N} + \ket{\downarrow}^{\otimes N}})/{\sqrt{2}}$ or $\ket{\epsilon_1(s=1)} = ({\ket{\uparrow}^{\otimes N} - 
\ket{\downarrow}^{\otimes N}})/{\sqrt{2}}$. These two states 
have equal all-up and all-down populations, which is why the AME simulation results do not distinguish them. Notice the importance of the energy eigenbasis (i.e., the states $\ket{\epsilon_0(s)}$ and $\ket{\epsilon_1(s)}$) in this argument; this is special to the AME.

\subsection{Strong coupling: PTRE}
\label{sec:PTRE}

Given the troubling discrepancy between the empirical partial success probability results and the AME simulations, in this subsection we use the Lindblad form polaron-transformed Redfield equation (PTRE)~\cite{chen_hoqst_2020,xu_non-canonical_2016}. The idea is to check whether a quantum model that is not subject to the weak coupling limit is capable of avoiding the discrepancy. This is motivated in part by previous studies, which showed that a hybrid AME/PTRE model works better than the pure AME in explaining the linewidth broadening phenomenon observed in a tunneling spectroscopy experiment~\cite{Albash:2015pd,chen_hoqst_2020}. The PTRE holds under intermediate to strong coupling to the environment~\cite{chen_hoqst_2020}. Unlike the AME, wherein decoherence is between energy eigenstates, decoherence in the PTRE is between computational basis states (eigenbasis of $\sigma^z$). But, unlike the singular coupling limit (SCL)~\cite{Albash:2015nx}, where decoherence is also between computational basis states, the PTRE is governed by a non-flat (and hence non-trivial) spectral density.
This means that, unlike the SCL, the PTRE is sensitive to the particulars of the bath model, a fact we take advantage of below when we combine low and high frequency components into a single noise spectrum; see Eqs.~\eqref{eq:convolutional_form} and~\eqref{eq:GL-GH} below. This type of hybrid spectrum approach has been used successfully~\cite{smirnov_theory_2018} to explain a $16$-qubit quantum annealing experiment with a small gap~\cite{DWave-16q}.

To be explicit, the PTRE is given by Eq.~\eqref{eq:15} 
with the polaron-frame system Hamiltonian
\begin{equation}
    \label{eq:polaron_H}
H(t) = \frac{1}{2}B(t)H_T \ ,
\end{equation}
the polaron-frame 
dissipator
\begin{align}
\label{eq:ptre_lindblad}
    \mathcal{D}\pqty{\rho}&=\sum_{\alpha\in\Bqty{+,-}}\sum_{\omega, i} \gamma_\mathrm{p}(\omega)\Big(L_{i,\omega}^\alpha(t)\rho L_{i,\omega}^{\alpha\dagger}(t)\notag \\
    &\quad -\frac{1}{2}\Bqty{L_{i,\omega}^{\alpha\dagger}(t)L_{i,\omega}^\alpha(t), \rho}\Big) \ ,
\end{align}
and the Lamb shift term
\begin{equation}
    \label{eq:H_LS}
    H_{\mathrm{LS}}(t)=\sum_{i,\alpha,\omega} L_{i,\omega}^{\alpha\dagger}(t) L_{i,\omega}^\alpha(t) S_{\mathrm{p}}(\omega) \ .
\end{equation}
Here $\gamma_\mathrm{p}\pqty{\omega}$ and $S_\mathrm{p}(\omega)$ denote the polaron-frame noise spectrum and the corresponding  principal value integral (discussed below), and the Lindblad operators are
\begin{equation}
    \label{eq:ptre_lindblad_operator}
    L_{i,\omega}^\alpha(t) = \frac{A(t)}{2}\sum_{\varepsilon_b-\varepsilon_a=\omega}\bra{a}{\sigma^\alpha_i}\ket{b}\ketb{a}{b} \ ,
\end{equation}
where $\ket{a}$ is the eigenstate of $H(t)$ with eigenenergy $\varepsilon_a$.  Because $H(t)$ is diagonal in the $\sigma^z$ basis, the eigenstates $\ket{a}$ are classical spin states. This should be contrasted with the AME, for which the Lindblad operators involve the instantaneous eigenstates of the system Hamiltonian that includes the transverse field, as in Eq.~\eqref{eq:21}.

Because the effective Hamiltonian $H(t)+H_{\mathrm{LS}}(t)$ [Eqs.~\eqref{eq:polaron_H} and~\eqref{eq:H_LS}] and the Lindblad operators in Eq.~\eqref{eq:ptre_lindblad_operator} are diagonal in the $\sigma^z$-basis, the PTRE can be simplified into an equation involving only the diagonal components of the density matrix, as shown in App.~\ref{app:PTRE}. We use this form henceforth.

In contrast to the Ohmic noise spectrum we used for the AME [Eq.~\eqref{eq:Ohmic}], here we choose a hybrid model with a high-frequency Ohmic bath and a low-frequency component~\cite{amin_macroscopic_2008,smirnov_theory_2018}. Namely, $\gamma_\mathrm{p}(\omega)$ has a convolutional form
\begin{equation}
\label{eq:convolutional_form}
    \gamma_\mathrm{p}(\omega)=\frac{1}{2 \pi} \int_{-\infty}^{\infty} G_{\mathrm{L}}(\omega-x) G_{\mathrm{H}}(x) d{x} \ ,
\end{equation}
where
\begin{subequations}
\label{eq:GL-GH}
\begin{align}
    G_{\mathrm{L}}(\omega)&=\sqrt{\frac{\pi}{2 W^{2}}} \exp \left[-\frac{\left(\omega-4 \epsilon_{L}\right)}{8 W^{2}}\right] \label{eq:G_low}\\
    G_{\mathrm{H}}(\omega)&=\frac{4 \gamma(\omega)}{\omega^{2}+4 \gamma^{2}(0)} \label{eq:G_high}
\end{align}
\end{subequations}
which describe the low and high-frequency components, respectively.
$W$ and $\varepsilon_L$ in Eq.~\eqref{eq:G_low} are known as the macroscopic resonant tunneling (MRT) linewidth and reorganization energy. They are connected through the fluctuation-dissipation theorem: $W^2 = 2\varepsilon_L T$. The quantity $\gamma(\omega)$ in Eq.~\eqref{eq:G_high} is the standard spectral function for an Ohmic bath, given by Eq.~\eqref{eq:Ohmic}.
We assume that the low-frequency and high-frequency components share the same temperature.

We note that the PTRE can be approximately thought of as a multi-level extension of the noninteracting-blip approximation (NIBA) method~\cite{Dekker1987-zk, smirnov_theory_2018}, which successfully modeled the open system dynamics in a multiqubit-cluster tunneling experiment~\cite{Boixo:2014yu}. Because in the low temperature limit NIBA only works for stronger coupling than PTRE~\cite{xu_non-canonical_2016, Grifoni1998-hc}, the success of NIBA indicates the existence of a strong coupling region during the anneal.

\subsubsection{4-qubit case.}
\label{sec:4_qubit_res}

\begin{figure}[t]
     \subfigure[]{\includegraphics[width = 0.4939\columnwidth]{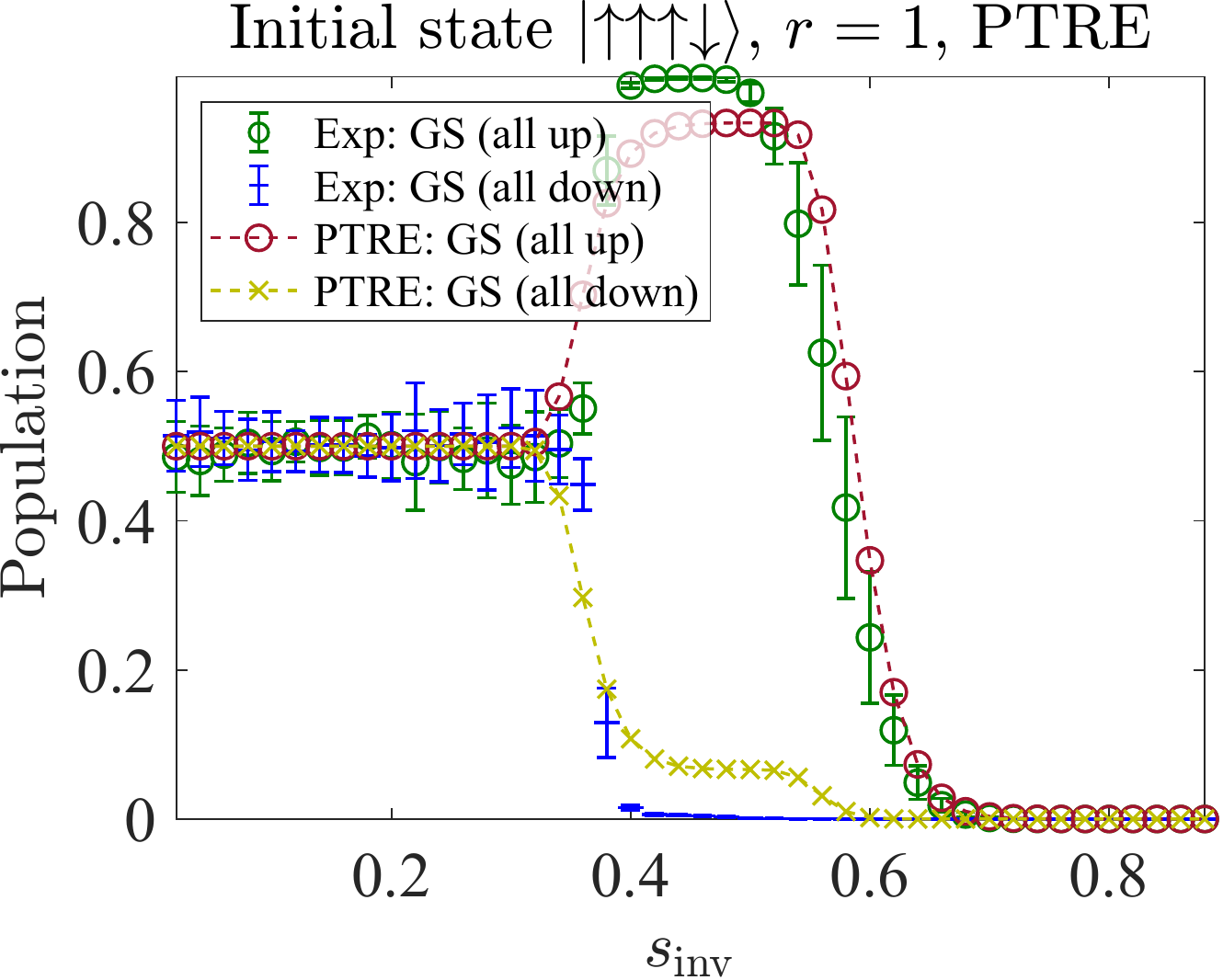}\label{fig:4qubit_0001}}
    \subfigure[]{\includegraphics[width = 0.4939\columnwidth]{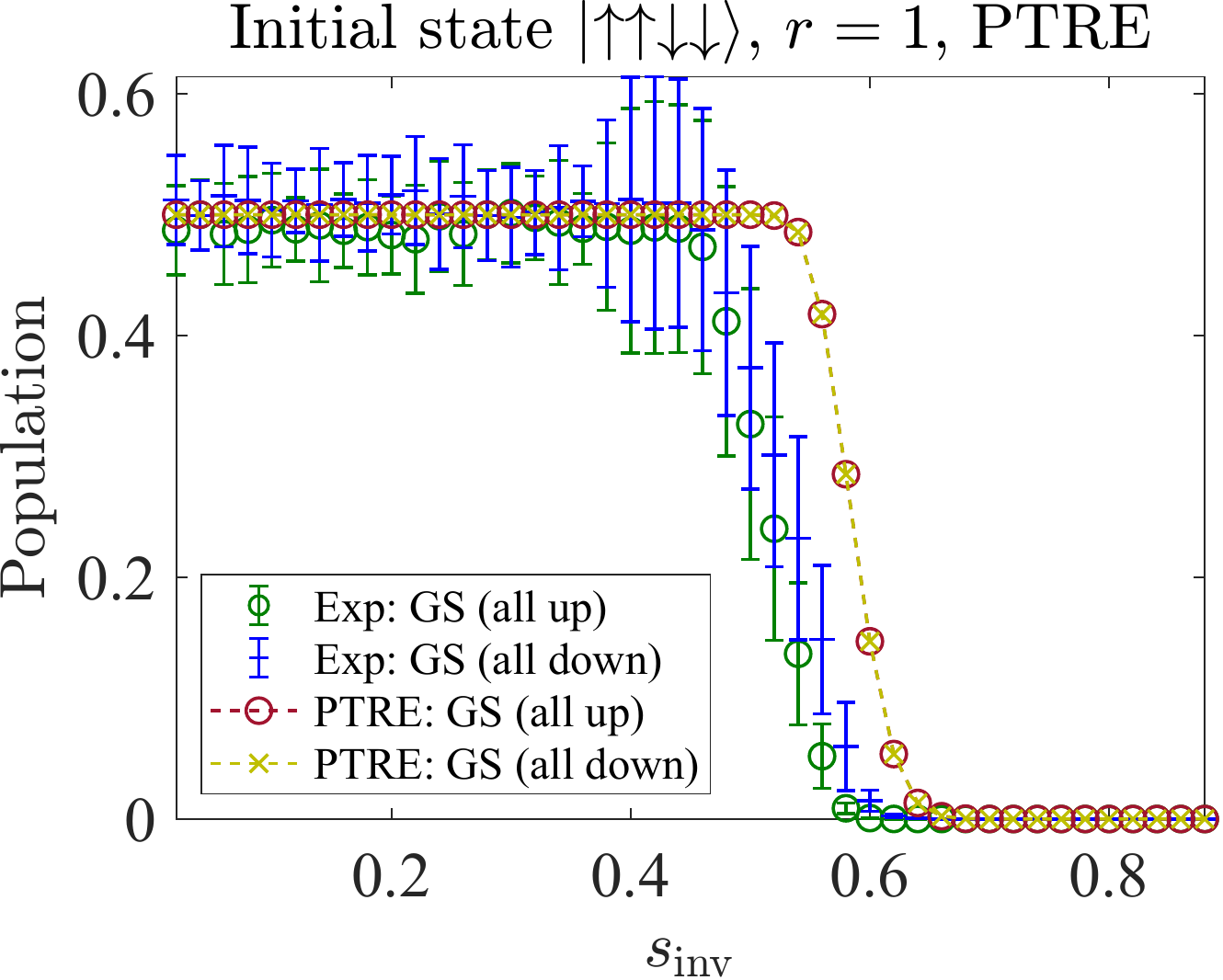}\label{fig:4qubit_0011}}
     \caption{PTRE simulation results versus the empirical results for the 4-qubit case. The initial states are (a) $\ket{\uparrow\uparrow\uparrow\downarrow}$ and (b) $\ket{\uparrow\uparrow\downarrow\downarrow}$. The best-fit simulation parameters are: $\eta g^2 = 2.5\times10^{-3}$, $T=25$\;mK, $\tau=5\mu$\;s, $W=8$mK, and $f_c = 1$THz. Here and in the subsequent figures the error bars represent $1\sigma$ confidence intervals. Overall the PTRE results are in good qualitative agreement with the empirical results for all cases shown. Note the different scale of the vertical axes.
     }
     \label{fig:optimal_n4}
\end{figure}

We first calibrate the simulation parameters using the empirical data from the $N=4$ case. We run the PTRE simulation with $\tau=5\;\mu$s and $f_c = 1$\;THz, and vary $T\in\{6,30\}$\;mK (step size of $1\;\mathrm{mK}$) and $W\in\{6,40\}$\;mK (step size of $2$\;mK), $\eta g^2\in \{2.5\times10^{-i}, 5\times10^{-i}\}_{i=1}^5$. We pick the optimal parameters from this set such that the simulation results have the smallest distance from the D-Wave data in the following six cases:  two cases where we start with $\ket{\uparrow\uparrow\uparrow\downarrow}$ and end up with the all-up and all-down states, and four cases where we start with $\ket{\uparrow\uparrow\downarrow\downarrow}$ and end up with the all-up, all-down, one-up and one-down states. Because all the numerical experiments are done on the same grid of inversion points $s_{\text{inv}}$, we denote the measured population (corresponding to the grid-points) by a vector $P_i$, where $i$ is the index for the six different cases. Then our parameter estimation procedure can be formally written as:
\begin{equation}
    \min_{W,T,\eta g^2}\sum_i \|{\tilde{P}_i-P_i}\| \ ,
\end{equation}
where $\tilde{P}_i$ is the simulation curve corresponding to the empirical data. The simulations results using the optimal parameters thus obtained are compared to the D-Wave data in Fig.~\ref{fig:optimal_n4}. Panel (a) of this figure reproduces the empirical $N=4$ data from Fig.~\ref{fig:RA_20_size}(a); panel (b) corresponds to the $m_0=0$ results shown in Figs.~\ref{fig:RA_20_initial}(a) and~\ref{fig:IRA}(c) (with $r=1$), though the latter are for $N=20$. Crucially, unlike the AME the PTRE correctly describes the asymmetry in the populations of the all-up and all-down ground states when the initial state has one spin down, as is clearly visible in Fig.~\ref{fig:optimal_n4}(a).

\subsubsection{8-qubit case.}
\begin{figure}[t]
     \subfigure[]{\includegraphics[width = 0.48\columnwidth]{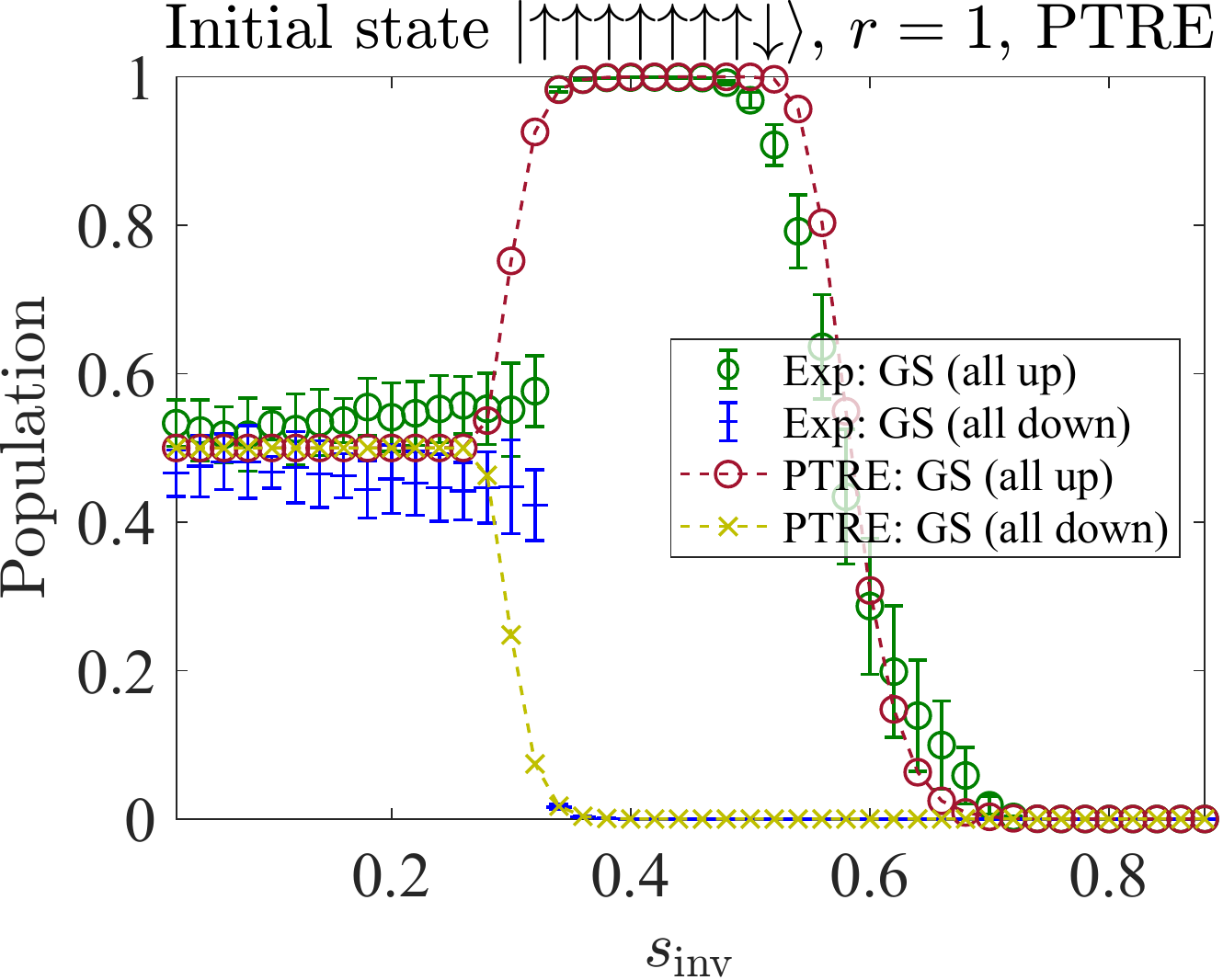}
         \label{fig:8qubit_0001}}
     \subfigure[]{\includegraphics[width = 0.48\columnwidth]{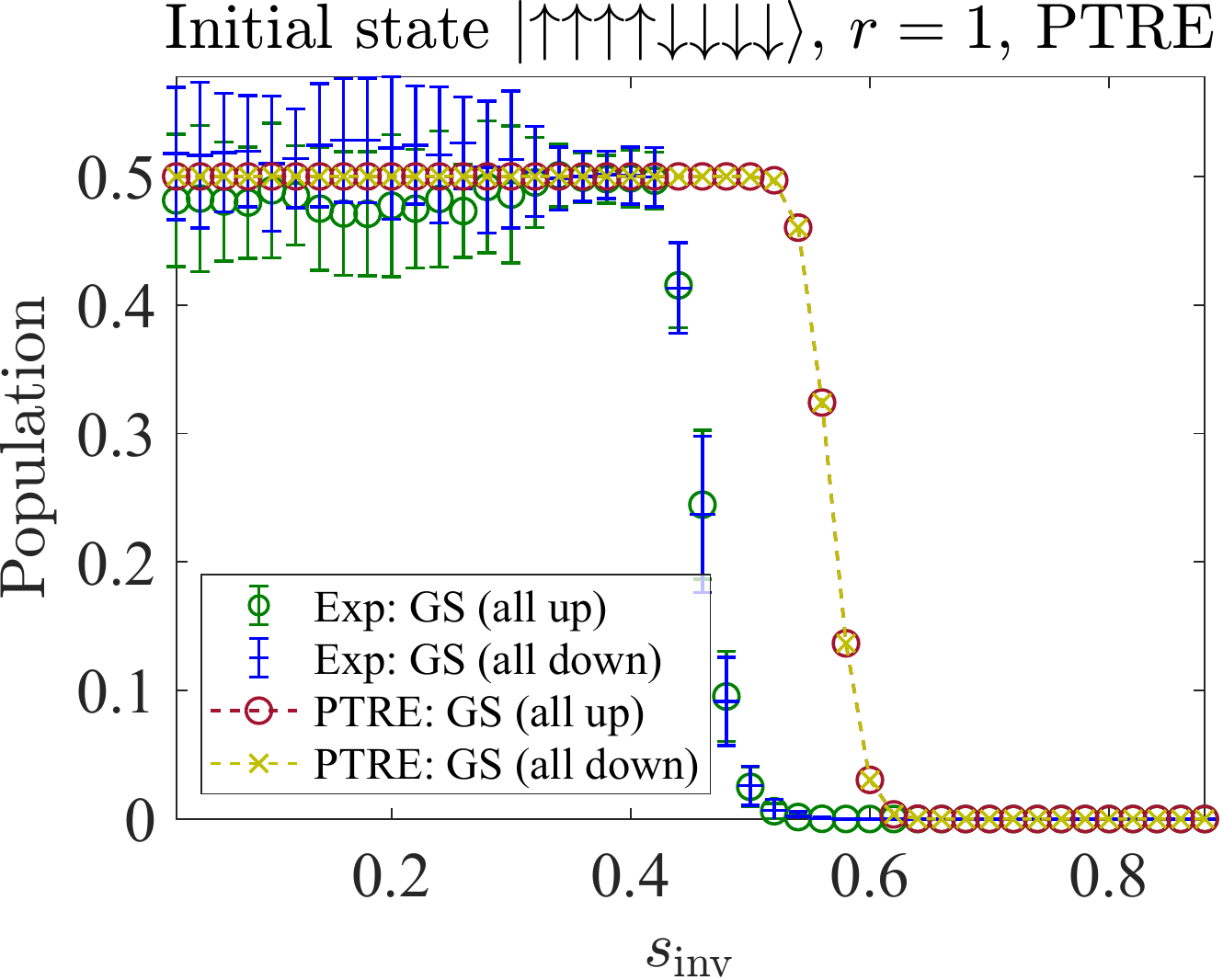}
\label{fig:8qubit_0011_0}}
     \caption{PTRE simulation results versus the experimental results for the 8-qubit case. The initial states are (a) the first excited state $\ket{\uparrow\cdots\uparrow\downarrow}$ and (b) the maximally excited state $\ket{\uparrow\uparrow\uparrow\uparrow\downarrow\downarrow\downarrow\downarrow}$. The simulation parameters are identical to the optimal ones obtained from the parameter estimation procedure in Fig.~\ref{fig:optimal_n4}. Note the different scale of the vertical axes.}
     \label{fig:compare_n8_ground}
\end{figure}

Next, we simulate the 8-qubit case using the optimal parameters obtained above. Again, we consider the experiments with two different initial states and present the results in Fig.~\ref{fig:compare_n8_ground}. Panel (a) of this figure reproduces the empirical $N=8$ data from Fig.~\ref{fig:RA_20_size}(a); panel (b) corresponds to the $m_0=0$ results shown in Figs.~\ref{fig:RA_20_initial}(a) and~\ref{fig:IRA}(c) (with $r=1$), though the latter are for $N=20$.

Similar trends are observed as in the 4-qubit case. The PTRE works best when the initial state has only a single spin flip from the ground state [Fig.~\ref{fig:compare_n8_ground}(a)]. When an equal number of spins is up as down (the maximally excited state), the PTRE is less accurate but the qualitative trend is correct [Fig.~\ref{fig:compare_n8_ground}(b)]. It is possible that fine-tuning of the PTRE parameters could further improve the quantitative agreement. However, we did not pursue this since the main goal has already been achieved: our simulation results demonstrate that the PTRE achieves a much better qualitative agreement with the experiments than the AME (Fig.~\ref{fig:4vs8open}). In particular, it correctly predicts the asymmetry in the populations of the two degenerate all-up and all-down ground states when the initial state is the first excited state, for both $N=4$ and $N=8$. This is the case because in the PTRE decoherence happens between the classical spin states [Eq.~\eqref{eq:ptre_lindblad_operator}] while in the AME, decoherence happens between the eigenstates of the full Hamiltonian [Eq.~\eqref{eq:AME-Lind}].

\section{Conclusions}
\label{section:Conc}

In this work, we presented a comprehensive study of reverse annealing of the $p$-spin model with $p=2$, and compared empirical results from the D-Wave 2000Q used as a quantum simulator of this model, with two quantum master equations in the weak and strong coupling limits (the AME and PTRE, respectively).
Results for one classical Monte Carlo based algorithm (SVMC-TF~\cite{albash2020comparing}) are presented in App.~\ref{sec:classical}; 
to keep the scope manageable we did not consider other popular models such as simulated quantum annealing (SQA)~\cite{sqa2,sqa1,Bapst:2013vt,q108,speedup,Bando2021}
(also known as Path-Integral Monte Carlo Quantum Annealing in Refs.~\cite{Santoro:2006hh,Morita2008} or Quantum Monte Carlo Annealing in Ref.~\cite{Das2008}), or the recent Spin-Vector Langevin model~\cite{subires2021benchmarking}; these are interesting topics for a future study. 

Our main observation is a stark failure of the AME,
which successfully captured the D-Wave annealers' behavior in a variety of previous studies~\cite{q-sig,albash2015consistency,ALMZ:12,Albash:2015pd,Mishra:2018,albash2020comparing}, in correctly describing the empirical ``partial" success probabilities, i.e., the probability of finding either the all-up or all-down ground state, as opposed to the sum of the probabilities of these two degenerate ground states of the $p=2$ $p$-spin model. As illustrated in Fig.~\ref{fig:4vs8open}, the AME predicts equal partial success probabilities, but the empirical data are strongly biased towards one or the other ground state, depending on the initial condition of the reverse anneal; see, e.g., Fig.~\ref{fig:RA_20_initial} and all subsequent figures displaying empirical partial success probability data. The reason for this failure of the AME is that in the weak coupling limit that it describes, the two degenerate ground states are energy eigenstates that are equal superpositions of the all-up and all-down states with opposite signs, and there is nothing about the dynamics described by the AME that breaks the symmetry between these two states for the $p=2$ problem under consideration. The failure of the AME thus indicates that the weak coupling limit itself is the problem in the setting of the D-Wave 2000Q device
subject to reverse annealing, and indeed, when we considered the PTRE (strong coupling) 
it exhibited the observed asymmetry; see, e.g., Fig.~\ref{fig:optimal_n4}. 
This represents strong evidence of the breakdown of the weak coupling limit in the D-Wave 2000Q device, at least for reverse annealing on timescales of a $\mu$sec or longer.

Our results demonstrate the importance of choosing the correct decoherence model when analyzing real-world devices, and that even when agreement is observed (as was the case for previous studies involving the AME), certain aspects that can reveal a way to distinguish between different decoherence models may remain hidden. The case in point is the difference between the partial probabilities of the two degenerate ground states, which forced us to conclude that strong coupling is the more appropriate decoherence model for the D-Wave 2000Q on the $\gtrsim 1\mu$sec timescale. Had we considered only the total success probability, we would not have been able to distinguish between decoherence models with weak \textit{vs} strong coupling.


Our work shows that the PTRE --  a first principles, fully quantum dynamical model with strong decoherence -- achieves the best agreement overall with the empirical data among the various models we have tested.
This does not necessarily imply that the D-Wave 2000Q behaves as a 
classical device on the $\gtrsim 1\mu$sec timescale. Instead, our results should be interpreted to mean that 
models 
with strong decoherence 
can be successful in predicting the outcome of quantum annealing experiments when the chosen observable is not sensitive to quantum effects. An earlier study using a previous generation of the D-Wave devices already established evidence for entanglement~\cite{DWave-entanglement}, which was verified using AME simulations~\cite{Albash:2015pd}, and this is a clear-cut example of the measurement of an observable that is sensitive to quantum effects. We anticipate that new experiments based on much shorter anneal times than were available to us in this work will provide further evidence of quantum effects in experimental quantum annealing \cite{king2022}.
Our work highlights the importance of testing  such new evidence using models that critically evaluate the strength of any claimed quantum effects.

\begin{acknowledgments}
We thank Mohammad Amin and Masayuki Ohzeki for useful comments and Sigma-i Co., Ltd. for providing part of the D-Wave 2000Q machine time.
The research is based partially upon work supported by the Office of the Director of National Intelligence (ODNI), Intelligence Advanced Research Projects Activity (IARPA) and the Defense Advanced Research Projects Agency (DARPA), via the U.S. Army Research Office contract W911NF-17-C-0050. The views and conclusions contained herein are those of the authors and should not be interpreted as necessarily representing the official policies or endorsements, either expressed or implied, of the ODNI, IARPA, DARPA, or the U.S. Government. The U.S. Government is authorized to reproduce and distribute reprints for Governmental purposes notwithstanding any copyright annotation thereon. Computation for some of the work described in this paper was supported by the University of Southern California Center for High-Performance Computing and Communications (hpcc.usc.edu).
\end{acknowledgments}
\appendix

\section{Optimal Value of Ferromagnetic Interactions within a Logical Qubit}
\label{append:optJF}

\begin{figure}
\includegraphics[width=.6\columnwidth]{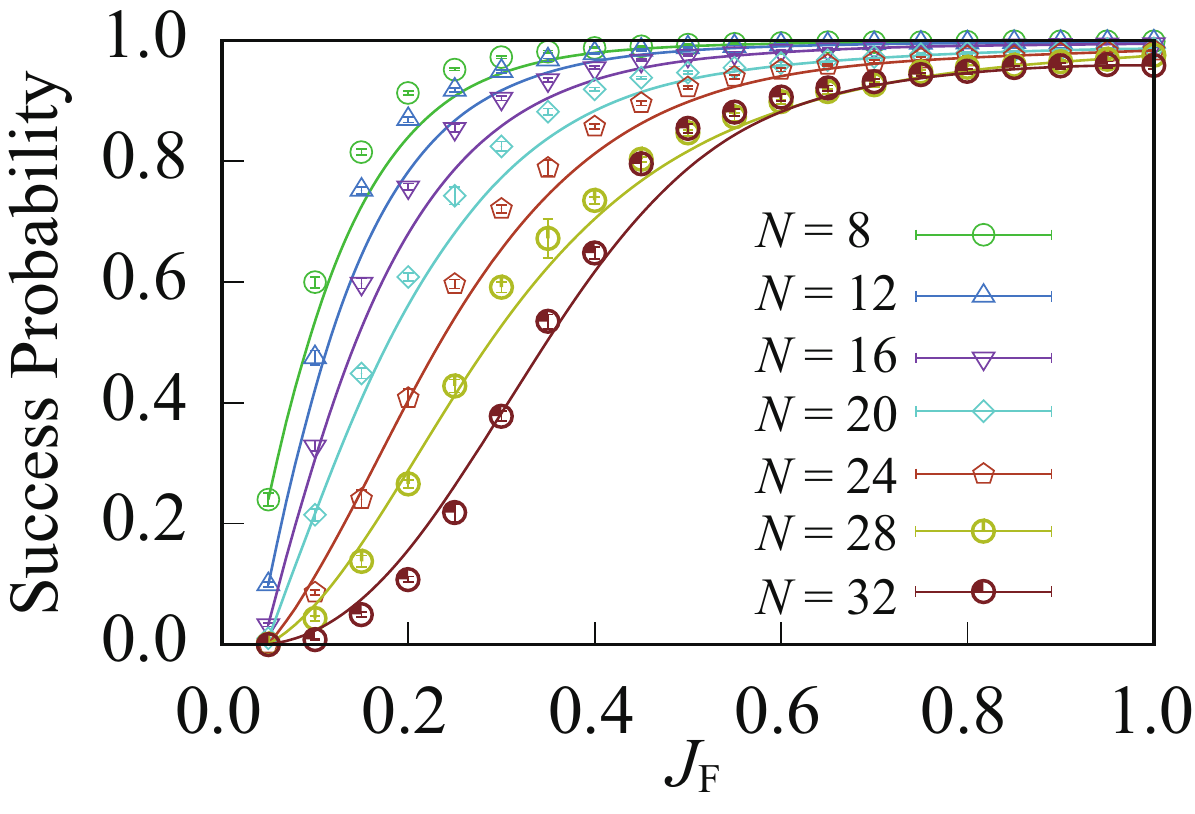}%
\caption{\label{append_fig:JFvsN} Success probability of forward annealing with annealing time $\tau=1.0~\mu s$ as a function of the coupling between physical qubits in a logical qubit for different system sizes. The solid lines are Bezier curves. The penalty value we used is $-J_\text{F}$, i.e., ferromagnetic.} 
\end{figure}

The ferromagnetic coupling strength between physical qubits for a given logical qubit on the D-Wave device affects the final success probability. The appropriate value of $J_{F}$ depends on the system size $N$~\cite{Venturelli2015}. We checked the success probability for various system sizes of the $p=2$ $p$-spin model under traditional forward annealing. The result is shown in Fig.~\ref{append_fig:JFvsN}.  It is clearly observed that the smallest value of coupling $J_{\text{F}}=-1.0$ we tried yields the best result for all $N$ values we tried, and we adopted this value in all our experiments. The smallest allowed value is $J_{\text{F}}=-2.0$, but saturation is already observed for $J_{\text{F}}=-1$.

\section{Reverse annealing details}
\label{app:RA}

In our experiments for single-iteration ($r=1$) reverse annealing, we constructed 15 instances (different hardware embeddings of $H_T$ on the Chimera graph of the device) and generated 10 random gauges for each set of parameter values. 
A gauge (originally known as a spin inversion transformation~\cite{q-sig}) is a given choice of $\{h_i,\,J_{ij}\}$ in the general Ising problem Hamiltonian $H_T = \sum_{i=1}^N h_i \sigma^z_i + \sum_{i<j}^N J_{ij} \sigma^z_i\sigma^z_j$; a new gauge is realized by randomly selecting $a_i =\pm 1$ and performing the substitution $h_i \mapsto a_i h_i$ and $J_{ij} \mapsto a_i a_j J_{ij}$. Provided we also perform the substitution $\sigma_i^z \mapsto a_i \sigma_i^z$, we map the original Hamiltonian to a gauge-transformed Hamiltonian with the same energy spectrum but where the identity of each energy eigenstates is relabeled accordingly.  In total, there are $2^N$ different gauges for an $N$-spin problem. See Ref.~\cite{Job:2017aa} for a review and more details.

1000 annealing runs were performed for a given random gauge. Thus, the total number of runs for each set of annealing schedule parameter values is 150,000. In iterated reverse annealing with $r>1$, we constructed 15 instances and generated 90 random gauges for each $r$. Thus, the total number of samples at each $r$ and each $s_{\mathrm{inv}}$ is 1350.

We computed $1\sigma$ error bars by cluster sampling over instances. Namely, we regarded each of the $15$ instances as as a sample of size 10000 [the number of random gauges (10) $\times$\ the number of iterations (1000)], and computed the standard error of the mean for each data point shown in our experimental figures.
 
We remark that the reverse annealing protocol we have adopted here for experiments on the D-Wave device 
is different from the one used in Ref.~\cite{Passarelli2019} for numerical computation. In particular, when $s_{\mathrm{inv}}\approx 0$, the second part of Ref.~\cite{Passarelli2019}'s protocol has a very sharp increase of $s$ as a function of $t$, which seems to have led to a significant drop of success probability.
This is not the case in the present protocol, where the time derivative of $s(t)$ is a constant $\pm 1/\tau$ (or $0$ during pausing) irrespective of $s_{\mathrm{inv}}$.  

\section{Spectrum of the $p$-spin problem with $p=2$ and scaling of the minimum gap with $N$}
\label{append:spectrum}

\begin{figure}[t]
    \centering
    \includegraphics[width=0.7\linewidth]{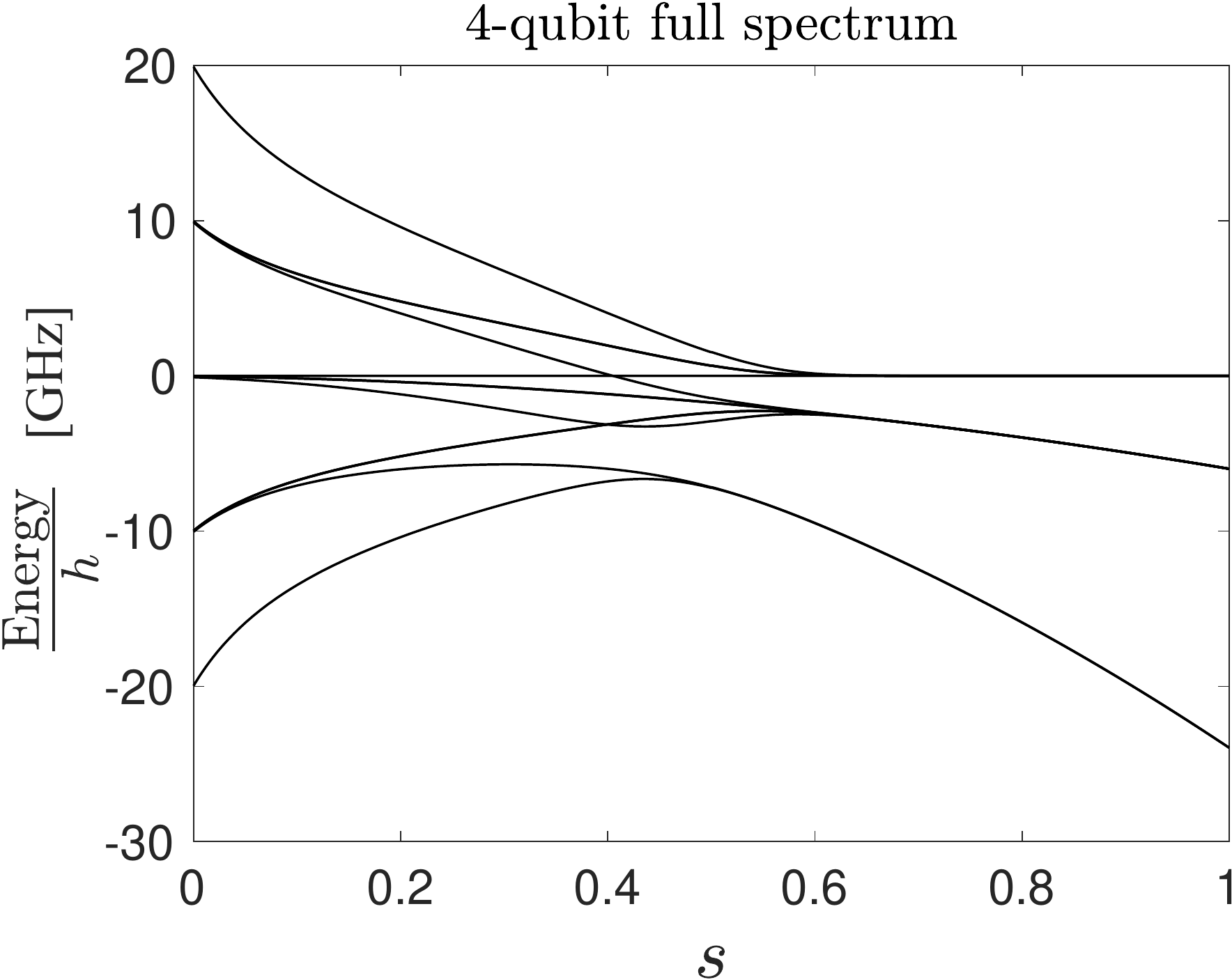}
    \caption{Full spectrum of four qubits for the $p=2$ $p$-spin model, subject to the annealing schedules shown in Fig.~\ref{fig:schedule}.}
    \label{fig:myspectrum}
\end{figure}

The spectrum of the $p$-spin problem and the ordering of energies of different spin sectors can be found in Ref.~\cite{Bapst2012}. For the particular case of $n=4, p=2$ and the D-Wave 2000Q annealing schedule, we plot the spectrum in Fig.~\ref{fig:myspectrum}. 

We denote the energy gap between the instantaneous eigenenergies $\epsilon_{i}(s)$ and $\epsilon_{j}(s)$ by $\Delta_{ij}(s) = \epsilon_{i}(s) - \epsilon_{j}(s)$, and the corresponding minimum energy gap by $\Delta_{ij} = \min_{s} \Delta_{ij}(s)$. For $p=2$, we are interested instead in the minimum energy gap between the ground state $\epsilon_{0}(s)$ and the second excited state $\epsilon_{2}(s)$, denoted by 
\begin{equation}
    \Delta = \Delta_{20} = \min_{s} \Delta_{20}(s) = \min_{s} \epsilon_{2}(s) - \epsilon_{0}(s) \,,
\end{equation}
since as can be seen from Fig.~\ref{fig:myspectrum}, the instantaneous first excited state $\epsilon_{1}(s)$ and the instantaneous ground state $\epsilon_{0}(s)$ converge at $s=1$. This, of course, is true for every $N$ due to the double degeneracy of the ground state of the $p=2$ $p$-spin model, which exhibits $\mathbb{Z}_2$ symmetry.

Figure~\ref{fig:gap} shows the value of $\Delta$ for $N \in \{4, \cdots, 22\}$, along with the position $s$ of the minimum gap, i.e.,
\begin{equation}
    s_\Delta = \mathrm{argmin} \Delta_{20}(s) \,.
\end{equation}

\section{Upper bound on the success probabilities for collective dephasing}
\label{append:successbound}

If the dynamics preserve spin symmetry (for example in the closed system Schr\"{o}dinger equation and in open-system simulations with collective system-bath coupling), then there exists a natural upper bound on the maximal success probabilities achievable in reverse annealing. The upper bound is the population of the initial state in the maximum-spin sector.

For the example of $N=4$, there are two degenerate ground states ($\ket{0000}$ 
and $\ket{1111}$) of the problem Hamiltonian $H_T$ and they both belong to the maximum-spin subspace of $S=S_{\max}={N}/{2}=2$. While the computational basis state with one spin down, for example, $\ket{0001}$ ($\ket{0} \equiv \ket{\uparrow}$, $\ket{1} \equiv \ket{\downarrow}$) is a first excited state of $H_T$, it does not belong to the maximum-spin subspace $S=2$. A uniform superposition of the computational basis states with one spin down, i.e., $(\ket{0001}+\ket{0010}+\ket{0100}+\ket{1000})/2$, however, does belong to the maximum-spin subspace $S=2$.
Unfortunately, the D-Wave device does not allow such an initialization. 

In general, suppose that the initial computational basis state is $\ket{\psi(t=0)}_{\text{comp}}$, with a particular magnetization $m_0$ [Eq.~\eqref{eq:mo}]. It can be represented as a linear combination of states with a fixed value of total spin $S$ and magnetization $m_0$:
\begin{equation}
\label{eq:compinitialstate}
    \ket{\psi(t=0)}_{\text{comp}} = \sum_{S=S_{\min}}^{S_{\max}}a_S\ket{S, m_{S}=m_0}. 
\end{equation}
From angular momentum addition theory~\cite{Rose:book}, we know that $S_{\min} = N/2-\lfloor N/2 \rfloor$ and $S_{\max} = N/2$. The total spin (integer or half-integer) $S \in \{S_{\min}, S_{\min} + 1, \dots, S_{\max}\}$. The (unnormalized) magnetization
is
$M_S \in \{-S, -S+1, \dots, S\}$, but we can also label the states using the normalized magnetization $m_S = \frac{2}{N}M_S$; then $\ket{S, m_{S}=m_0}$ is a simultaneous eigenstate of $\mathbf{S}^2$ and $S^z$ with eigenvalues
\begin{align}
    \mathbf{S}^2\ket{S, m_{S}=m_0} &= S(S+1)\ket{S, m_{S}=m_0} \,,\\
    S^z\ket{S, m_{S}=m_0} &= \Big(\frac{N}{2}m_0\Big)\ket{S, m_{S}=m_0}\,.
\end{align}
In Eq.~\eqref{eq:compinitialstate}, $|a_S|^2$ is the initial state's population in that spin subspace. 

For the $p$-spin Hamiltonian, in a closed system the state in each spin subspace evolves independently due to the preservation of spin symmetry of the Hamiltonian~\cite{Yamashiro:2019aa}. At the end of a single cycle of reverse annealing ($r=1$), we have from Eq.~\eqref{eq:compinitialstate}) that 
\begin{equation}
    U(2t_\text{inv},0)\ket{S, m_{S}=m_0} = \sum_{M_S=-S}^{S} c_{M} \ket{S, m_{S}=2M_S/N},
\end{equation}
where $c_{M}$ is the amplitude developed by the other basis elements of the spin $S$ subspace at time $t=2t_\text{inv}$. 

Therefore, the final state of $1$-cycle reverse annealing can be expressed as:
\begin{align}
    \ket{\psi}_{\text{fin}} &= U(2t_\text{inv},0)\ket{\psi(t=0)}_{\text{comp}} \nonumber\\
    &= \sum_{S=S_{\min}}^{S_{\max}}a_S\left(\sum_{M_S=-S}^{S} c_M \ket{S, m_{S}=2M_S/N}\right)\,.
\end{align}
The all-up state ($\ket{0}^{\otimes N} = \left|\uparrow\right>^{\otimes N} = \ket{\text{up}}$) and all-down state ($\ket{1}^{\otimes N} = \left|\downarrow\right>^{\otimes N} = \ket{\text{down}}$) are both ground states of $H_T$, and moreover they lie in the maximum-spin subspace $S=S_{\max}$. In particular, $\ket{\text{up}} = \ket{S_{\max}, 1}$ and  $\ket{\text{down}} = \ket{S_{\max}, -1}$. Therefore, projecting $\ket{\psi}_{\text{fin}}$ onto $\ket{\text{up}}$ gives:
\begin{align}
    &\phantom{==}\braket{\text{up}|\psi}_{\text{fin}}\nonumber\\
    &=  \sum_{S=S_{\min}}^{S_{\max}}a_S\left(\sum_{M_S=-S}^{S} c_{M} \braket{\text{up}|S, m_{S}=2M_S/N}\right)\\
    &= a_{S_{\max}}c_{S_{\max}}\,.
\end{align}
Similarly, $\braket{\text{down}|\psi}_{\text{fin}} = a_{S_{\max}}c_{-S_{\max}}$. 

The total success probability is thus bounded by: 
\begin{align}
p(r=1) &= |\braket{\text{up}|\psi}_{\text{fin}}|^2 + |\braket{\text{down}|\psi}_{\text{fin}}|^2 \nonumber\\
&= |a_{S_{\max}}|^2(|c_{S_{\max}}|^2+|c_{-S_{\max}}|^2) \nonumber\\
&\leq |a_{S_{\max}}|^2 \,,
\end{align}
with the bound saturated when 
$(|c_{S_{\max}}|^2+|c_{-S_{\max}}|^2)=1$.
\newline \rightline{$\square$}

The upper bound $|a_{S_{\max}}|^2$ is, as claimed above, the population of the initial state in the maximum-spin subspace. For the initial state $\ket{0001}$ we have $a_{S_{\max}} = 1/2$ since $\ket{S=S_{\max}=2, m_{S}=0.5} = (\ket{0001}+\ket{0010}+\ket{0100}+\ket{1000})/2$. Therefore, the total success probability is bounded by $|a_{S_{\max}}|^2 = 1/4$. For the other examples in the main text, the initial state $\ket{0011}$ has $a_{S_{\max}} = 1/{\sqrt{{{4}\choose{2}}}} = 1/{\sqrt{6}}$; while the initial state of $\ket{00000001}$ has $a_{S_{\max}} = 1/{\sqrt{8}}$.

\begin{figure}[t]
\includegraphics[width=0.65\columnwidth]{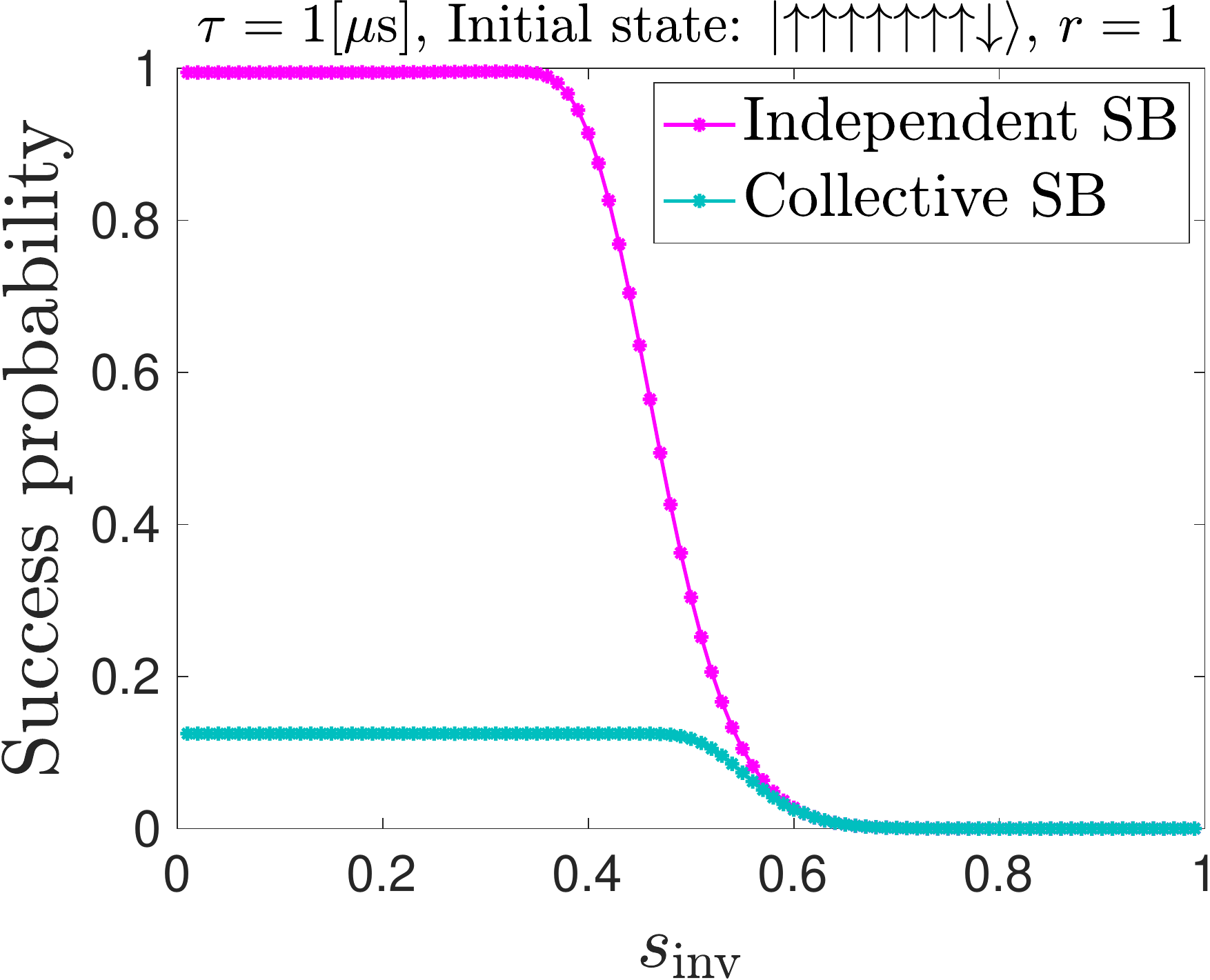}
\caption{Success probability as a function of $s_{\mathrm{inv}}$ for independent and collective dephasing, for $N=8$ spins. The initial state has a single spin flipped and $\tau =1\;\mu$s.}
\label{fig:8open}
\end{figure}

This conclusion
is
directly generalized to the open-system case under collective dephasing, where the population of each spin-sector is also preserved during the dynamics. This is illustrated in the main text for $N=4$ in Fig.~\ref{fig:4open_0001_r1}(b), and also in Fig.~\ref{fig:8open}, which shows the analogous result for $N=8$, for which the initial state is  $\ket{00000001}$. In this case the maximum success probability of collective dephasing is  $1/8$, which is the population of the initial state in the maximal spin subspace. We remark that this case is different from Ref.~\cite{Passarelli2019}, where the initial state (a superposition of computational basis state) lies completely inside the maximal spin subspace, in which case the collective dephasing model has a maximum success probability of $1$.

\section{The PTRE equation involves only the diagonal components of the density matrix}
\label{app:PTRE}

As argued in Sec.~\ref{sec:PTRE}, the PTRE equation involves only the diagonal components of the density matrix. To show this we first consider
\begin{equation}
    \label{eq:HM}
    M = [H_\mathrm{e},{\rho}] \ ,
\end{equation}
where $H_\mathrm{e} \equiv H+H_\mathrm{LS}$ is a diagonal matrix, i.e., $H_\mathrm{e}^{ab}=\delta_{ab}H^{aa}_\mathrm{e}$. The explicit form of $M$ can be written as $M_{ab}=\pqty{H_\mathrm{e}^{aa}-H_\mathrm{e}^{bb}}{\rho}^{ab}$, which implies that the diagonal elements of $M$ ($M_{aa}$) vanish.
Next, we consider
\begin{equation}
    \label{eq:Lindblad_N}
    N^{\omega,\alpha}_i = L_{i,\omega}^{\alpha}
    {\rho}L_{i,\omega}^{\alpha\dagger}-\frac{1}{2}\Bqty{L_{i,\omega}^{\alpha\dagger}L_{i,\omega}^{\alpha}, {\rho}} \ ,
\end{equation}
where the Lindblad operators $L_{i,\omega}^\alpha(t)$ were defined in Eq.~\eqref{eq:ptre_lindblad_operator}.
To simplify the notation, we omit the indices $\omega$, $\alpha$ and $i$ in the following discussion. Let us denote $A_{ab}\equiv\frac{A\pqty{t}}{2}\bra{a}{\sigma_i^\alpha}\ket{b}=A^*_{ab}$. The first term in Eq.~\eqref{eq:Lindblad_N} can be written as
\begin{align}
    L{\rho}L 
    & = \sum_{a,b,a',b'}A_{ab}A_{a'b'}{\rho}^{bb'}\ketb{a}{a'} \ ,
    \label{eq:LrhoL}
\end{align}
where ${\rho}^{bb'} = \bra{b}{\rho}\ket{b'}$.
The second term in Eq.~\eqref{eq:Lindblad_N} is
\begin{subequations}
\begin{align}
    -\frac{1}{2}\Bqty{L^\dagger L, {\rho}} & = -\sum_{a,b,b'}\frac{1}{2}A_{ab'}A_{ab}\{\ketb{b'}{b},{\rho}\} \\
    & = -\sum_{a,b,b'}\frac{1}{2}A_{ab}A_{ab'}\pqty{\ketb{b'}{b}{\rho}+{\rho}\ketb{b'}{b}} \ .\label{eq:LLrho}
\end{align}
\end{subequations}
If we assume ${\rho}$ is diagonal, then all the diagonal elements of $\dot{\rho}$ depend only on the diagonal elements of ${\rho}$, with an explicit form 
\begin{equation}
    \label{eq:diagonal_eom}
    \dot{\rho}^{aa} = \sum_{b\neq a}\gamma_\mathrm{p}(\omega_{ba})Z_{ab}\rho^{bb} - \sum_{b\neq a}\gamma_\mathrm{p}(\omega_{ab})Z_{ba}\rho^{aa} \ ,
\end{equation}
where 
$Z_{ab}=A^2(t)\sum_{\alpha, i}| \bra{a}{\sigma_i^\alpha}\ket{b}|^2/4$, and
$\omega_{ab}=\omega_a - \omega_b$. 
Thus, if the initial state is diagonal (a classical spin state), we can consider the Pauli master equation~\cite{Lidar:2019aa} for the diagonal elements only [Eq.~\eqref{eq:diagonal_eom}] to speed up the computation. The matrix form of this equation is
\bes
    \label{eq:diag_ptre}
\begin{align}
&    \begin{pmatrix}
        \dot{\rho}_{00} \\
        \dot{\rho}_{11}\\
        \vdots
    \end{pmatrix} = T \begin{pmatrix}
        \rho_{00} \\
        \rho_{11}\\
        \vdots
    \end{pmatrix} \\
&T \equiv     \begin{pmatrix}
        -\sum_{b\neq 0}\gamma_\mathrm{p}(\omega_{0b})Z_{b0} & \gamma_\mathrm{p}(\omega_{10})Z_{01} & \cdots\\
        \gamma_\mathrm{p}(\omega_{01})Z_{10} & -\sum_{b\neq 1}\gamma_\mathrm{p}(\omega_{1b})Z_{b1} & \cdots\\
        \vdots & \vdots & \ddots \\
    \end{pmatrix}
      \ .
\end{align}
\ees

Before proceeding, we note that the sparsity of $T$ is determined by the sparsity of $Z_{ab}$, whose full size is $2^N \times 2^N$. To calculate the number non-zero elements in $Z_{ab}$, we first notice that for each $\sigma_i^\alpha$ operator, $2^{N-1}$ elements are non-zero (recall that $\alpha\in\{+,-\}$). The number of $\sigma_i^\alpha$ operators is $2N$. So there are $N2^N$ non-zero elements in $Z_{ab}$. If we add the number of diagonal elements in $T$, the total number of non-zero elements in $T$ is $(N+1)2^N$. As a result, the sparsity of the transfer matrix $T$ is $(N+1)/2^N$.

The transfer matrix $T$ in Eq.~\eqref{eq:diag_ptre} provides the incoherent ``tunneling'' rate between classical spin states. Because the total Hamiltonian is symmetric under permutations of qubits, we may also want to calculate the tunneling rate between the spin coherent states  \cite{muthukrishnan_tunneling_2016}
\begin{equation}
    |\theta, \phi\rangle =\bigotimes_{i=1}^{n}\left[\cos \left(\frac{\theta}{2}\right)|0\rangle_{i}+\sin \left(\frac{\theta}{2}\right) e^{i \varphi}|1\rangle_{i}\right]\ .
\end{equation}
However, because in the current form of the PTRE the coherence between computational basis decays exponentially, the spin coherent state cannot survive. 

\begin{figure}[t]
     \subfigure[]{\includegraphics[width = 0.48\columnwidth]{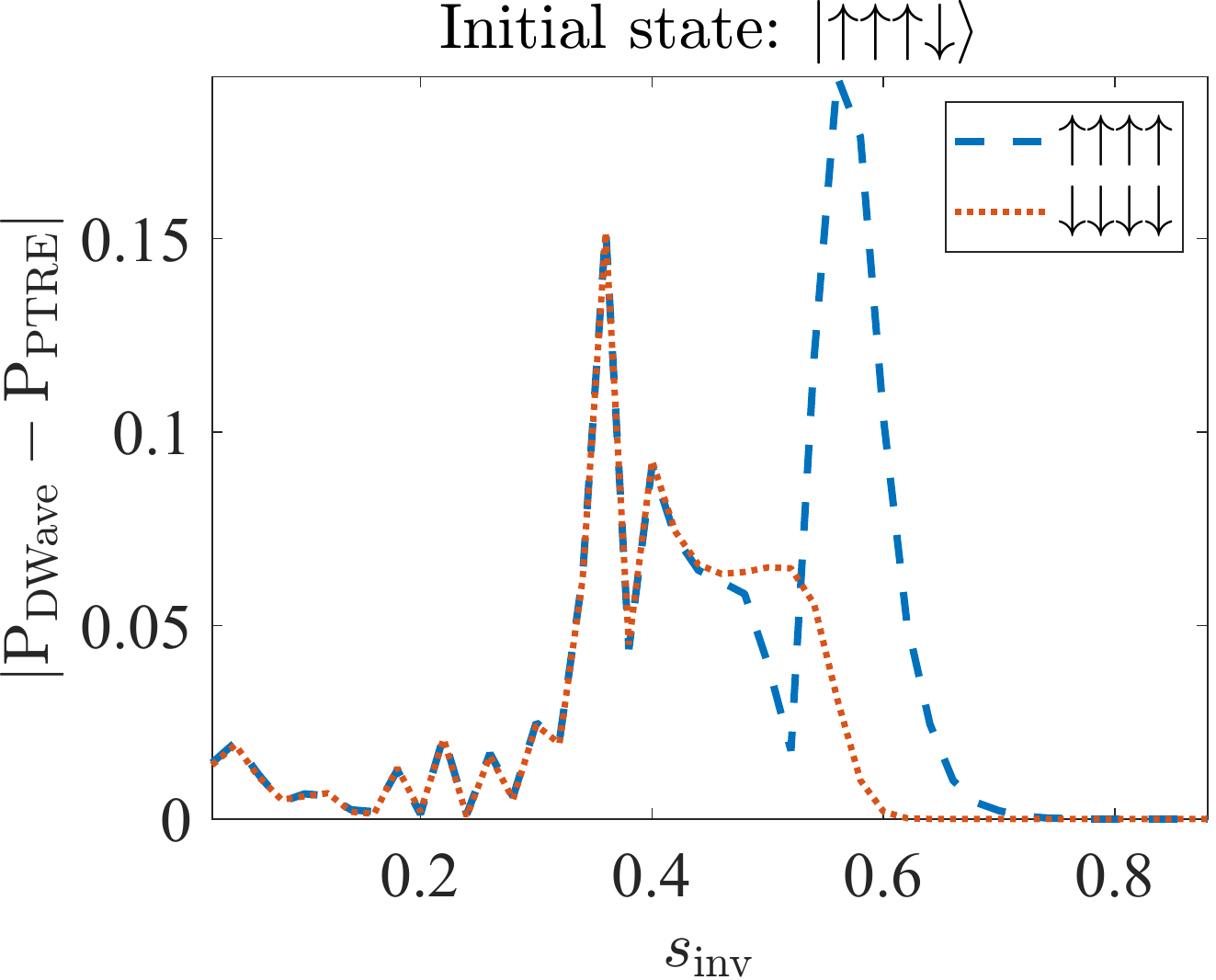}        \label{fig:4qubit_0001_diff}}
     \subfigure[]{\includegraphics[width = 0.48\columnwidth]{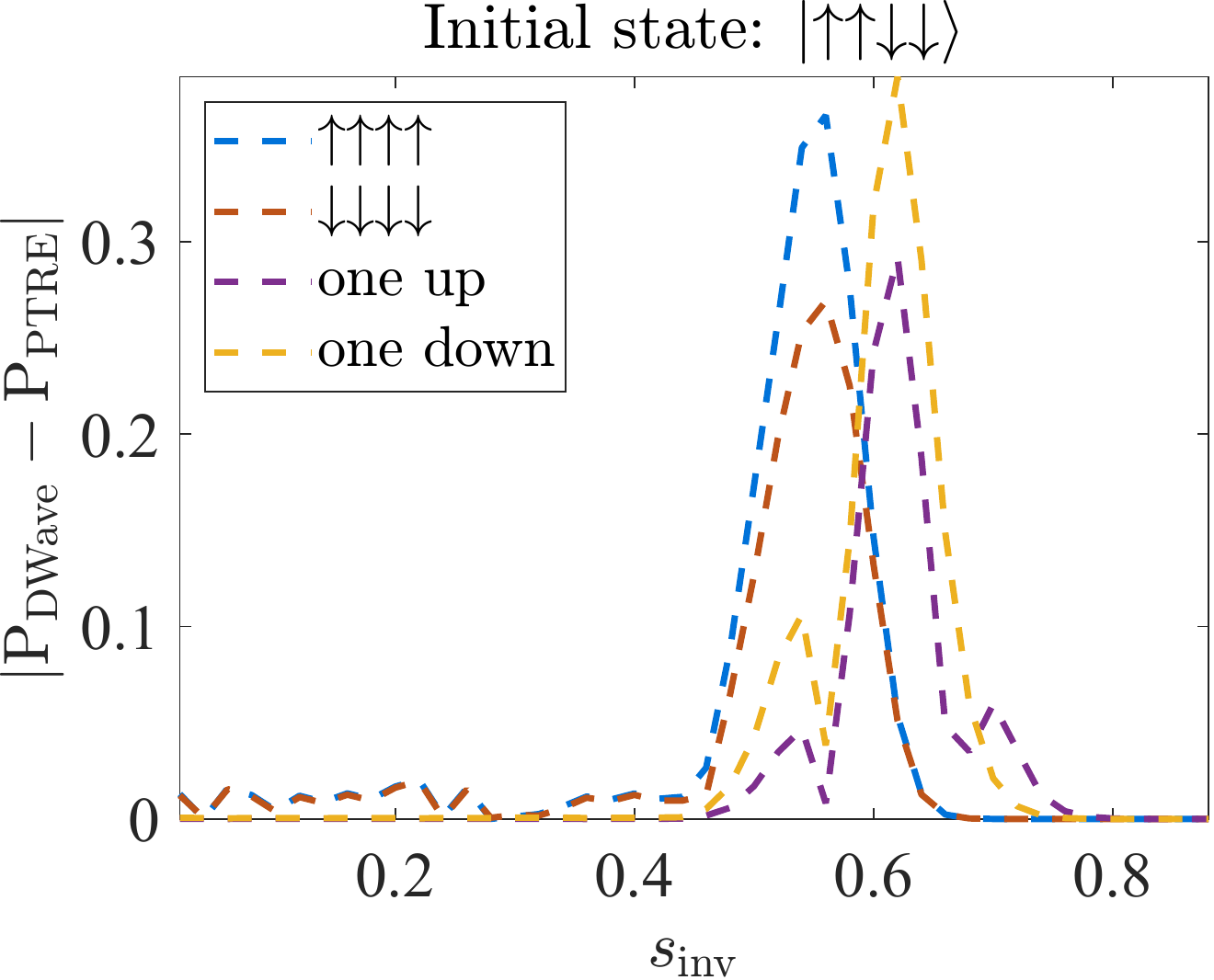}         \label{fig:4qubit_0011_diff}}\\
     \subfigure[]{\includegraphics[width = 0.48\columnwidth]{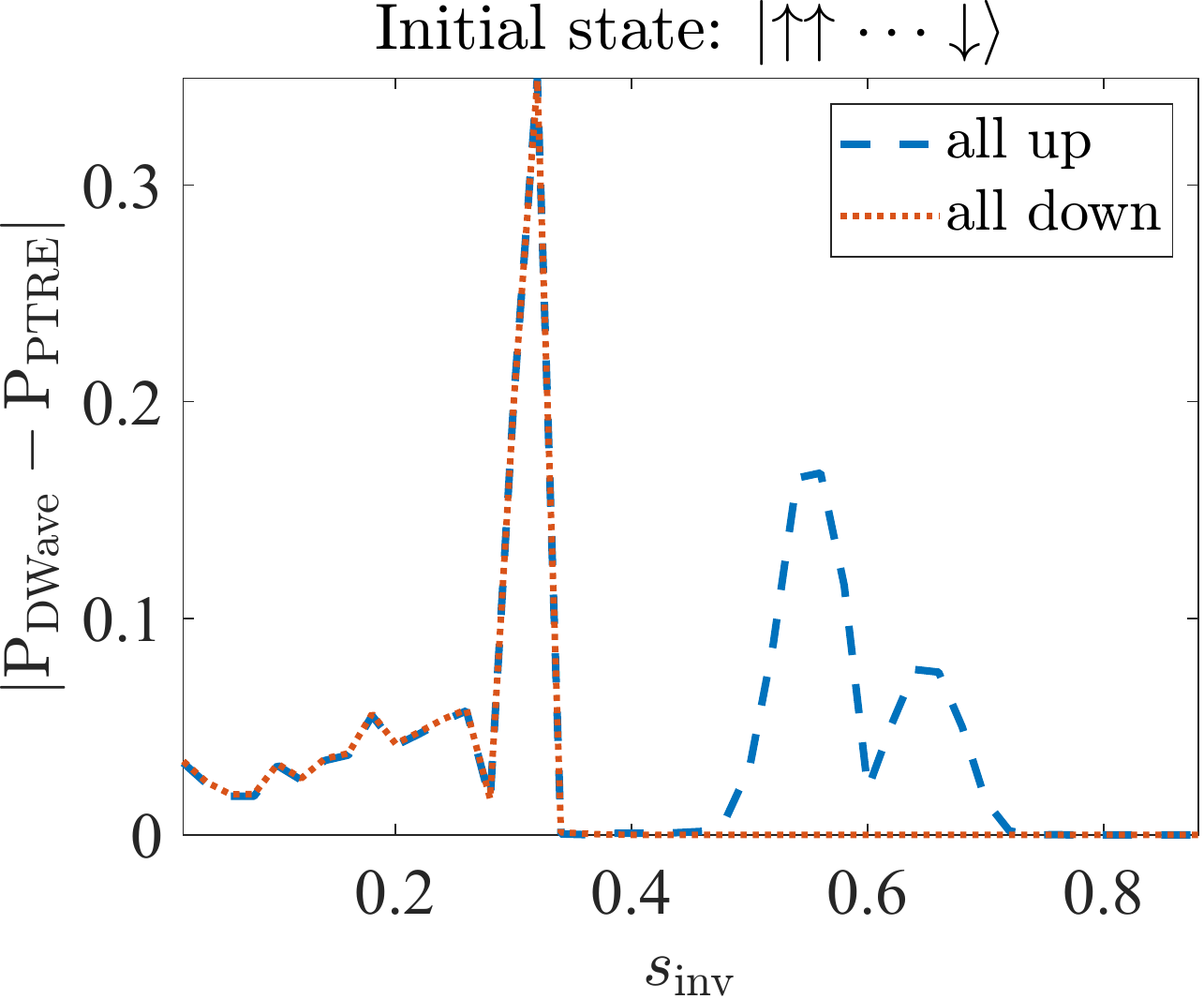}
          \label{fig:8qubit_0001_diff}}
     \subfigure[]{\includegraphics[width = 0.48\columnwidth]{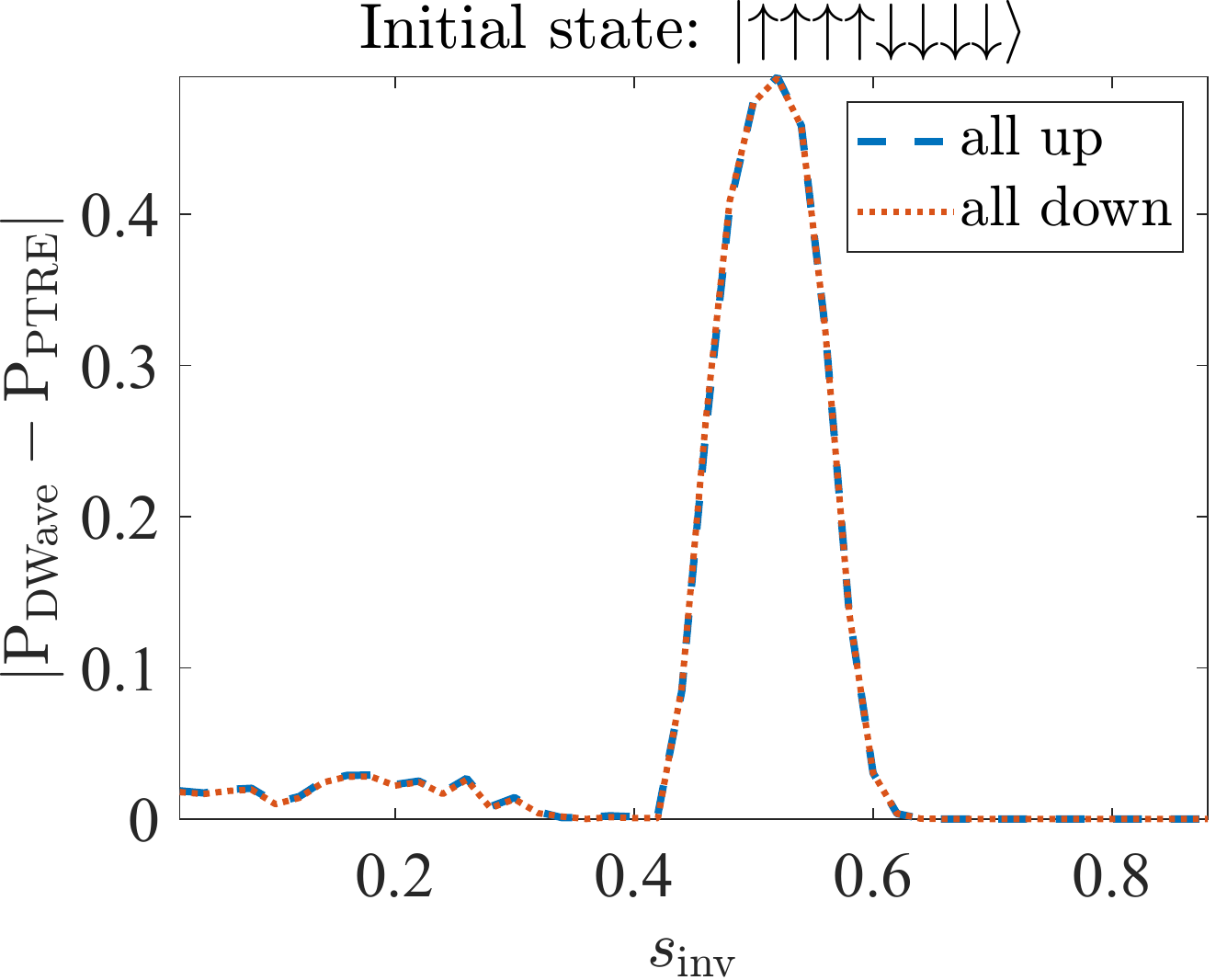}
         \label{fig:qubit_8_0011_diff_0}}
     \caption{(a) and (b): The population differences between the experimental and PTRE simulation results shown in Fig.~\ref{fig:optimal_n4} for $N=4$ qubits. (c) and (d): The population differences between the experimental and simulation results shown in Fig.~\ref{fig:compare_n8_ground} for $N=8$ qubits. Note the different scale of the vertical axes.}
     \label{fig:compare_n4_diff}
\end{figure}

Complementing the results shown in the main text, the differences (in absolute value) between the simulation and experimental results are shown in Fig.~\ref{fig:compare_n4_diff}.

\section{Fully classical simulations using the spin-vector Monte Carlo algorithm}
\label{sec:classical}

In an effort to better understand the asymmetric partial success probability observed in our experiments, we also performed fully classical simulations of the same problem using the spin-vector Monte Carlo (SVMC)~\cite{SSSV} algorithm and a new variant with transverse-field-dependent
updates (SVMC-TF)~\cite{albash2020comparing}, where this variant was successfully used to explain empirical D-Wave results for a particular $12$-qubit instance. For the $p$-spin problem, we replace the Hamiltonian of Eq.~\eqref{eq:H} by a classical Hamiltonian:
\begin{equation}
\mathcal{H}(s) = -\frac{A(s)}{2} \left(\sum_{i}^{N} \sin \theta_i\right) - \frac{B(s) N}{2}\left(\frac{1}{N}\sum_{i}^{N} \cos \theta_i\right)^{p}\,.
\label{eq:Hising}
\end{equation}
Each qubit $i$ is replaced by a classical $O(2)$ spin $\vec{M}_i=(\sin\theta_i,0,\cos\theta_i)$, $\theta_i \in [0, \pi]$. For the purpose of reverse annealing, we also need to specify the $t$ dependence of $s(t)$. The concept of time is here replaced by the number of Monte Carlo sweeps: we replace $\tau$ by a specified number of total sweeps. The total number of sweeps is then $2\tau(1-s_{\text{inv}})$, in analogy to the total annealing time in Eq.~\eqref{eq:t_a}. 

To simulate the effect of thermal hopping through this semi-classical landscape with inverse temperature $\beta$, we perform at each time step a spin update according to the Metropolis rule. In SVMC, a random angle $\theta_i^{'} \in [0, \pi]$ is picked for each spin $i$. Updates of the spin angles $\theta_i$ to $\theta_i^{'}$ are accepted according to the standard Metropolis rule associated with the change in energy ($\Delta E$) of the classical Hamiltonian. For the $p$-spin problem, $\Delta E$ cannot be expressed in a simple form as in case of the Ising problem Hamiltonian~\cite{SSSV}. 

In SVMC-TF, the random angle $\theta_i^{'} =\theta_i + \epsilon_i(s)$ is picked in a restricted range
\begin{align}
    \epsilon_i(s) \in \left[-\min \left(1,\frac{A(s)}{B(s)}\right) \pi, \min \left(1,\frac{A(s)}{B(s)}\right) \pi\right]. 
\end{align}
The goal of SVMC-TF is to restrict the angle update for $A(s) < B(s)$, and imitate the freeze-out effect discussed in Sec.~\ref{sec:freezeout}. The full SVMC and SVMC-TF algorithms for reverse annealing are summarized in App.~\ref{append:svmc}, including the expression for $\Delta E$. 

\begin{figure}[t]
\includegraphics[width = 0.9\columnwidth]{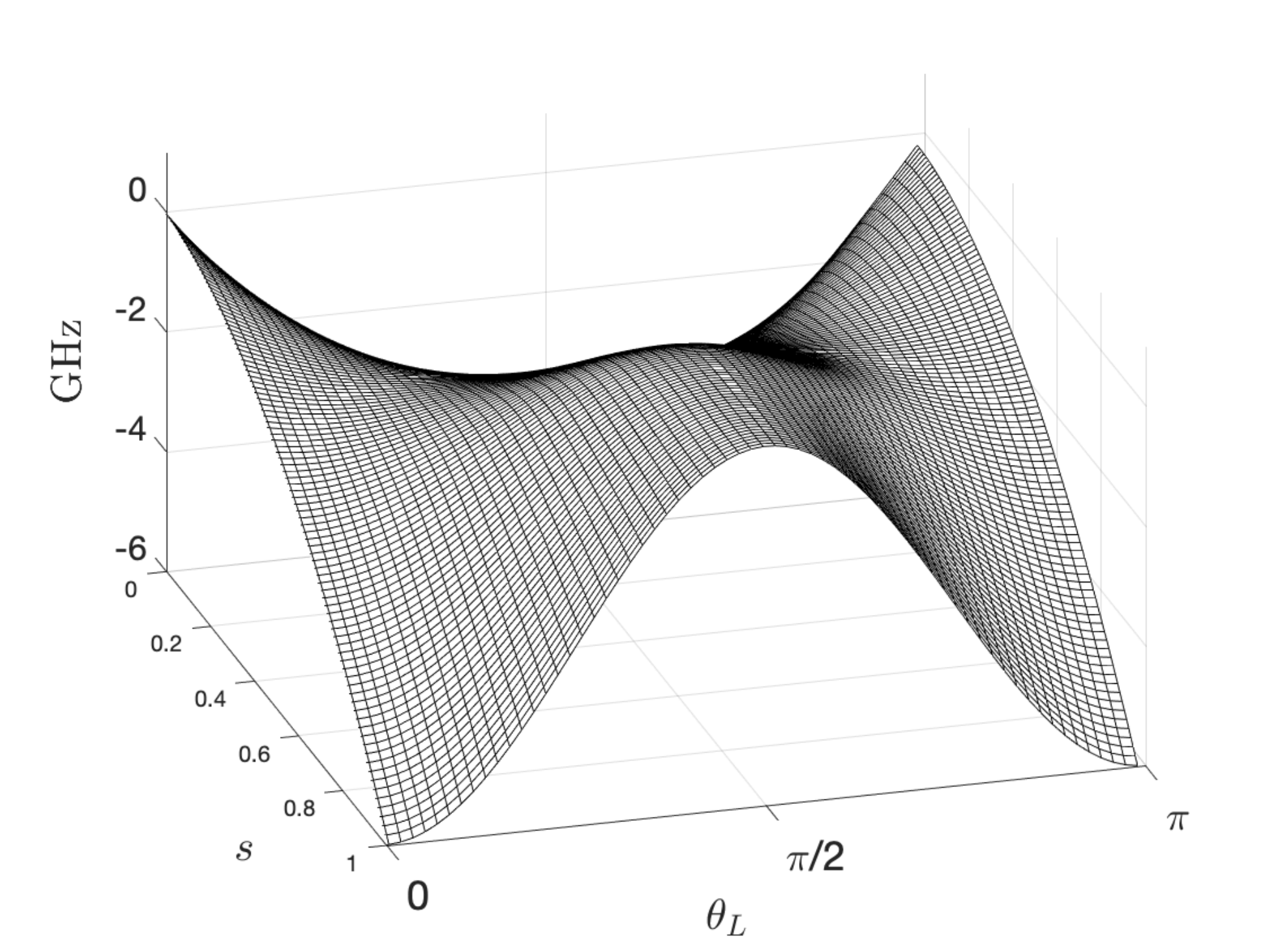}
\caption{The semiclassical potential landscape corresponding to the Hamiltonian of Eq.~\eqref{eq:Hising} for $N=1$. Ground states correspond to $\theta = 0, \pi$. The two saddle points are at $s=0.3948$ and $\theta = \pi/2\pm 0.31518\pi$.}
\label{fig:surface}
\end{figure}

A simple intuitive way to visualize the semiclassical dynamics described by the SMVC and SVMC-TF algorithms is to consider the energy landscape defined by $\mathcal{H}(s)$ [Eq.~\eqref{eq:Hising}] when equating all angles ($\theta_i \equiv \theta$). We plot the resulting surface in Fig.~\ref{fig:surface} for the simple case of $N=1$. For $s_{\mathrm{inv}}< 0.3948$, Metropolis updates to either spin direction happen frequently and are equally likely since there is no potential barrier separating them.  For $1 \gg s_{\mathrm{inv}}> 0.3948$, under Metropolis updates the system prefers staying in the original well and escaping to the opposite well is unlikely, due to the potential barrier. This preference explains the asymmetry part of partial success probabilities in the experimental results. For $1 > s_{\mathrm{inv}} \gg 0.3948$, Metropolis updates are very rare due to the rate suppression in SVMC-TF.

\subsection{Total success probability}

We report on SVMC and SVMC-TF simulations for $N=4$ and $8$. We choose the initial condition with a single spin down. In terms of angles, for $N=4$ the initial angles are $\{0,0,0,\pi\}$. We again use the temperature $T=12.1$\;mK. The classical analogue of the annealing time is chosen to be $\tau = 10^3$ and $10^4$ sweeps. 

\begin{figure}[t]
\subfigure[]{\includegraphics[width = 0.4939\columnwidth]{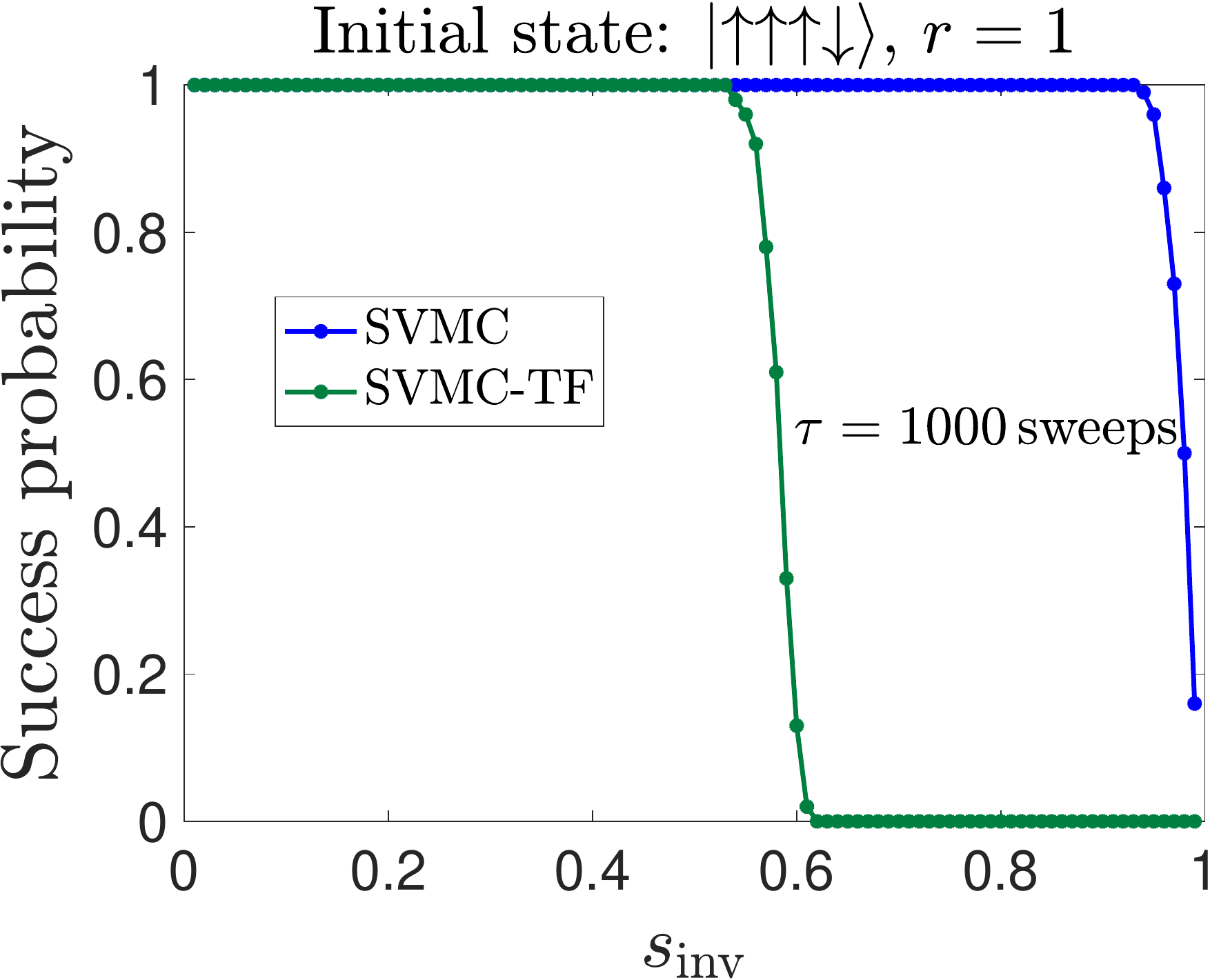}}
\subfigure[]{\includegraphics[width = 0.4939\columnwidth]{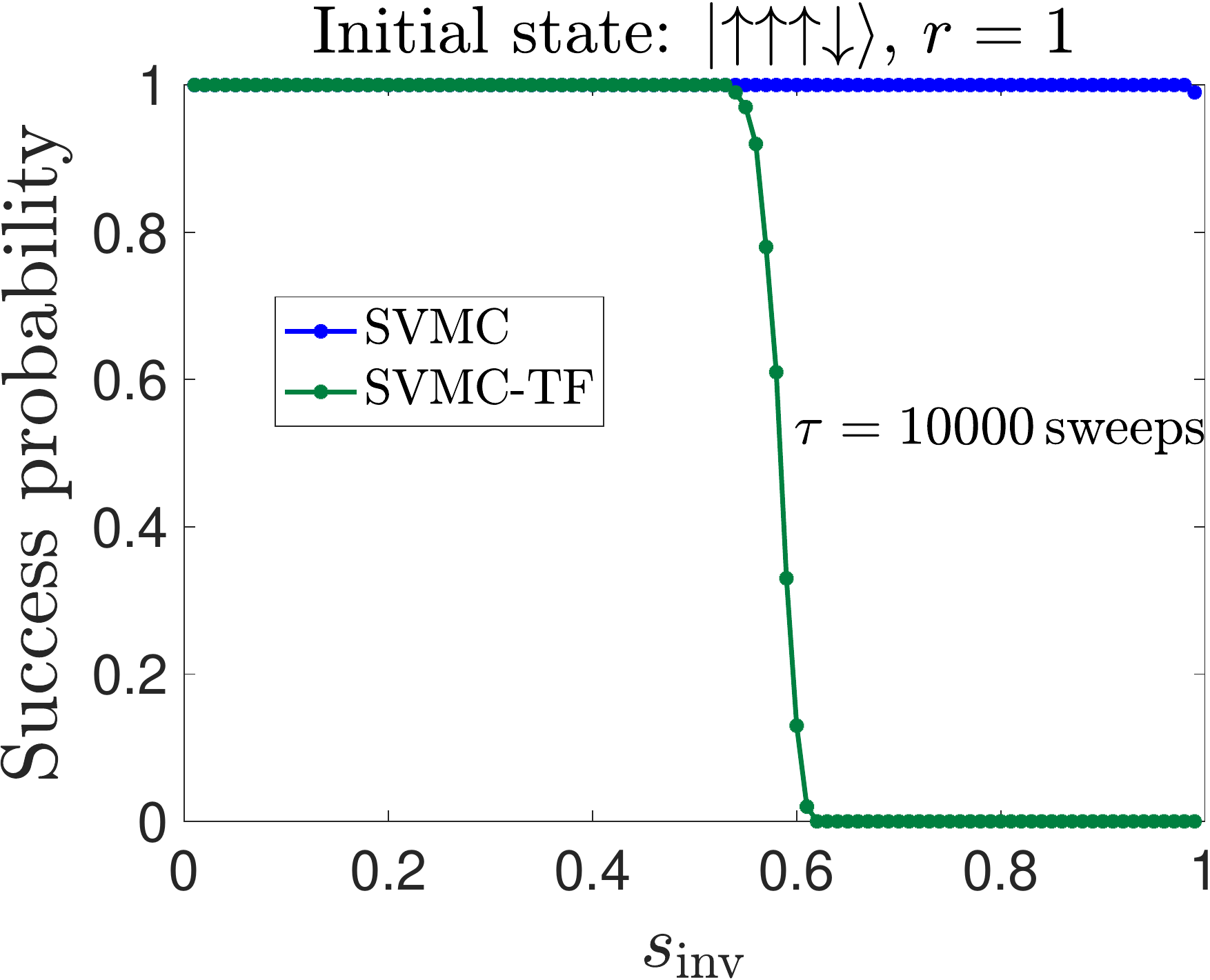}}
\caption{Total success probabilities as computed by SVMC and SVMC-TF. (a) $\tau = 10^3$ sweeps, (b) $\tau = 10^4$ sweeps. Error bars are 2$\sigma$ over $10^4$ samples.}
\label{fig:comparealltotal}
\end{figure}

In Fig.~\ref{fig:comparealltotal} we display the simulation results for the total success probability using SVMC and SVMC-TF. The number of samples is $10^4$.
SVMC gives high total success probabilities even with large inversion points $s_{\text{inv}}$. A large $s_{\text{inv}}$ value means that during the whole reverse annealing process the ratio $A(s)/B(s)$ is small. In the D-Wave device, it is expected that for small $A(s)/B(s)$ the dynamics freezes, which makes it difficult to reach the ground state(s) when the initial state is excited. With the number of sweeps increased from $\tau=10^3$ to $10^4$, we observe that even for very large inversion points $s_{\text{inv}}$ the total success probability of SVMC can be as high as $1$. This is because the angle updates in SVMC are completely random and thus with a sufficient number of sweeps, it is possible for the state to flip to the correct solutions.

However, in SVMC-TF, the range of angle updates is restricted for $A(s)/B(s) < 1$. The restricted angle updates (freeze-out effect) prevent the state from flipping to the correct solutions. Therefore the total success probability for large inversion points $s_{\text{inv}}$ is basically zero in SVMC-TF, regardless of the number of sweeps. This is also what we observe in the empirical data and in the adiabatic master equation simulations, as discussed in Sec.~\ref{sec:AME-tau}.

\begin{figure}[t]
\subfigure[]{\includegraphics[width = 0.4939\columnwidth]{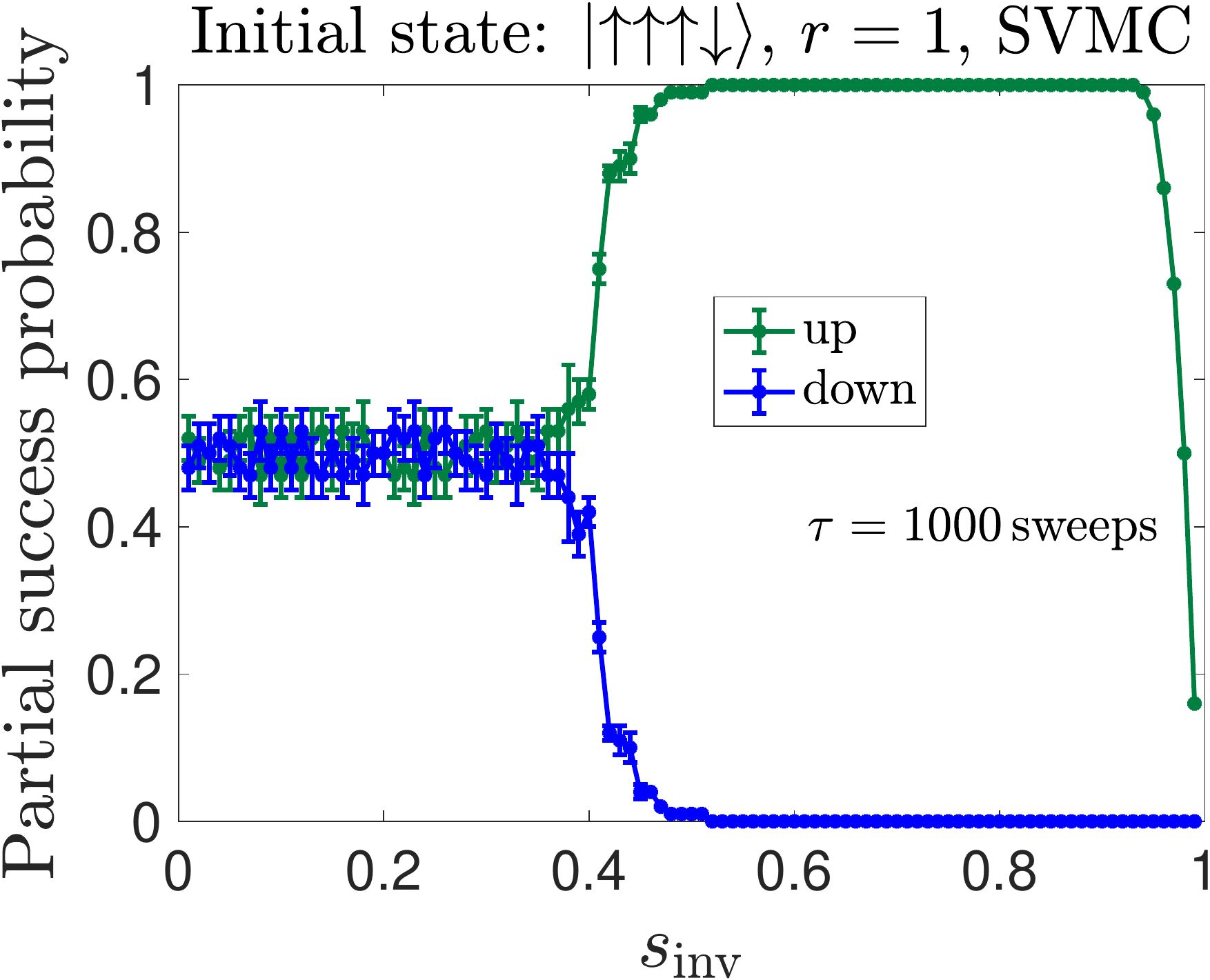}}
\subfigure[]{\includegraphics[width = 0.4939\columnwidth]{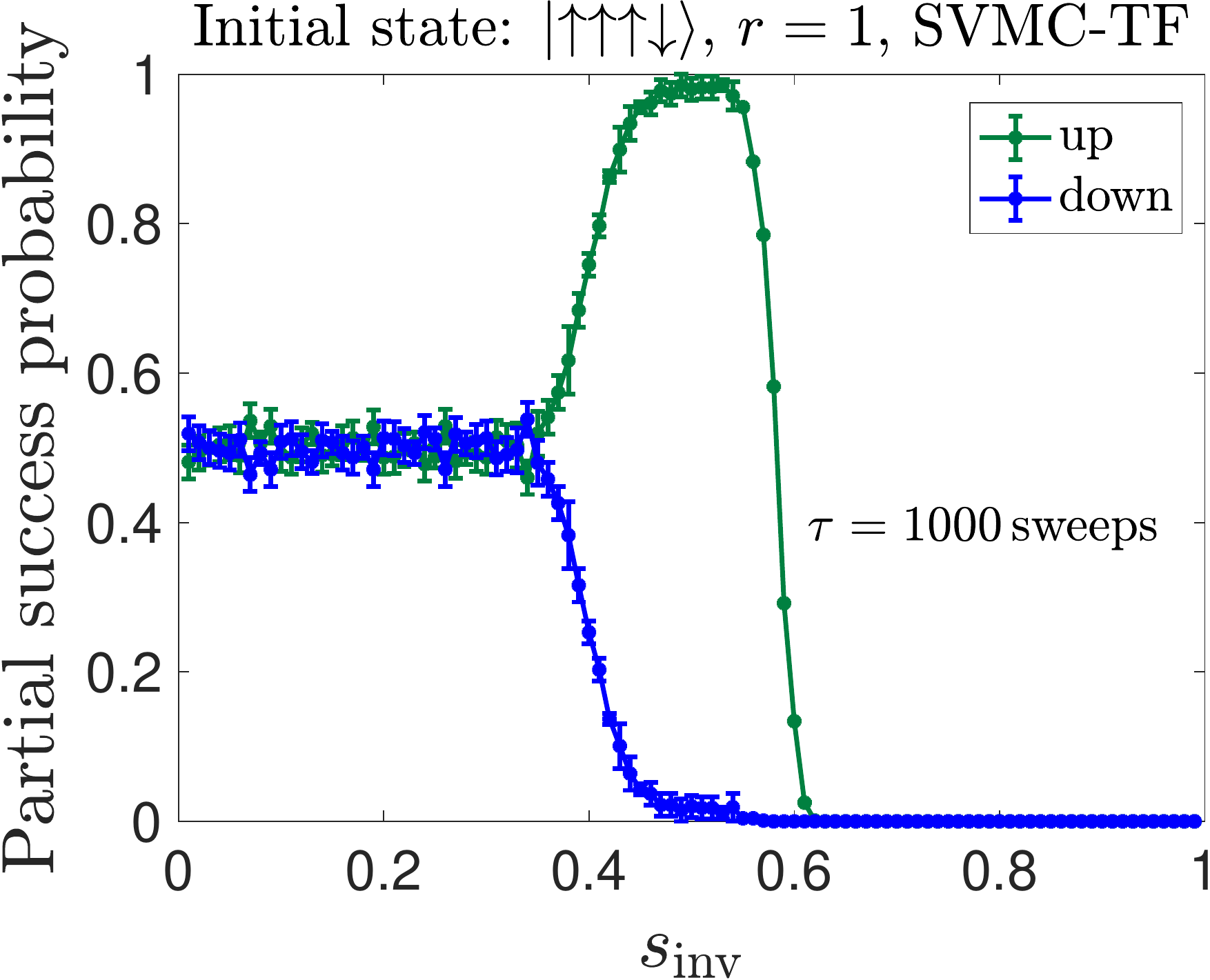}}
\subfigure[]{\includegraphics[width = 0.4939\columnwidth]{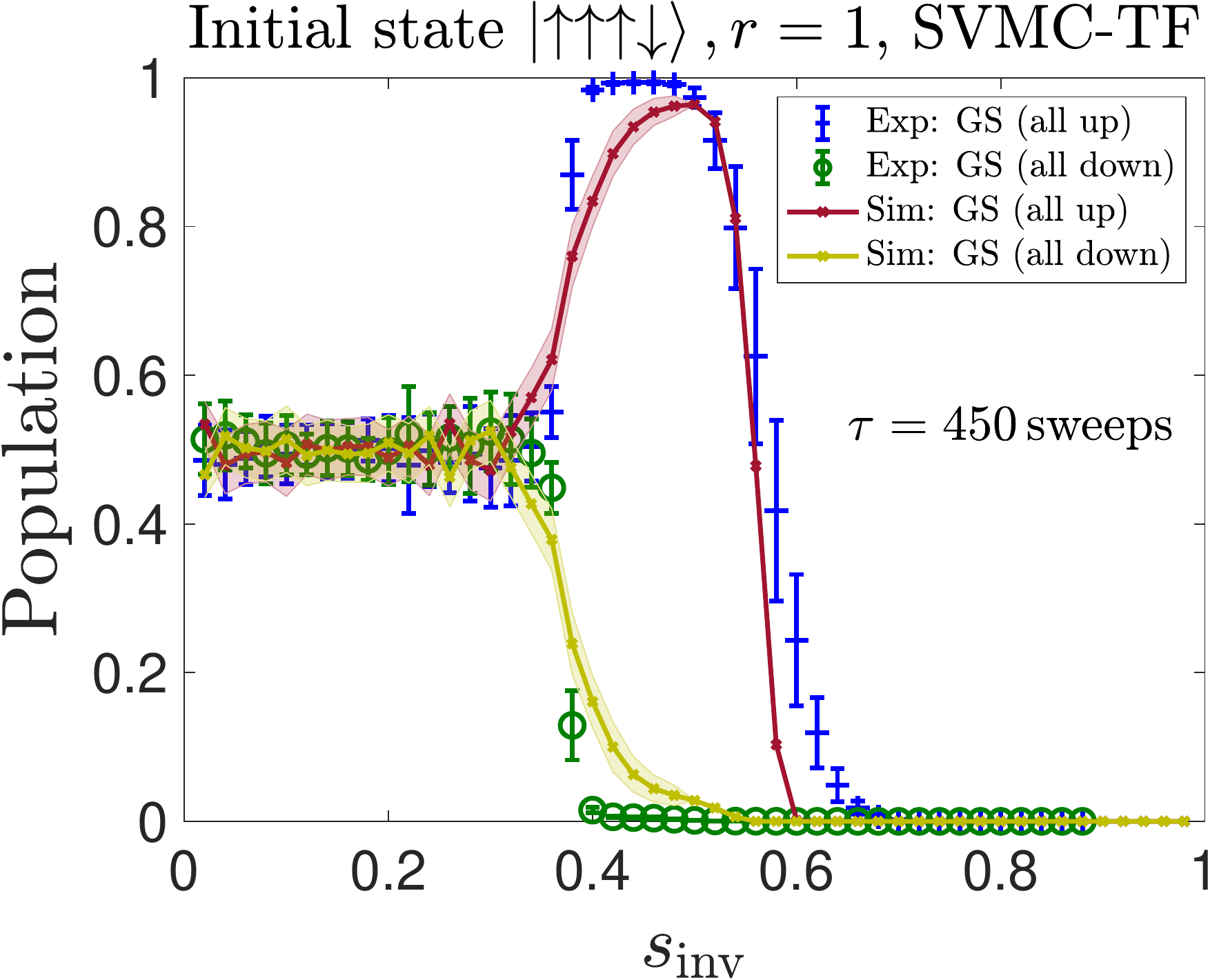}}
\subfigure[]{\includegraphics[width = 0.4939\columnwidth]{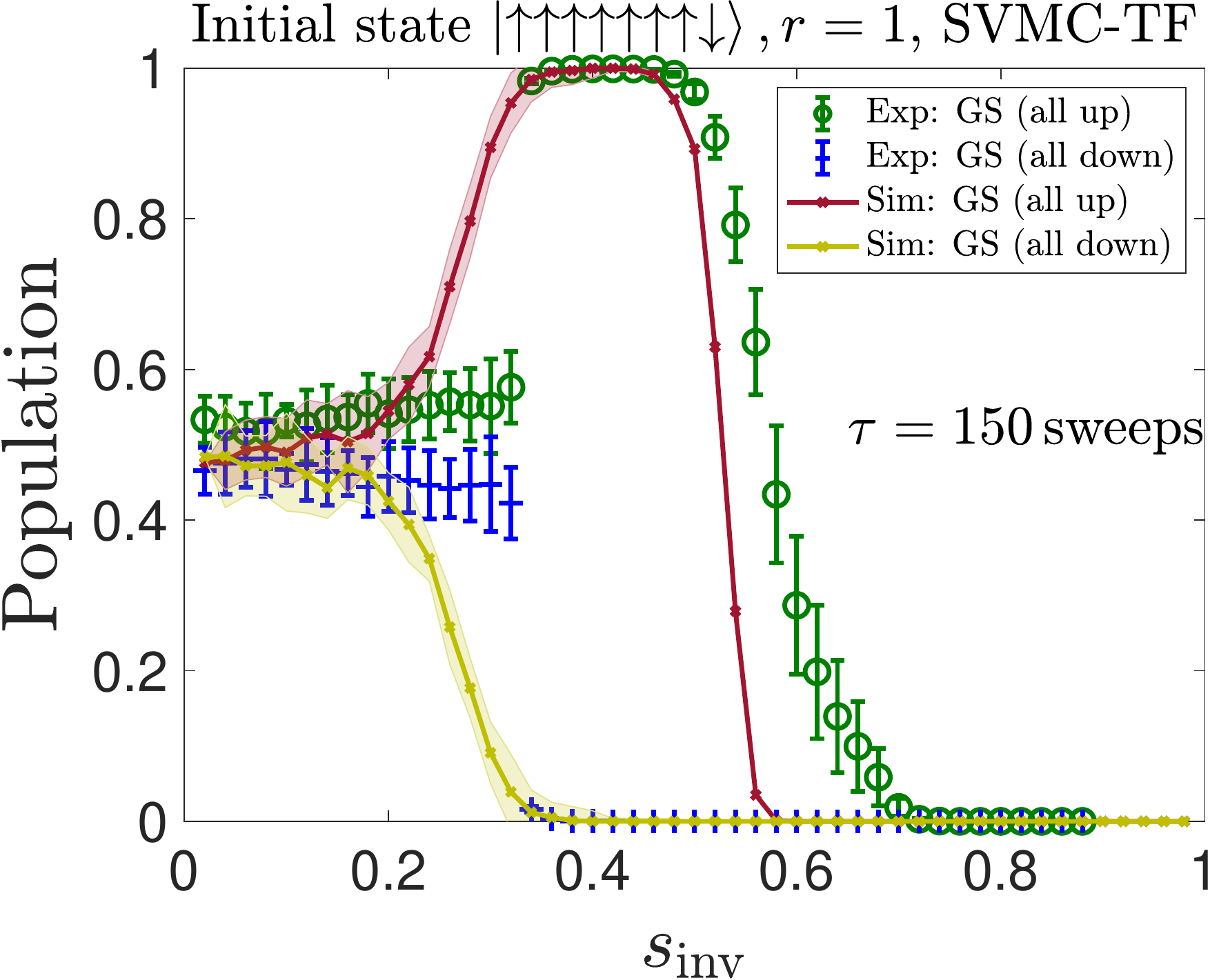}}
\caption{Partial success probabilities for the one-down initial state as computed by (a) SVMC for $N=4$, (b) SVMC-TF for $N=4$. Panels (c) and (d) show the partial ground state (GS) success probabilities for SVMC-TF (``Sim.'') at optimized sweeps numbers of (c) $\tau=450$ for $N=4$ and (d) $\tau=150$ for $N=8$, which yields a qualitatively good agreement with the experimental data (``Exp.''). 
In panels (a) and (b) $\tau = 10^3$ sweeps and error bars are $2\sigma$ over $10^4$ samples. In panels (c) and (d) error bars are $1\sigma$ over $10^4$ samples.}
\label{fig:compareallpartial}
\end{figure}

\begin{figure}[t]
\subfigure[]{\includegraphics[width = 0.4939\columnwidth]{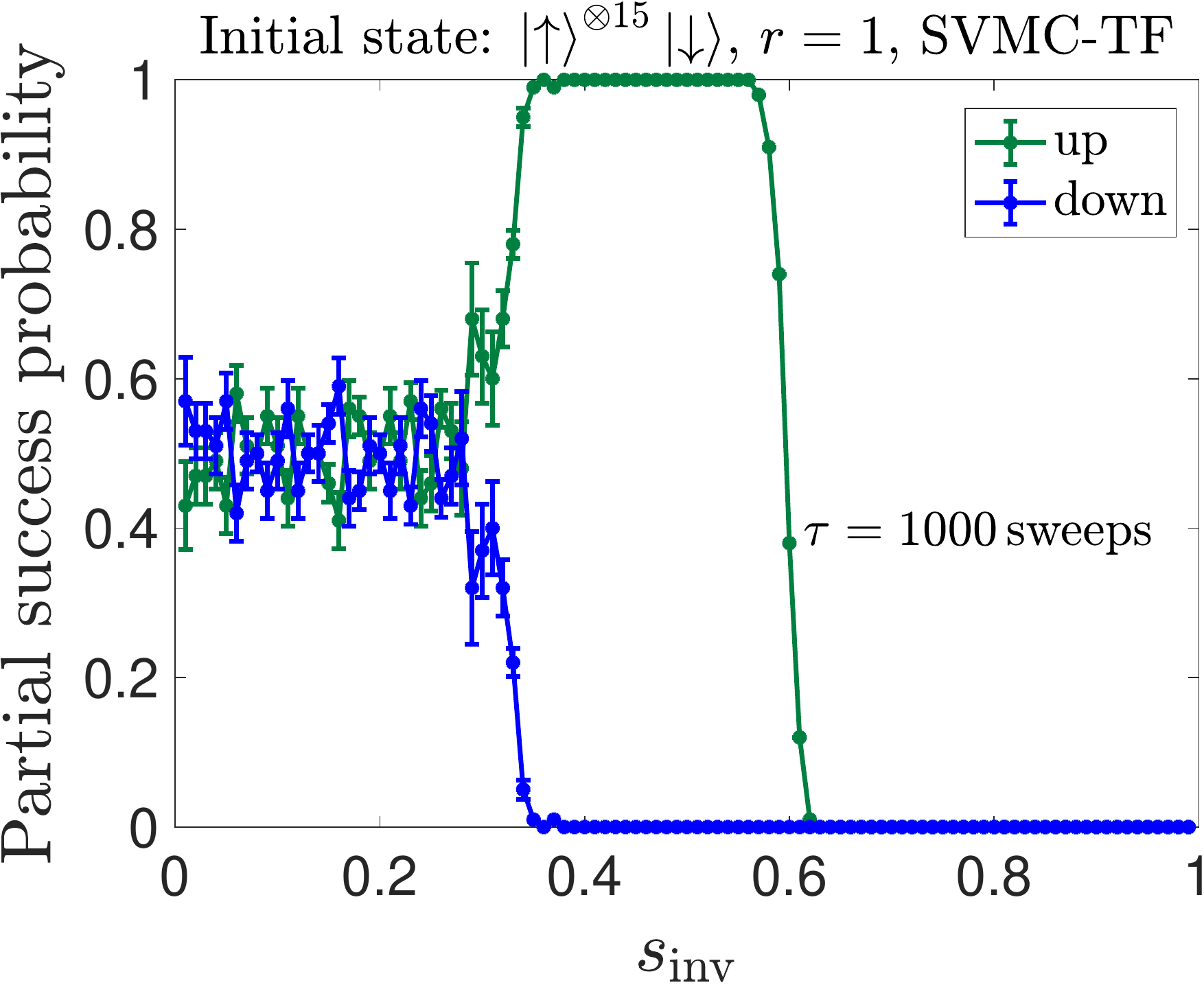}}
\subfigure[]{\includegraphics[width = 0.4939\columnwidth]{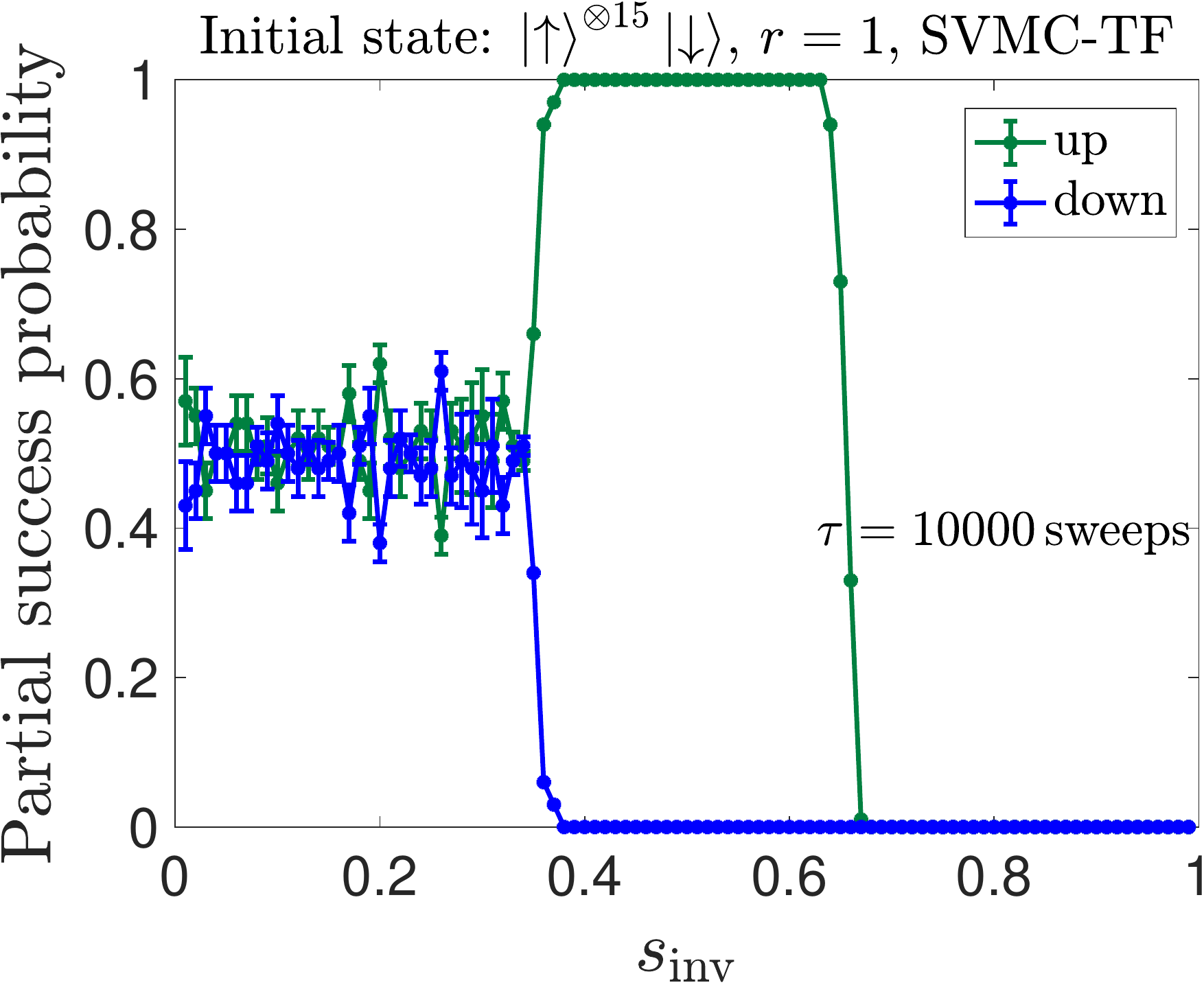}}
\subfigure[]{\includegraphics[width = 0.4939\columnwidth]{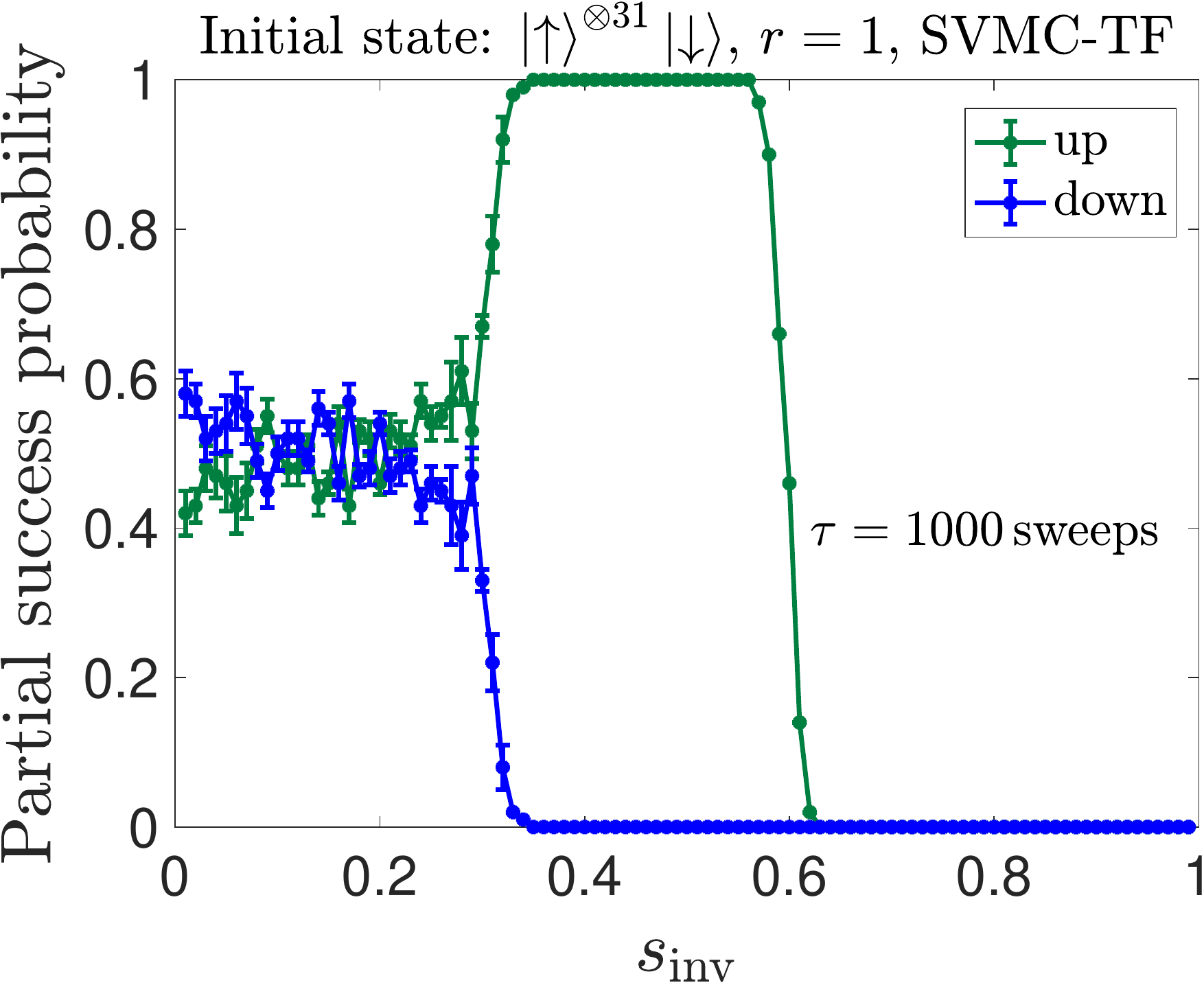}}
\subfigure[]{\includegraphics[width = 0.4939\columnwidth]{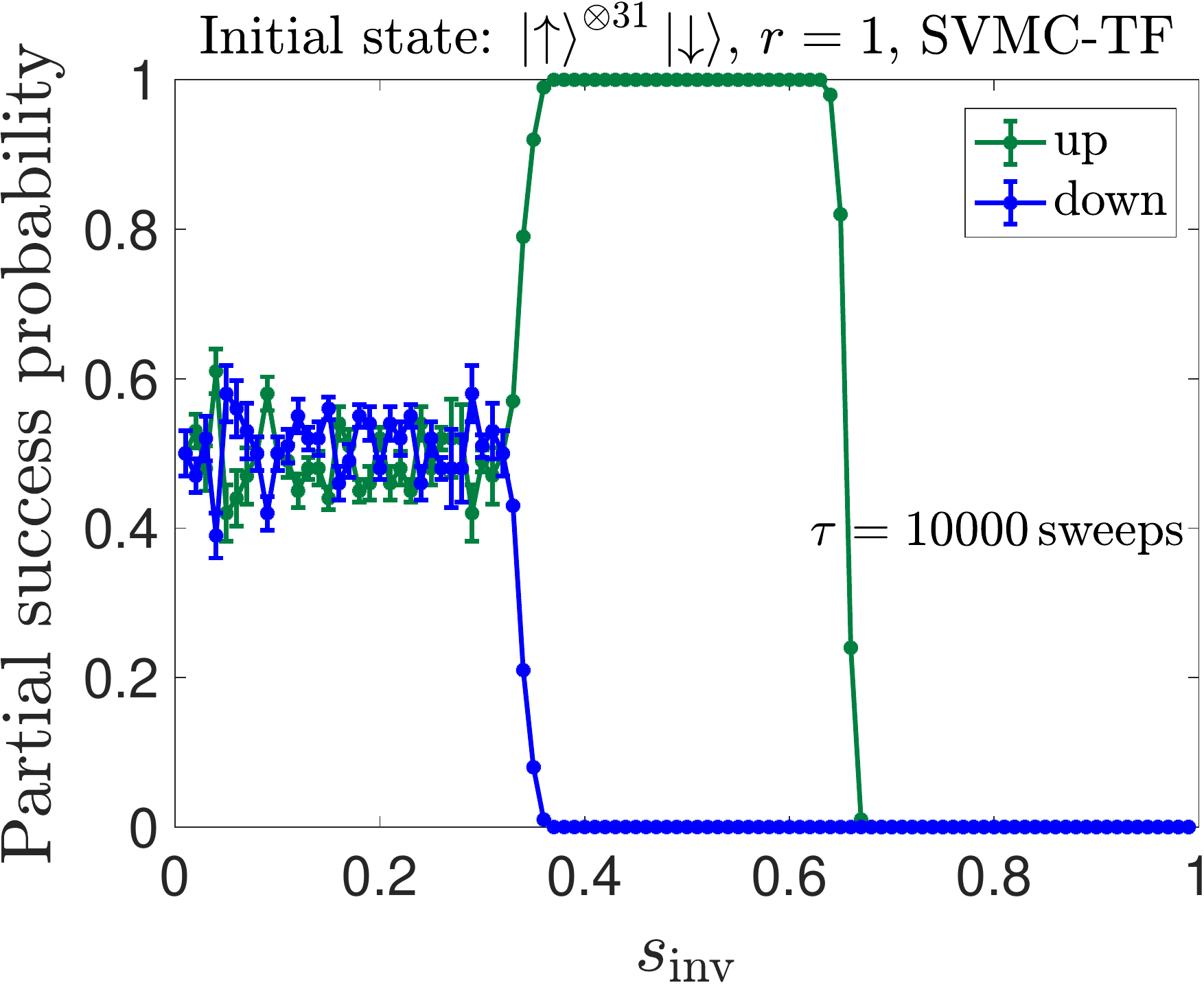}}
\caption{Partial success probabilities as computed by SVMC-TF. (a) $N=16$, $\tau = 10^3$ sweeps, (b) $N=16$, $\tau = 10^4$ sweeps, (c) $N=32$, $\tau = 10^3$ sweeps, (d) $N=32$, $\tau = 10^4$ sweeps. Error bars are 2$\sigma$ over $10^4$ samples.}
\label{fig:partial16}
\end{figure}

\begin{figure*}[t]
\subfigure[]{\includegraphics[width = 0.32\textwidth]{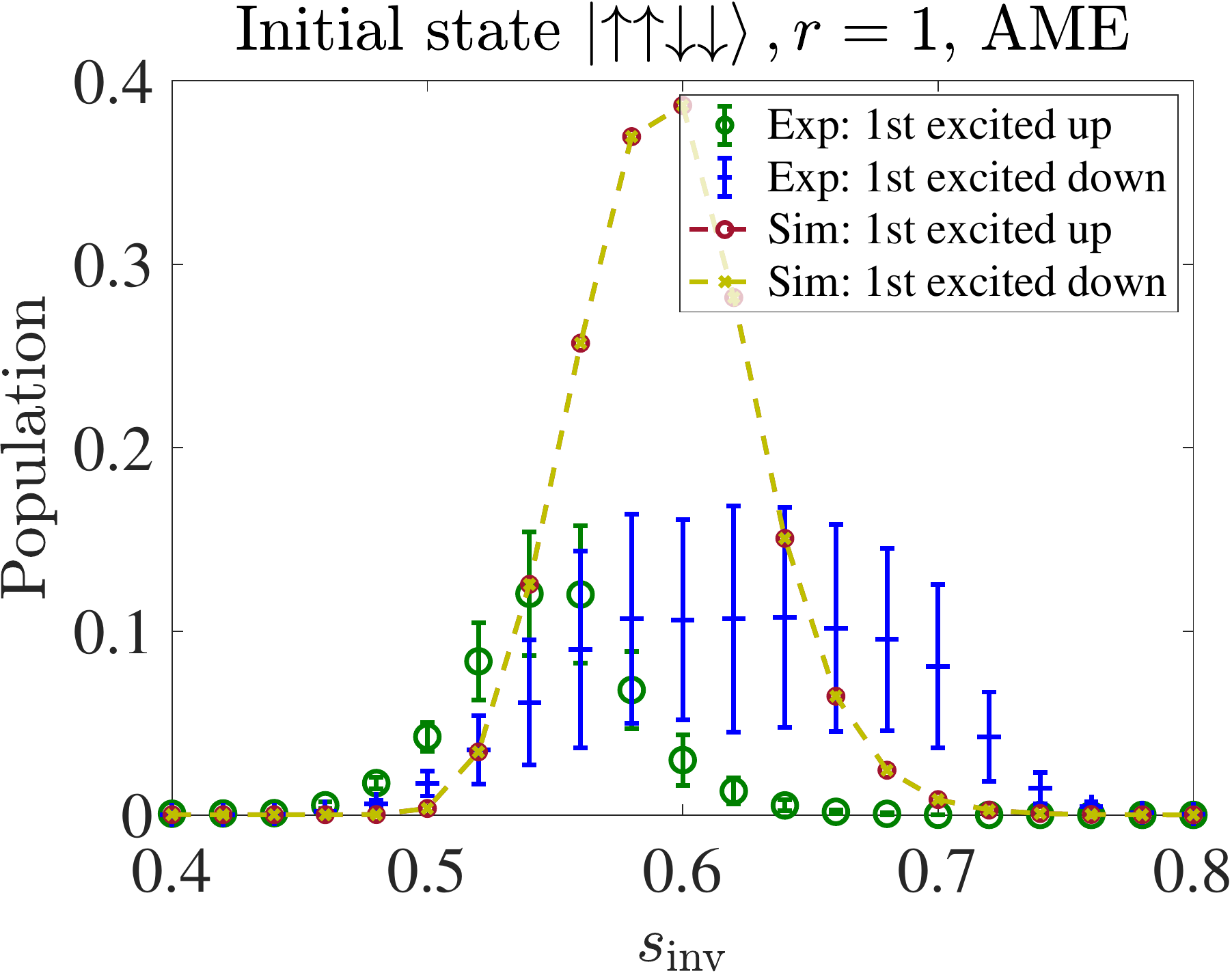}}
\subfigure[]{\includegraphics[width = 0.32\textwidth]{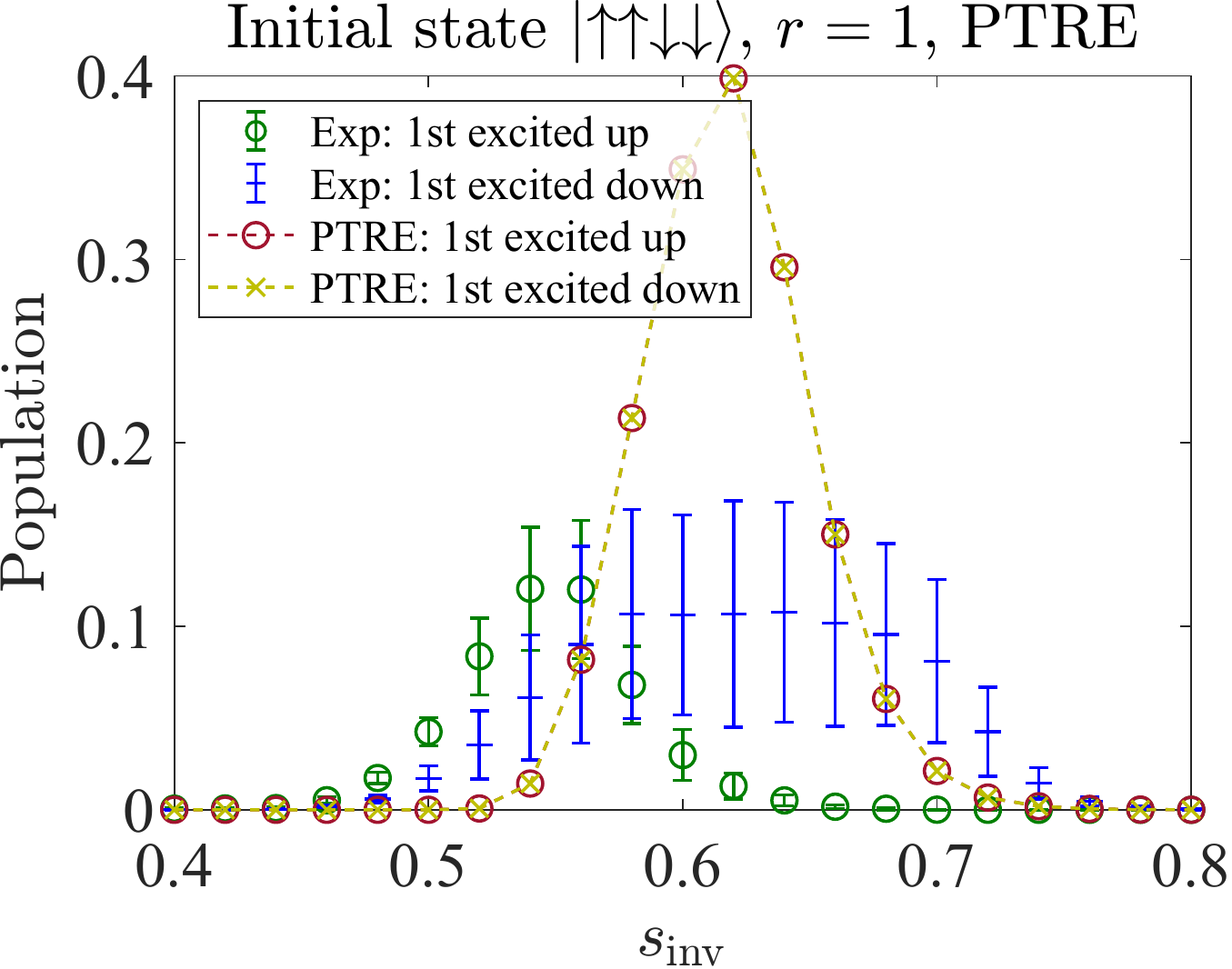}}
\subfigure[]{\includegraphics[width = 0.32\textwidth]{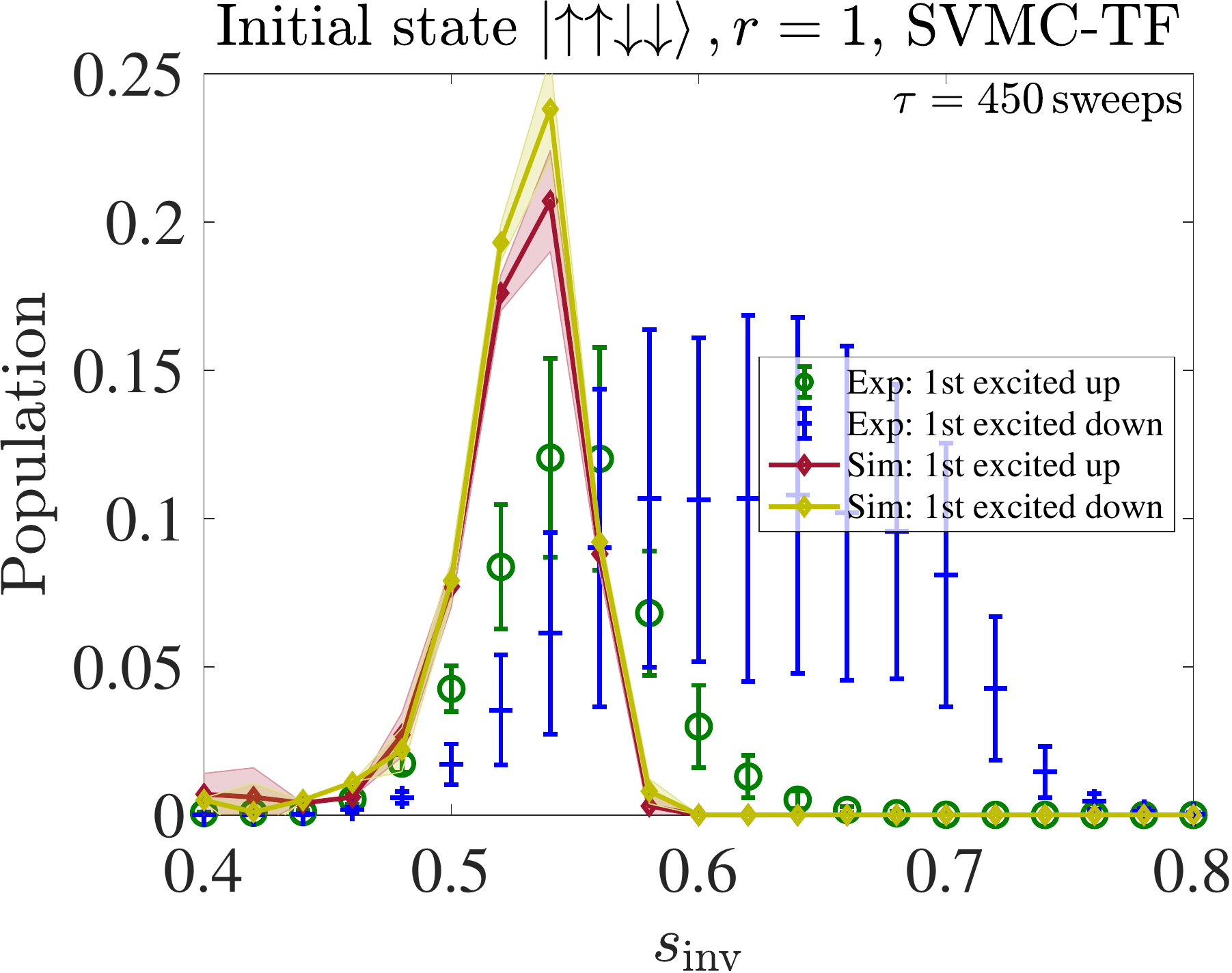}}
\caption{Population in the first excited state as a function of $s_{\text{inv}}$ for D-Wave (Exp) \textit{vs} simulations (Sim): (a) the AME, (b) the PTRE, and (c) SVMC-TF with $\tau=450$ sweeps. The initial state is  $\ket{\uparrow\uparrow\downarrow\downarrow}$. Error bars denote $1\sigma$ over $10^4$ samples. The slight difference in SVMC-TF peak heights in panel (c) is a numerical artifact. Note the different scale of the vertical axis of panel (c).}
\label{fig:AMEexcited_0011}
\end{figure*}

\begin{figure*}[t]
\subfigure[]{\includegraphics[width = 0.32\textwidth]{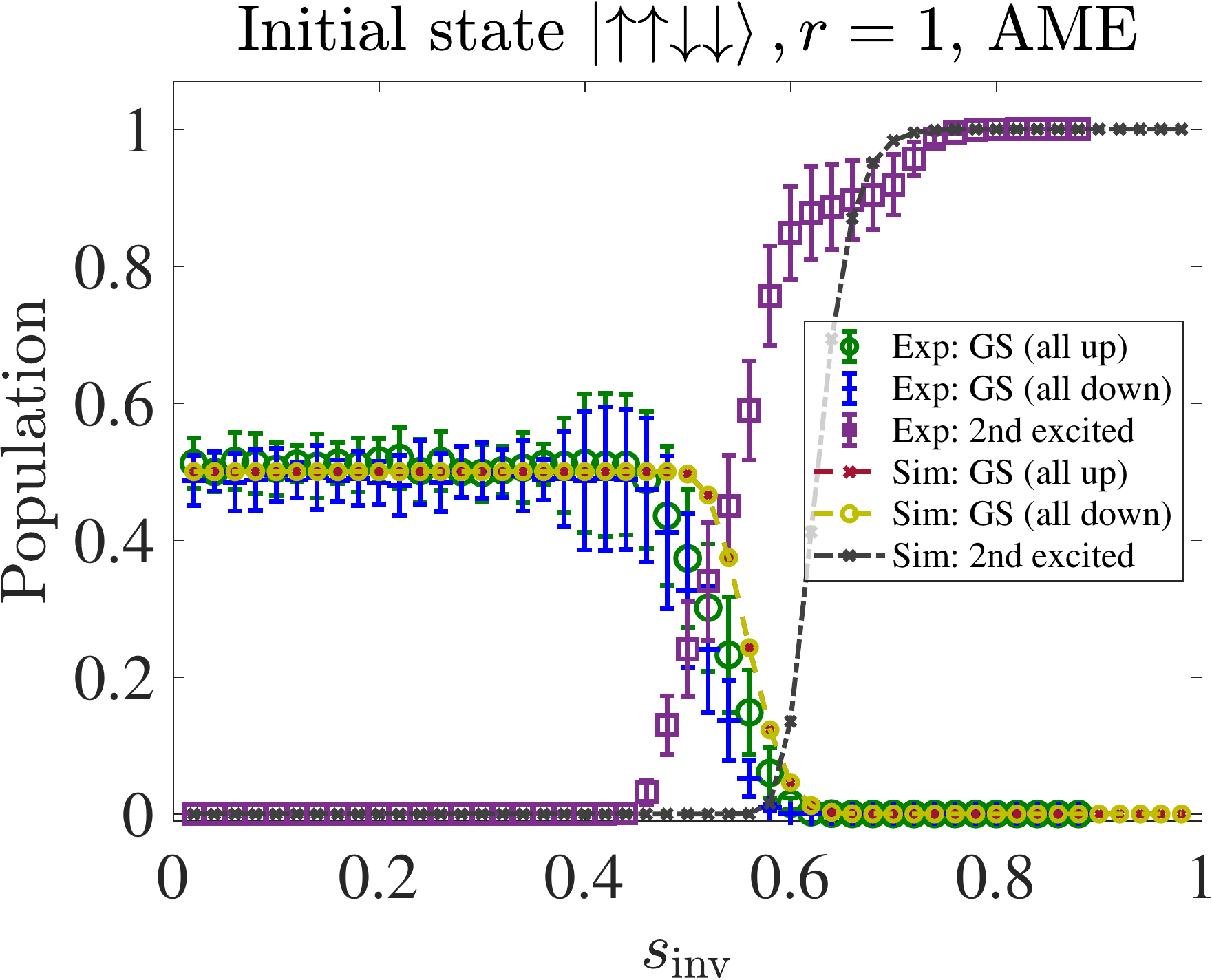}}
\subfigure[]{\includegraphics[width = 0.32\textwidth]{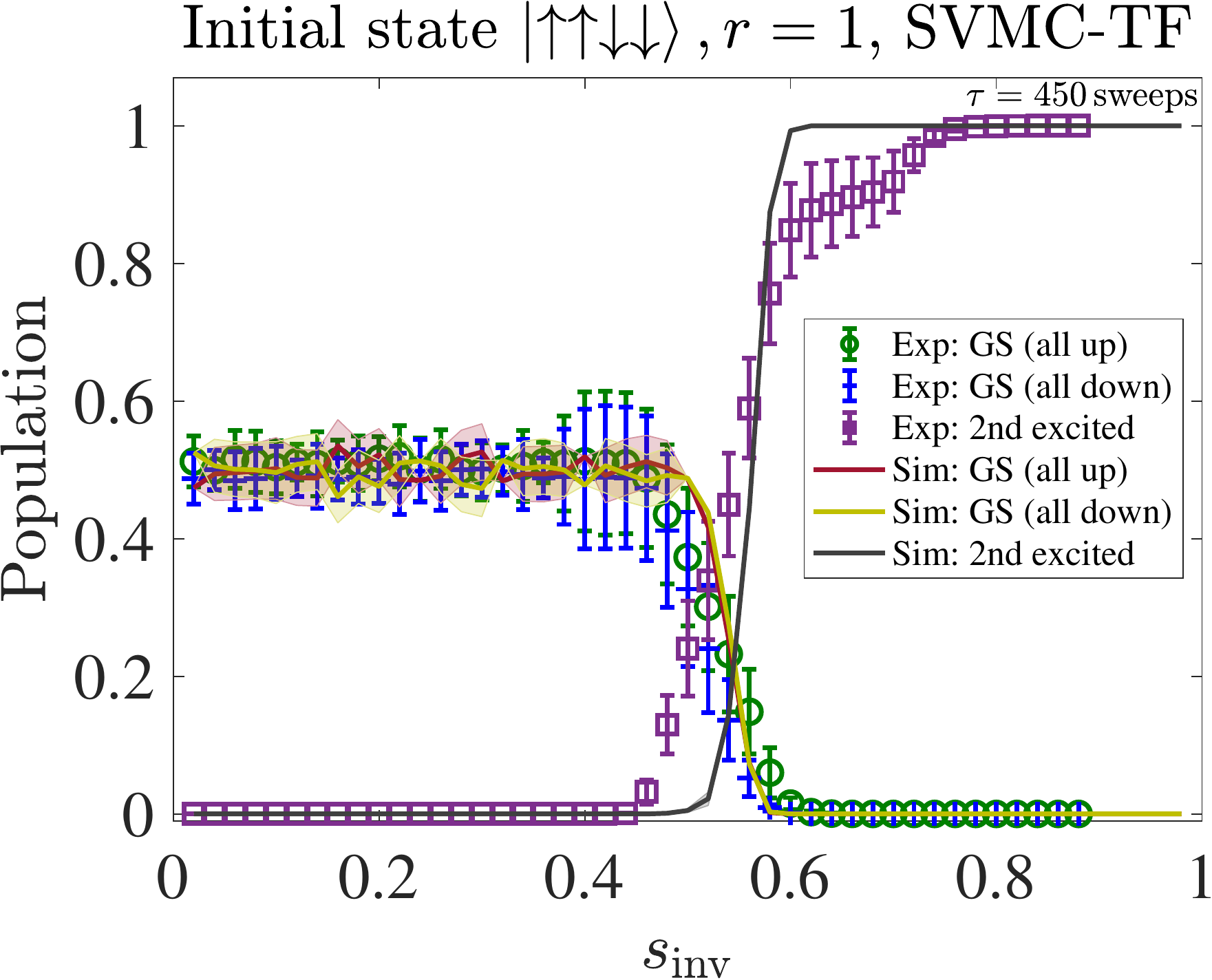}}
\subfigure[]{\includegraphics[width=0.32\textwidth]{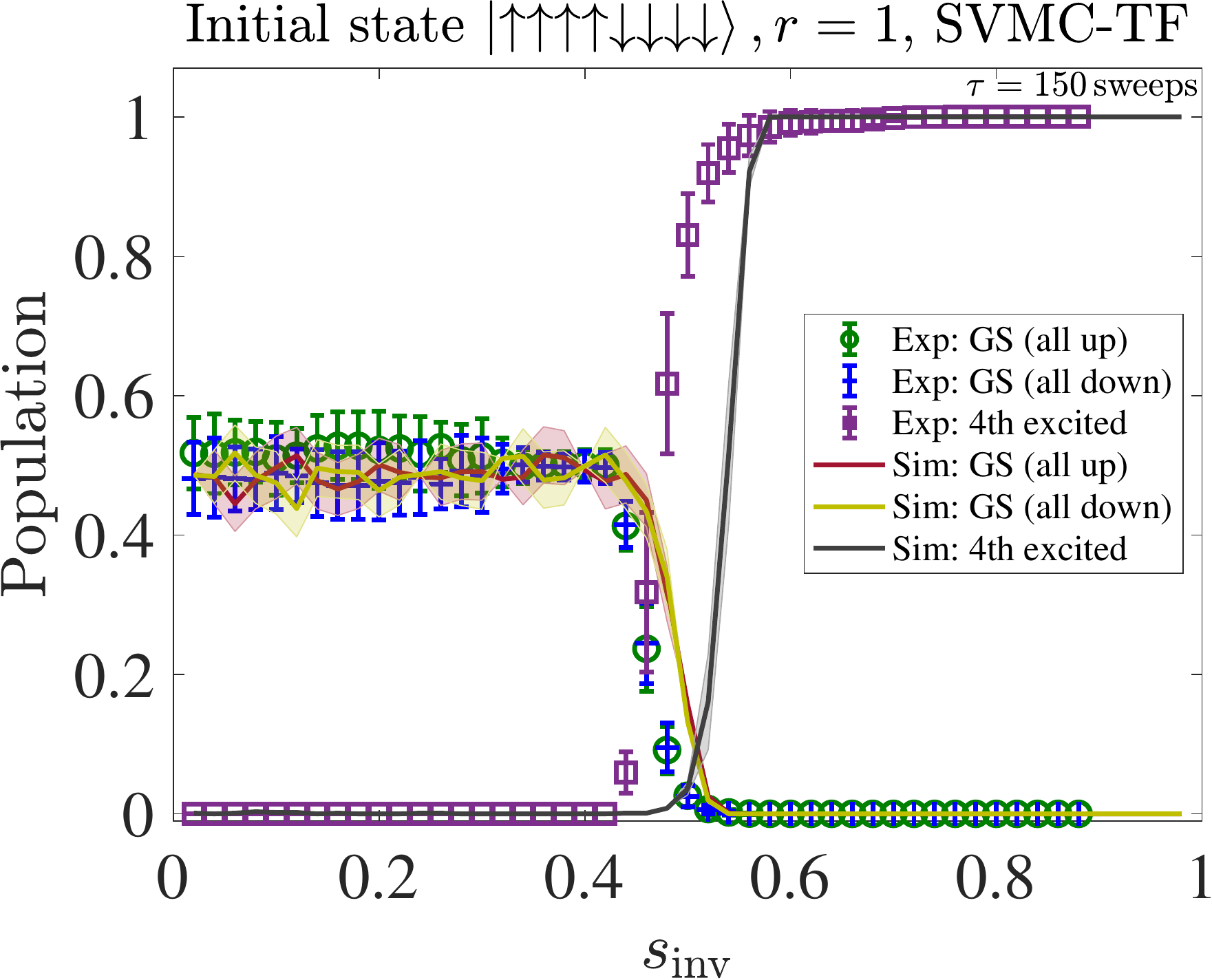}}
\caption{Population in the ground and second excited states as a function of $s_{\text{inv}}$ for D-Wave (Exp) \textit{vs} simulations (Sim): (a) the AME and (b) SVMC-TF with $\tau=450$ sweeps, for the initial state $\ket{\uparrow\uparrow\downarrow\downarrow}$. (c) Ground and fourth excited states for D-Wave \textit{vs} SVMC-TF with $\tau=150$ sweeps, for the initial state $\ket{\uparrow\uparrow\uparrow\uparrow\downarrow\downarrow\downarrow\downarrow}$. Error bars denote $1\sigma$ over $10^4$ samples.}
\label{fig:SVMC-exc2}
\end{figure*}

\begin{figure*}[t]
     \subfigure[]{\includegraphics[width = 0.32\textwidth]{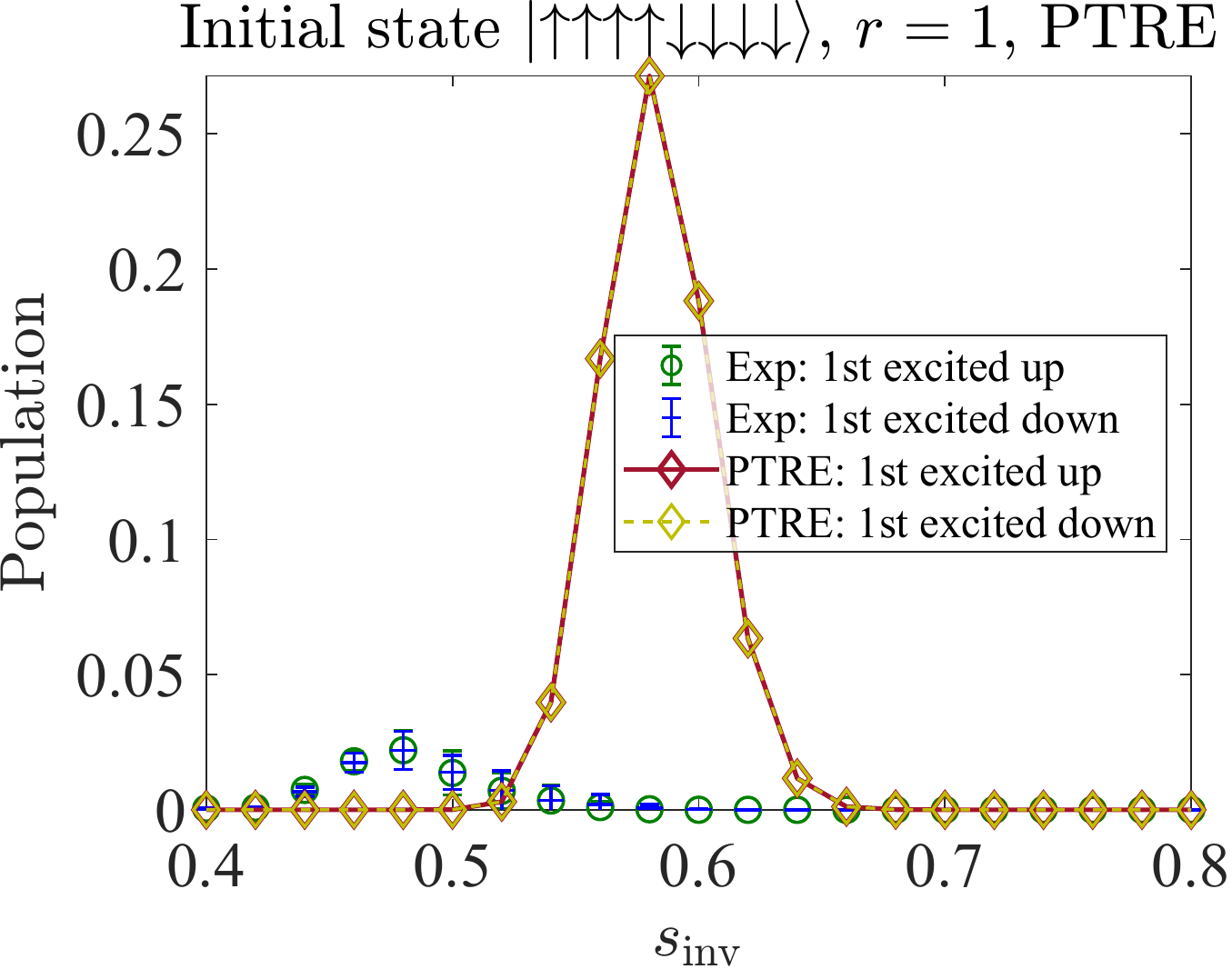}
         \label{fig:8qubit_0001_1}}
     \subfigure[]{\includegraphics[width = 0.32\textwidth]{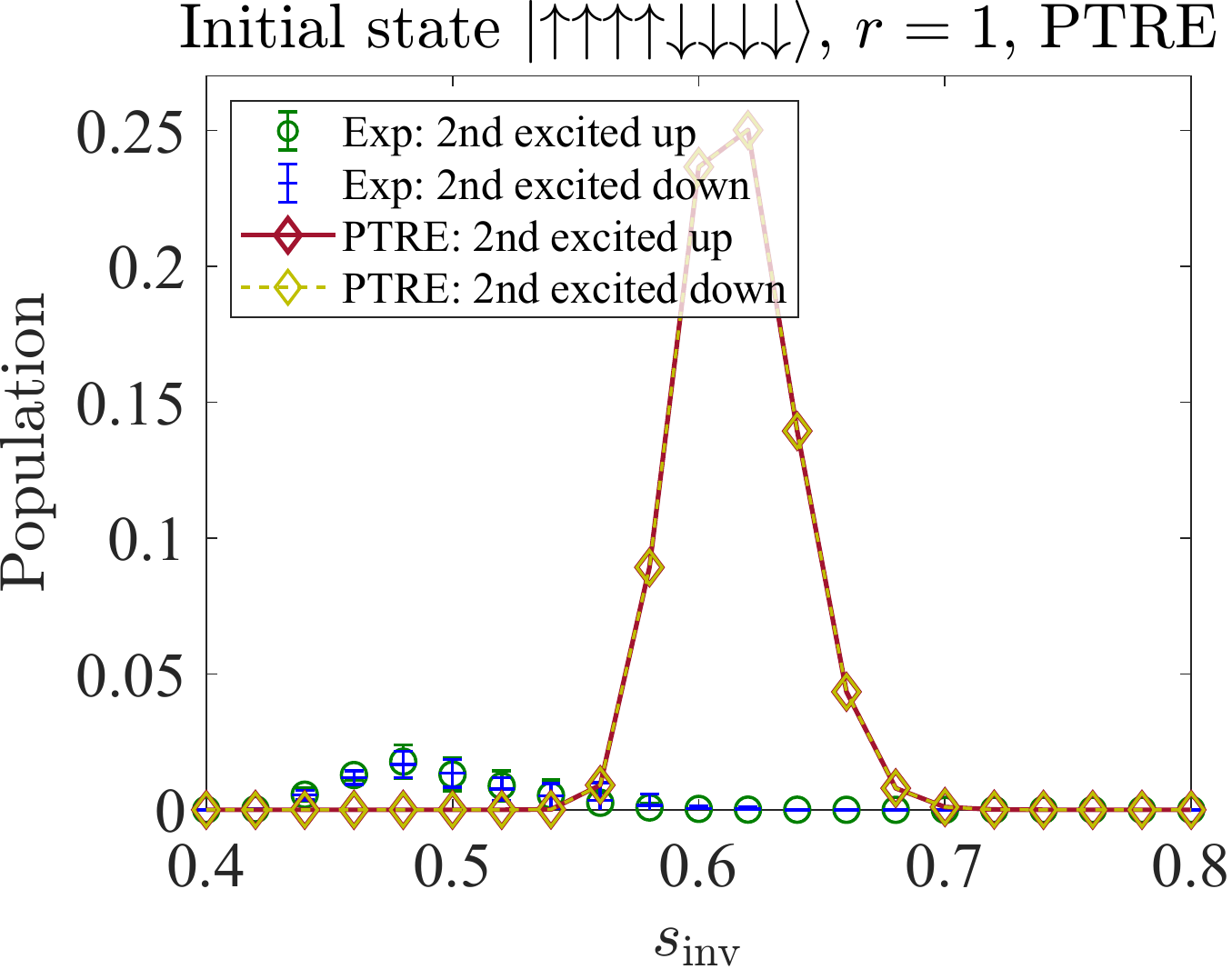}
         \label{fig:8qubit_0011_2}}
     \subfigure[]{\includegraphics[width = 0.32\textwidth]{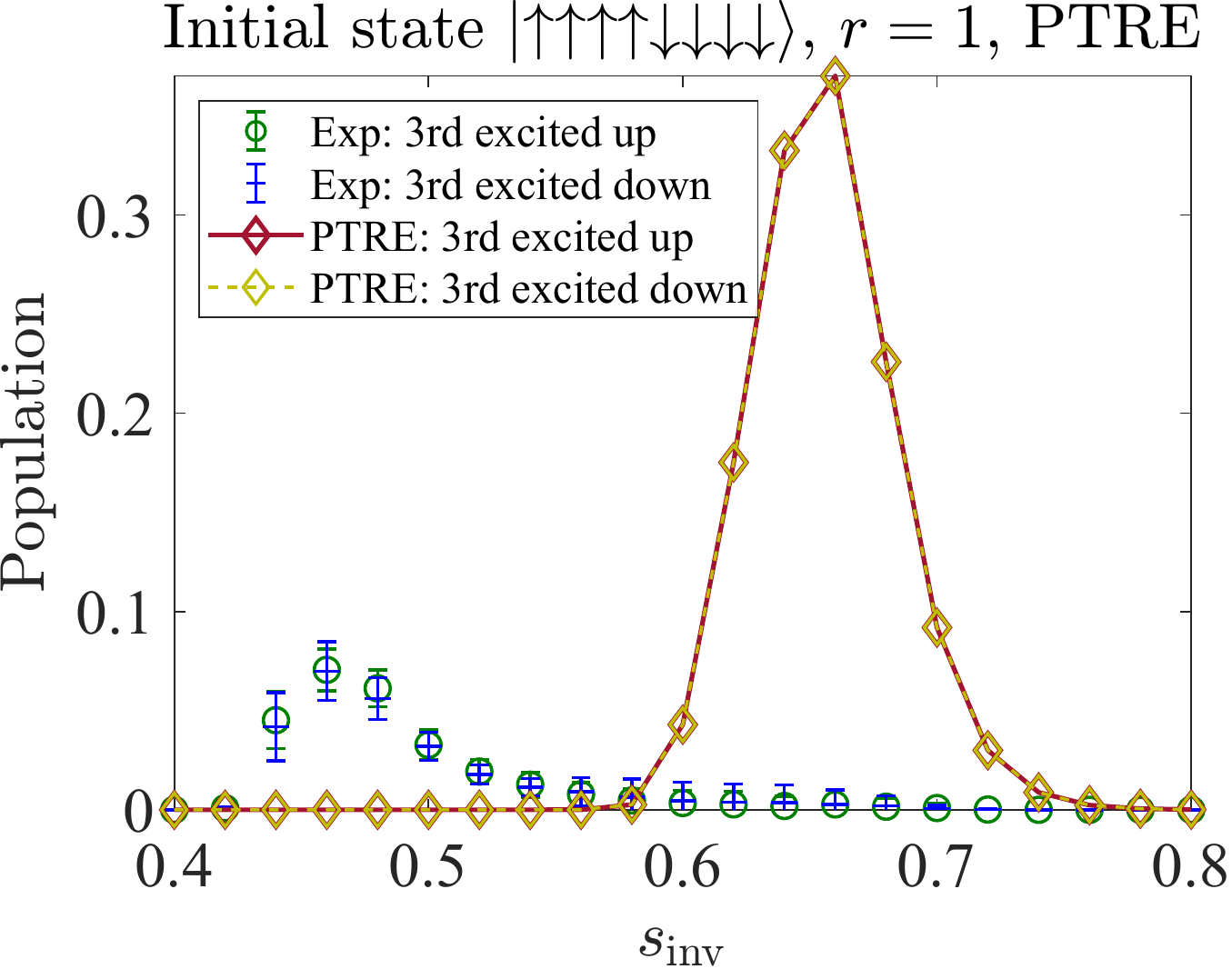}
         \label{fig:8qubit_0011_3}}
         \subfigure[]{\includegraphics[width = 0.32\textwidth]{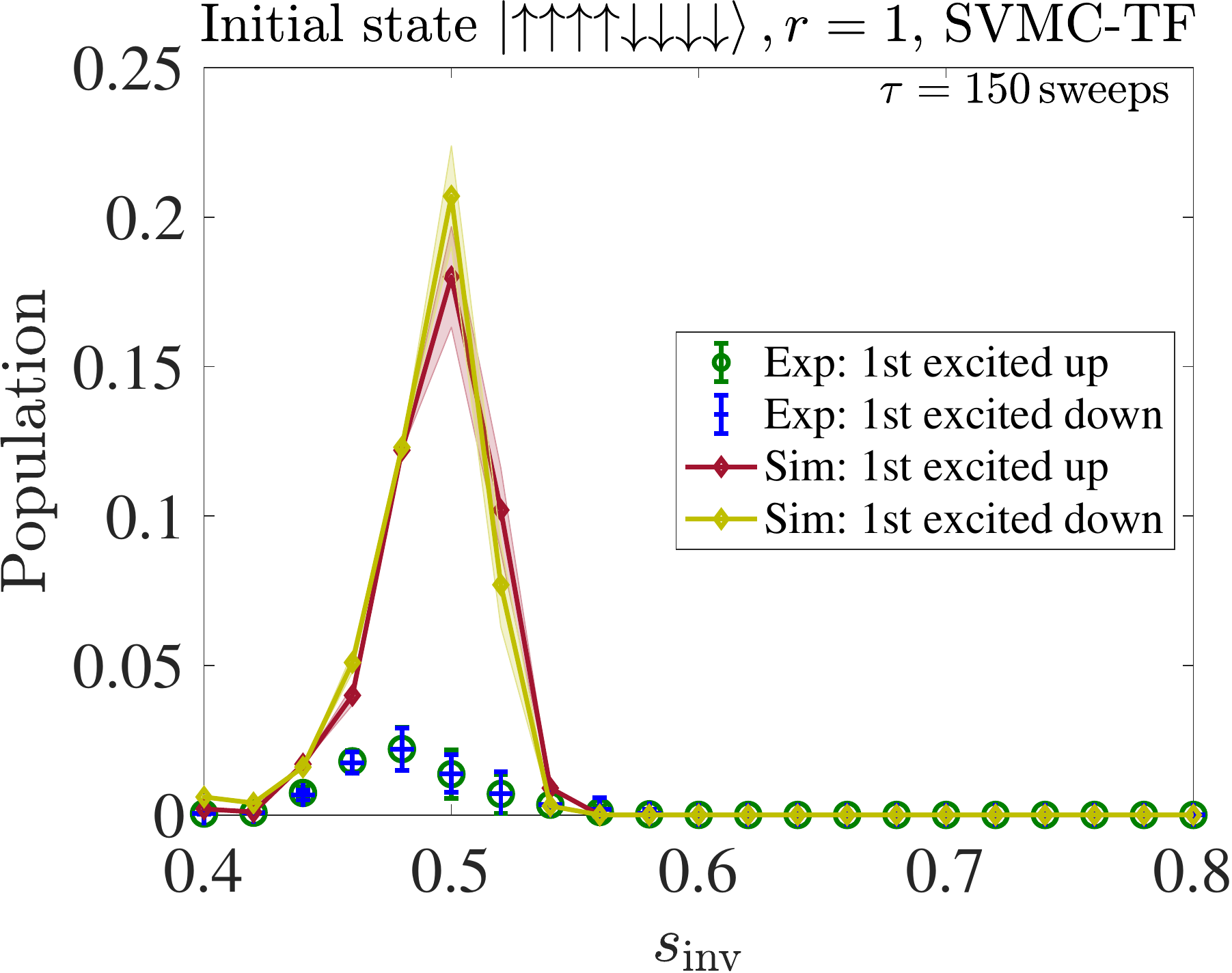}}
\subfigure[]{\includegraphics[width = 0.32\textwidth]{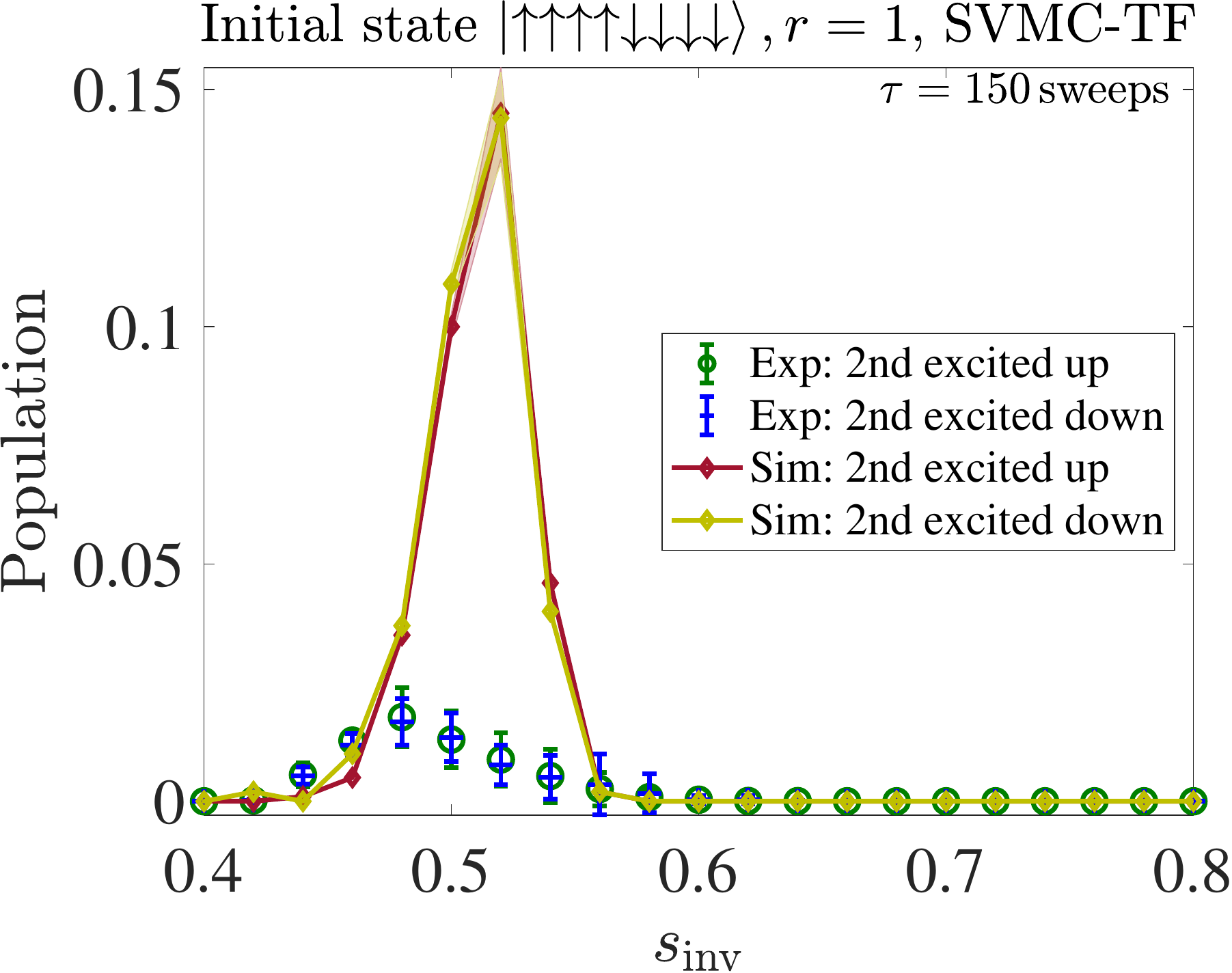}}
\subfigure[]{\includegraphics[width = 0.32\textwidth]{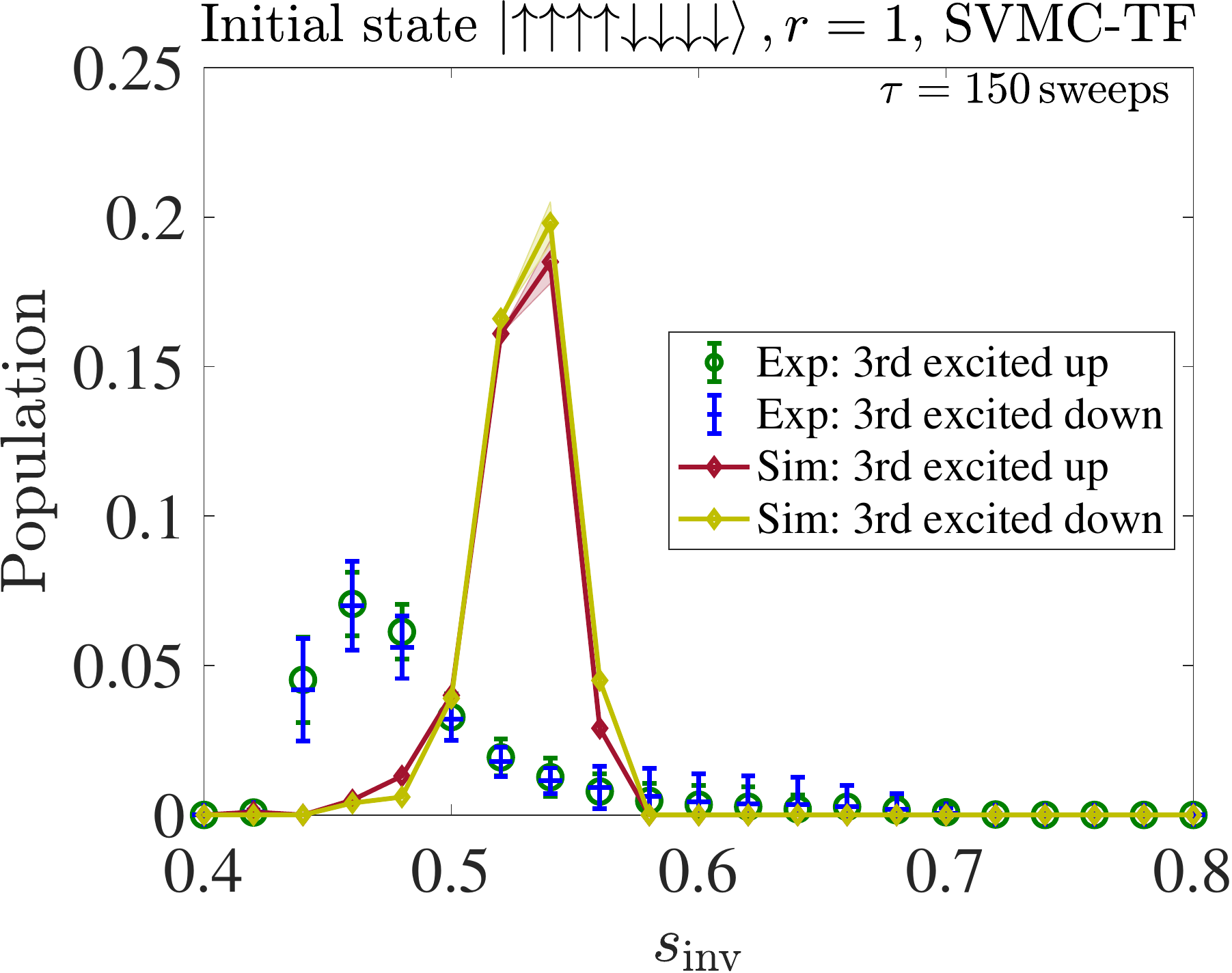}}
     \caption{Population in the first, second, and third excited states \textit{vs} $s_{\text{inv}}$ for D-Wave \textit{vs} the PTRE [(a)-(c)] and SVMC-TF with $\tau=150$ sweeps [(d)-(f)]. The initial state is  $\ket{\uparrow\uparrow\uparrow\uparrow\downarrow\downarrow\downarrow\downarrow}$. Error bars denote $1\sigma$ over $10^4$ samples.
     For the PTRE the simulation parameters are identical to the optimal ones obtained from the parameter estimation procedure in Fig.~\ref{fig:optimal_n4}. Note the different scale of the vertical axes.}
     \label{fig:compare_n8_excited}
\end{figure*}

\subsection{Partial success probability}
\label{sec:PSP-SVMC}

We compare the partial success probability obtained from SVMC and SVMC-TF in Figs.~\ref{fig:compareallpartial}(a) and (b), respectively, for $N=4$ and $\tau = 10^3$. It is again seen that SVMC-TF more accurately captures the early freezing than SVMC. 
The empirical data from Fig.~\ref{fig:RA_20_size}(a) is reproduced in Fig.~\ref{fig:compareallpartial}(c) and (d), where is it compared to SVMC-TF at an optimized number of sweeps (explained in App.~\ref{sec:optimalsweeps}). This yields a semi-quantitative agreement, in particular the correct trend and transition locations for the unequal up and down partial success probabilities. However, the agreement is clearly better for $N=4$ than for $N=8$ despite the optimization, which suggests the interesting possibility that SVMC-TF becomes less accurate at higher numbers of spins. Also noteworthy is that for $N=8$ we observe a deviation from the empirical data for $s_{\text{inv}} \lesssim 0.3$, where there exists a small but clear difference in the probabilities of all-up and all-down states, whereas the SVMC-TF data do not show such a trend. As discussed in Sec.~\ref{sec:spinbathpol}, we attribute the difference for $s_{\text{inv}} \lesssim 0.3$ to the spin-bath polarization effect, which is not modeled in our SVMC-TF simulations.

In Fig.~\ref{fig:partial16} we display SVMC-TF reverse annealing simulation results of partial success probabilities for $N=16,32$ with $\tau = 10^3,10^4$ sweeps. For both sizes shown, the regime of high partial success probability for the all-up state is shifted slightly to higher $s_{\textrm{inv}}$ for $\tau = 10^4$ sweeps than for $\tau = 10^3$. This is consistent with the trend in the experimental results observed in Fig.~\ref{fig:RA_20_diff_tau}(a). However, the trend with system size is inconsistent with the empirical partial success probabilities for the up case shown in Fig.~\ref{fig:RA_20_size}(a): the numerical results for $N=16$ and $N=32$ are virtually indistinguishable (apart from statistical fluctuations), while the empirical data shows that $N=16$ is not yet large enough for convergence. This (small) failure of the SVMC-TF model may hint at an interesting way in which to identify a ``quantum signature" in experimental quantum annealing~\cite{q-sig,albash2015consistency}. However, we did not pursue this direction further since we cannot rule out that further fine-tuning of the SVMC parameters will result in a closer match with the empirical data. To further explore ways in which a classical model such as SVMC-TF, or a model with strong coupling to the bath such as PTRE, might fail in describing the empirical data, we focus on excited state populations in the next Appendix.

\section{Excited states}
In this Appendix we present results comparing D-Wave data to simulations for the population in low-lying excited states. Our goal is not to be comprehensive, but rather to highlight agreements and discrepancies between the empirical and the numerical results.

The overall conclusion of this Appendix is that because of a persistence of small discrepancies even for the PTRE and SVMC-TF, in particular their failure to accurately predict the excited-state populations as shown below, further work is needed in order to  improve open-system models.

In both the empirical and the simulation data, the population of the $i$'th excited up (down) state is obtained after summation over all permutations of computational basis states with $N-i$ and $i$ up, or $i$ and $N-i$ down spins, where $N$ is the number of spins. We chose $\tau = 5\mu s$ for the D-Wave experiments.

\subsection{$N=4$ with a maximally excited initial state}
We compare in Fig.~\ref{fig:AMEexcited_0011} the AME, PTRE, and SVMC-TF simulation results to the empirical data for the initial state $\ket{0011}$. The open-system parameter settings are the same as in the ground state simulations reported above, for all three simulation methods, but the number of sweeps used in SVMC-TF is optimized for the closest agreement with the D-Wave data, as explained in App.~\ref{sec:optimalsweeps}. 

Since the initial state has no bias toward spin up or down, it is surprising that the D-Wave data exhibits an asymmetry between probability of ending in states with one up or one down spin. Since the same anomaly is not observed for $N=8$ (see Fig.~\ref{fig:compare_n8_excited}), we attribute it to an unexplained peculiarity associated with the embedding of the $N=4$ problem. All three simulation methods correctly predict no distinction between the up and down states. The predicted position of the peak in the first excited state population is different for the three simulation methods; the AME [panel (a)] predicts a position that is roughly the average of the empirically observed peaks for the case where the final state has a down or an up spin, while the PTRE [panel (b)] and SVMC-TF [panel (c)] are shifted to the right and left, respectively. It is not possible to say which prediction is correct due to the aforementioned anomaly.

The partial success probability results are shown in Fig.~\ref{fig:SVMC-exc2} for the second excited state as well as the ground state, for the AME and SVMC-TF. From (a) we see that the AME is qualitatively but not quantitatively in agreement with the empirical results for the second excited state, where it predicts an $s_{\text{inv}}$ value that is too large for the onset of the rise in the population of this state, and this rise is also somewhat too steep. 
The same is true for SVMC-TF [panel (b)] but it is in slightly closer agreement than AME for the second excited state. We also show results for $N=8$ [panel (c)], where SVMC-TF continues to exhibit good qualitative agreement with the empirical results.

\subsection{$N=8$ with a maximally excited initial state}
\label{sec:PTREfexcited}

In Fig.~\ref{fig:compare_n8_excited} we display the results for $N=8$ with $\ket{\uparrow\uparrow\uparrow\uparrow\downarrow\downarrow\downarrow\downarrow}$ as the initial state, for the probability of ending in the first, second, or third excited state, using the PTRE and SVMC-TF.\footnote{We do not present AME results since the computational cost of $N=8$ highly excited states using the AME is prohibitive: all $256$ eigenstates need to be taken into account.} This time the empirically observed population in the states with $i$ spins up or $i$ spins down ($1\leq i \leq 3$) is identical, as expected, i.e., we do not observe the anomaly mentioned above for $N=4$.

The top row shows the results for the PTRE, with the same set of optimal parameters as explained in Sec.~\ref{sec:PTRE}. The agreement is relatively poor, in that both the magnitude and the position of the population peak is missed, both being systematically overestimated.

The bottom row shows the results for SVMC-TF, with the optimal number of sweeps as determined in App.~\ref{sec:optimalsweeps}. The agreement is somewhat better than for the PTRE, in that both the peak's position and magnitude are closer to the empirical data, though the agreement in the peak's position deteriorates with increasing excitation level.

\subsection{First excited state as initial state}
\begin{figure}[t]
\subfigure[]{\includegraphics[width=0.4939\columnwidth]{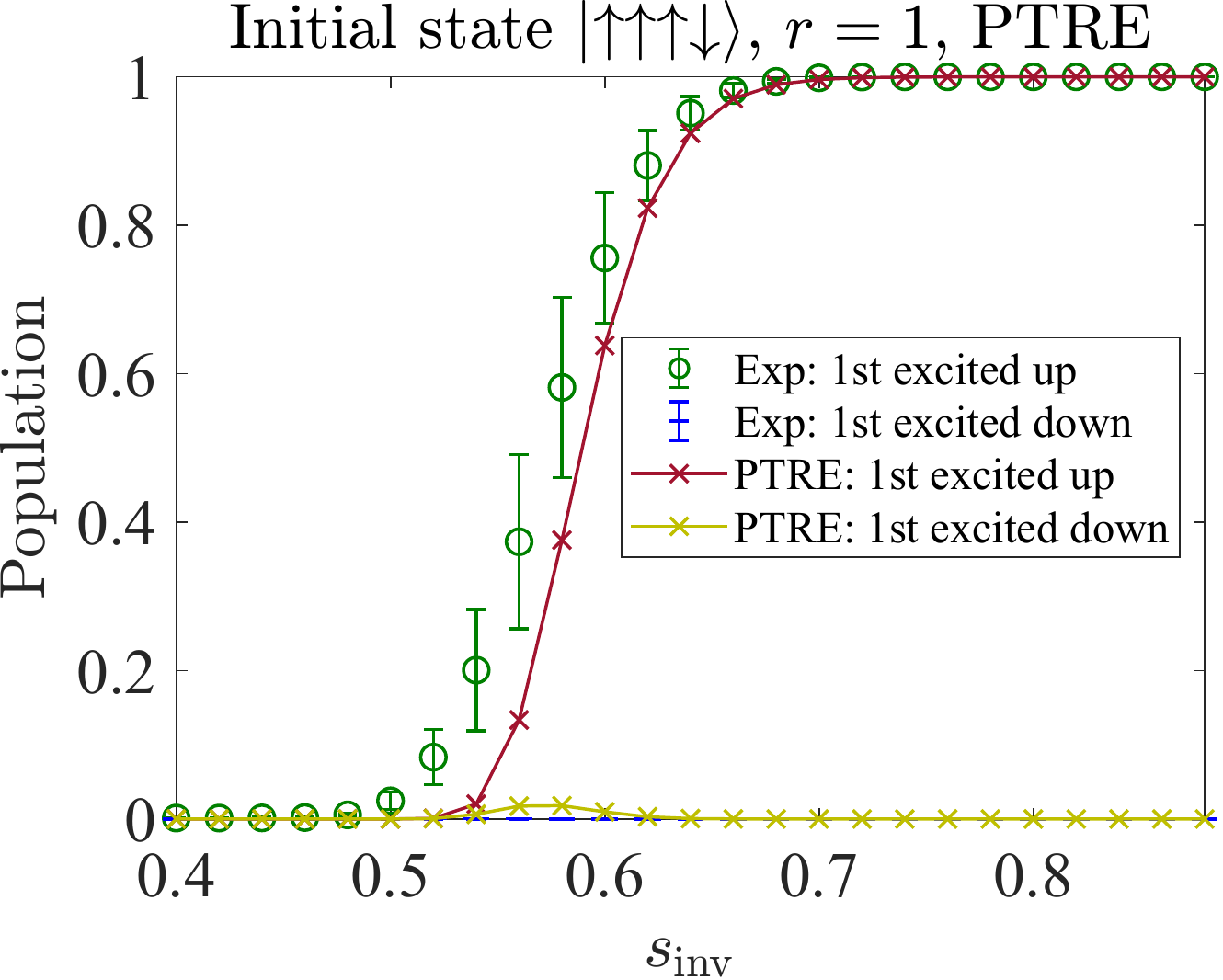}}
\subfigure[]{\includegraphics[width=0.4939\columnwidth]{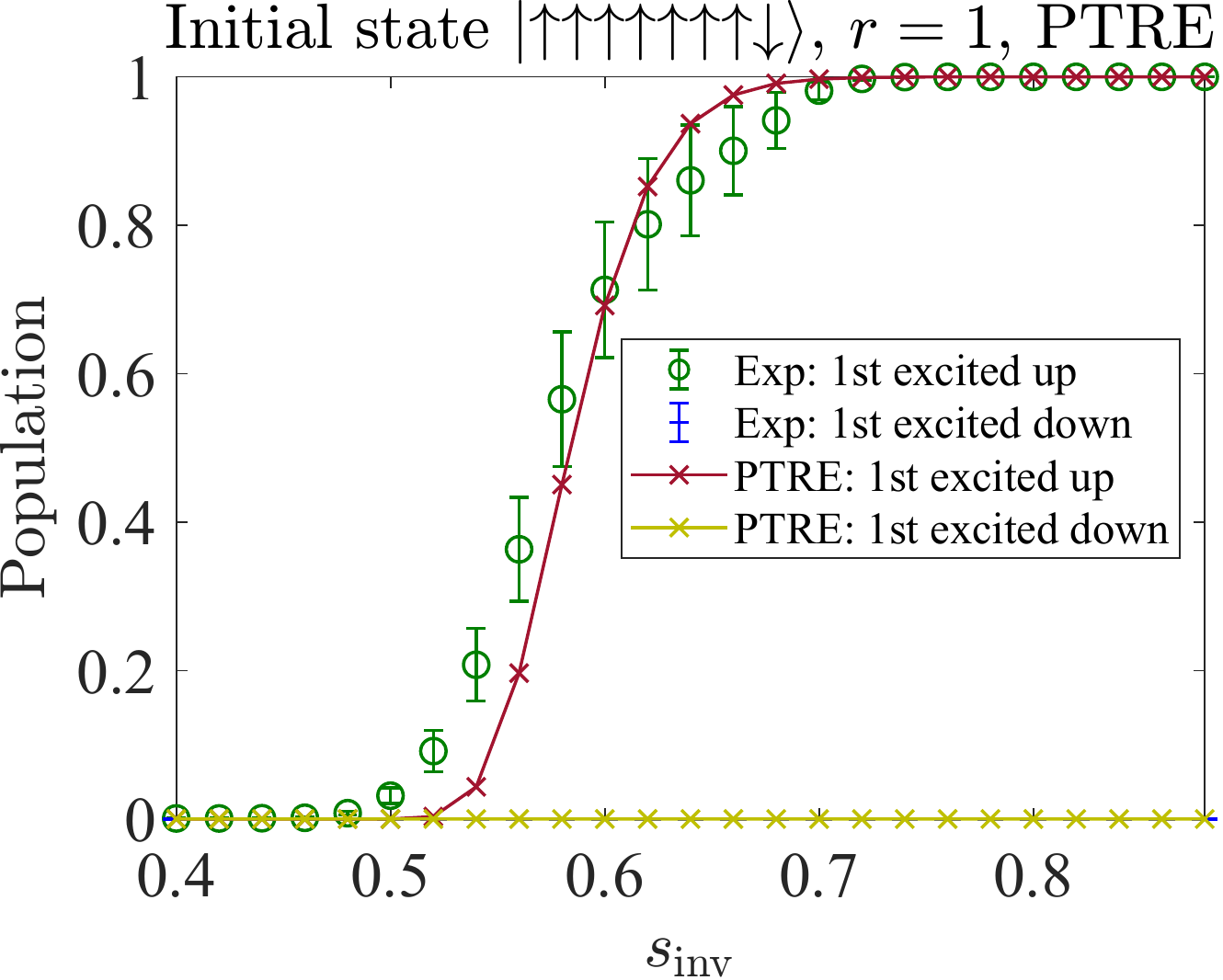}}
\subfigure[]{\includegraphics[width=0.4939\columnwidth]{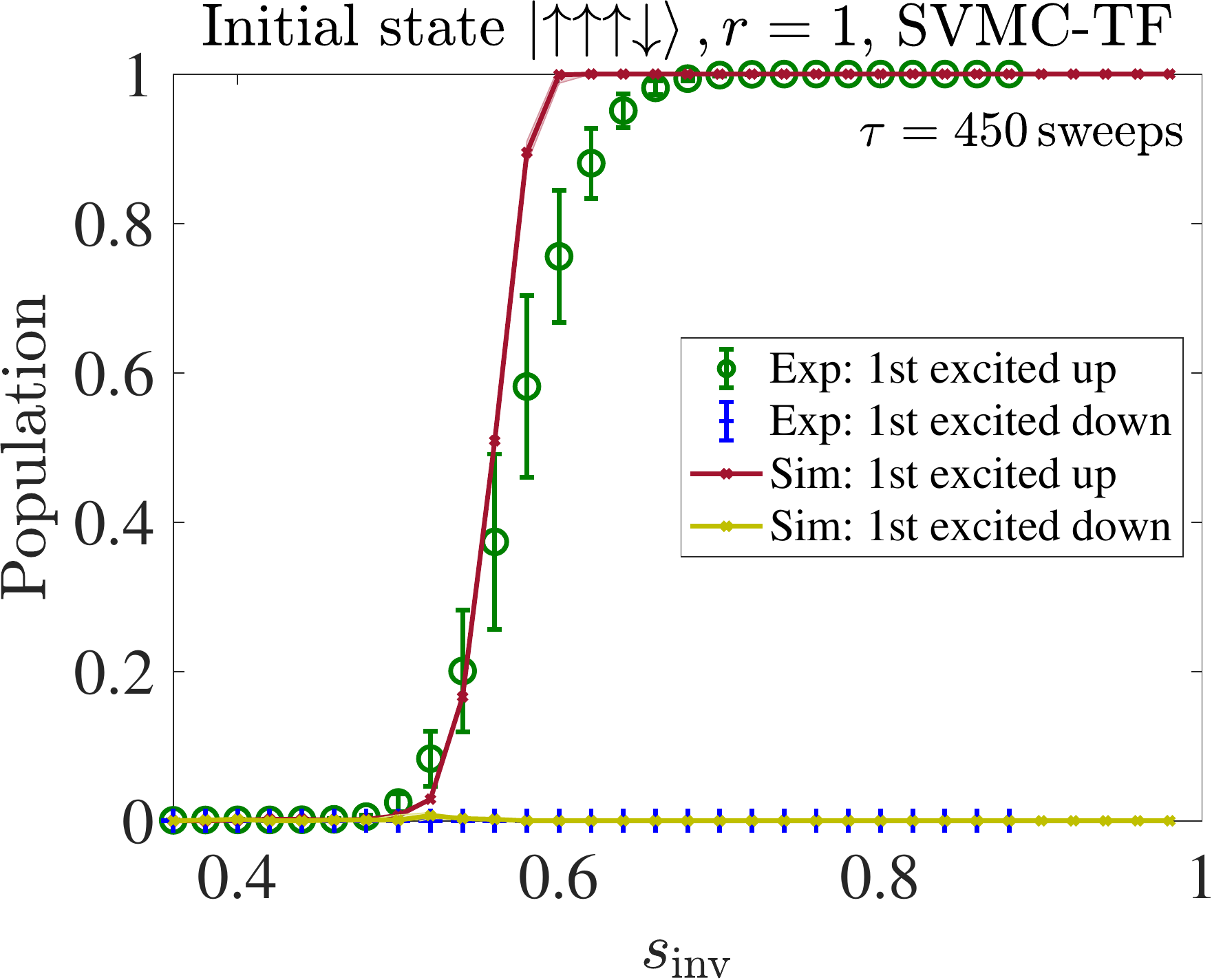}}
\subfigure[]{\includegraphics[width=0.4939\columnwidth]{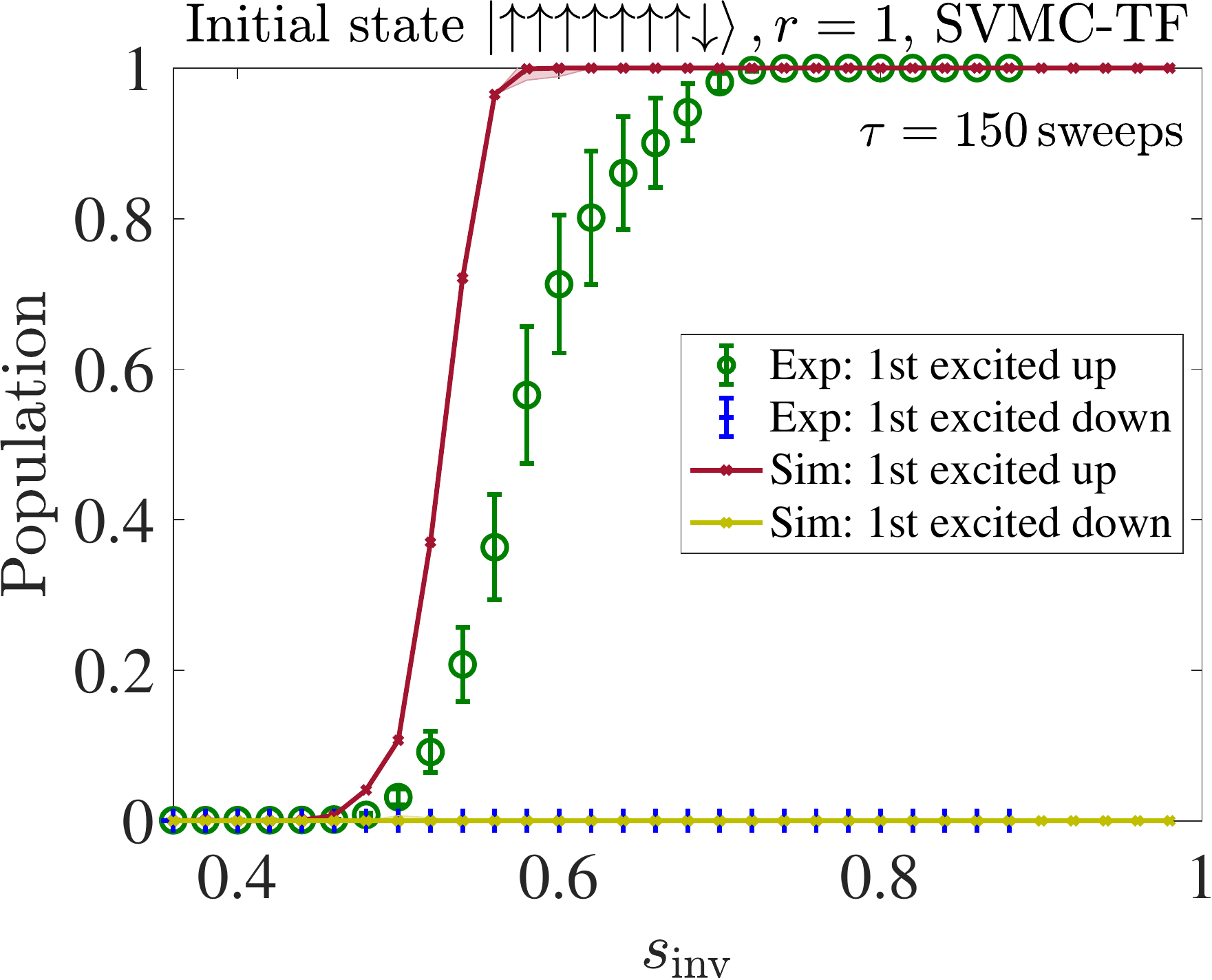}}

\caption{Comparison of experimental data (Exp) and simulation data (Sim) for the initial state $\ket{\uparrow\uparrow\uparrow\downarrow}$. (a) PTRE, (b) SVMC-TF with $450$ sweeps. (c) PTRE, (d) SVMC-TF with $150$ sweeps, for the initial state $\ket{\uparrow\uparrow\uparrow\uparrow\uparrow\uparrow\uparrow\downarrow}$ . The population shown is for the state with all but one spin up (``1st excited up'') and all but one spin down (``1st excited down''). Error bars denote $1\sigma$ over $10^4$ samples.}
\label{fig:SVMC-TFexcited_0001}
\end{figure}

As a final test, we checked the PTRE and SVMC-TF for initial states with one spin down, i.e., the first excited state. The results for $N=4$ and $N=8$ are shown in Fig.~\ref{fig:SVMC-TFexcited_0001}, and are in reasonable qualitative agreement. The agreement is overall somewhat better for the PTRE, especially for $N=8$. For $N=4$ the PTRE predicts a small non-zero probability for the first excited down state near the minimum gap point, which is absent in the empirical data and in the SVMC-TF results. This suggests that the PTRE slightly overestimates the incoherent tunneling rates for small $N$.

\section{SVMC-TF's dependence on the number of sweeps}
\label{sec:optimalsweeps}

For the $i$'th eigenstate population, the $\ell_2$ norm between the experiment and simulation data series (over a range of $s_{\text{inv}}$) is:
\begin{equation}
\ell^i_2 = \sqrt{\sum_{k=1}^{44} (x^{i, \text{exp}}_k - x^{i, \text{sim}}_k)^2}\ ,
\end{equation}
where for the $k$'th data point, $x^i_k$ is the population of the $i$'th eigenstate at $s_{\text{inv}} = 0.02k$ at the end of the reverse anneal (recall that the experimental data is sampled at every point from $\{0.02, 0.04, \dots, 0.88\}$).

\begin{figure}[t]
\subfigure[]{\includegraphics[width = 0.4939\columnwidth]{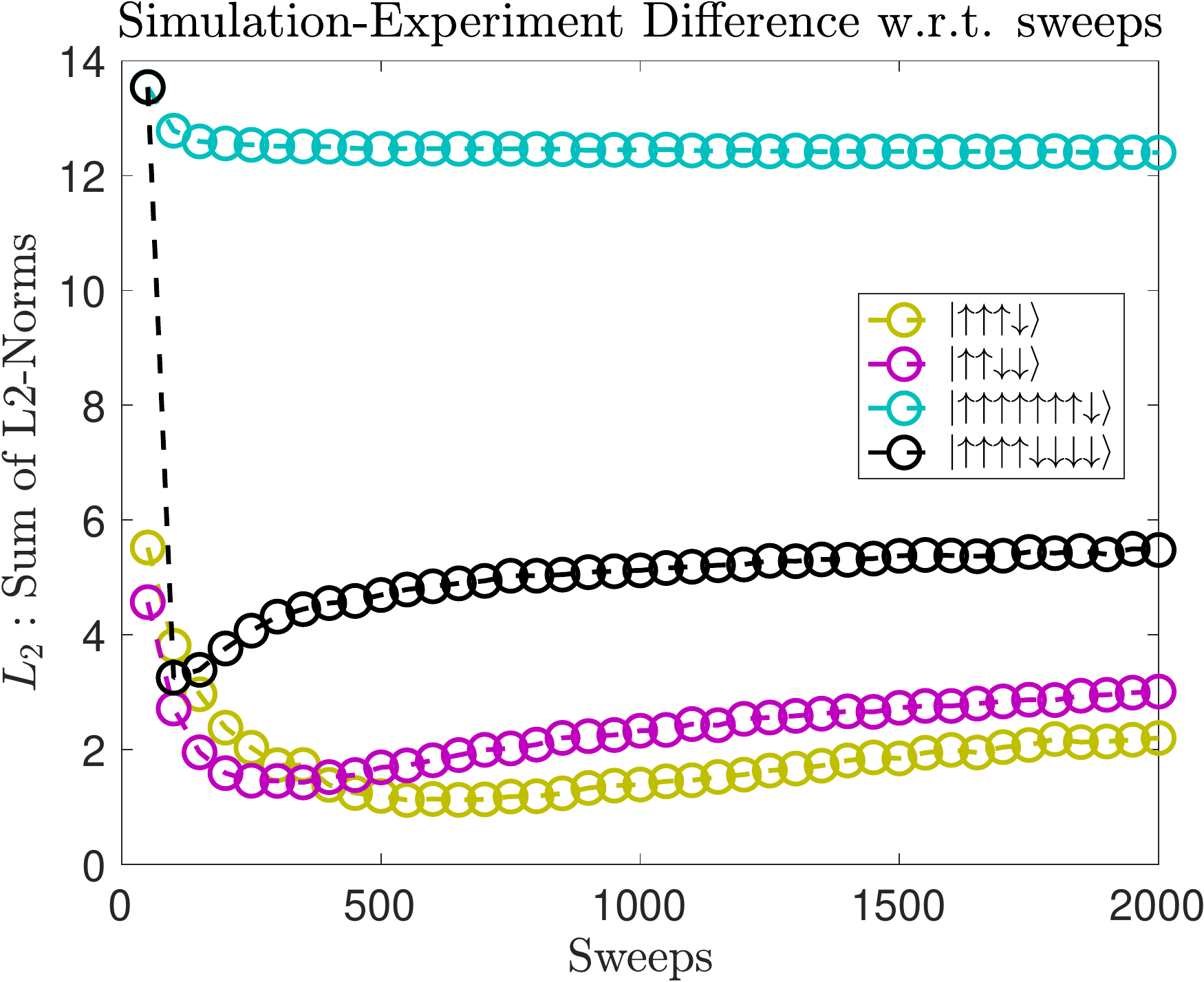}}
\subfigure[]{\includegraphics[width = 0.4939\columnwidth]{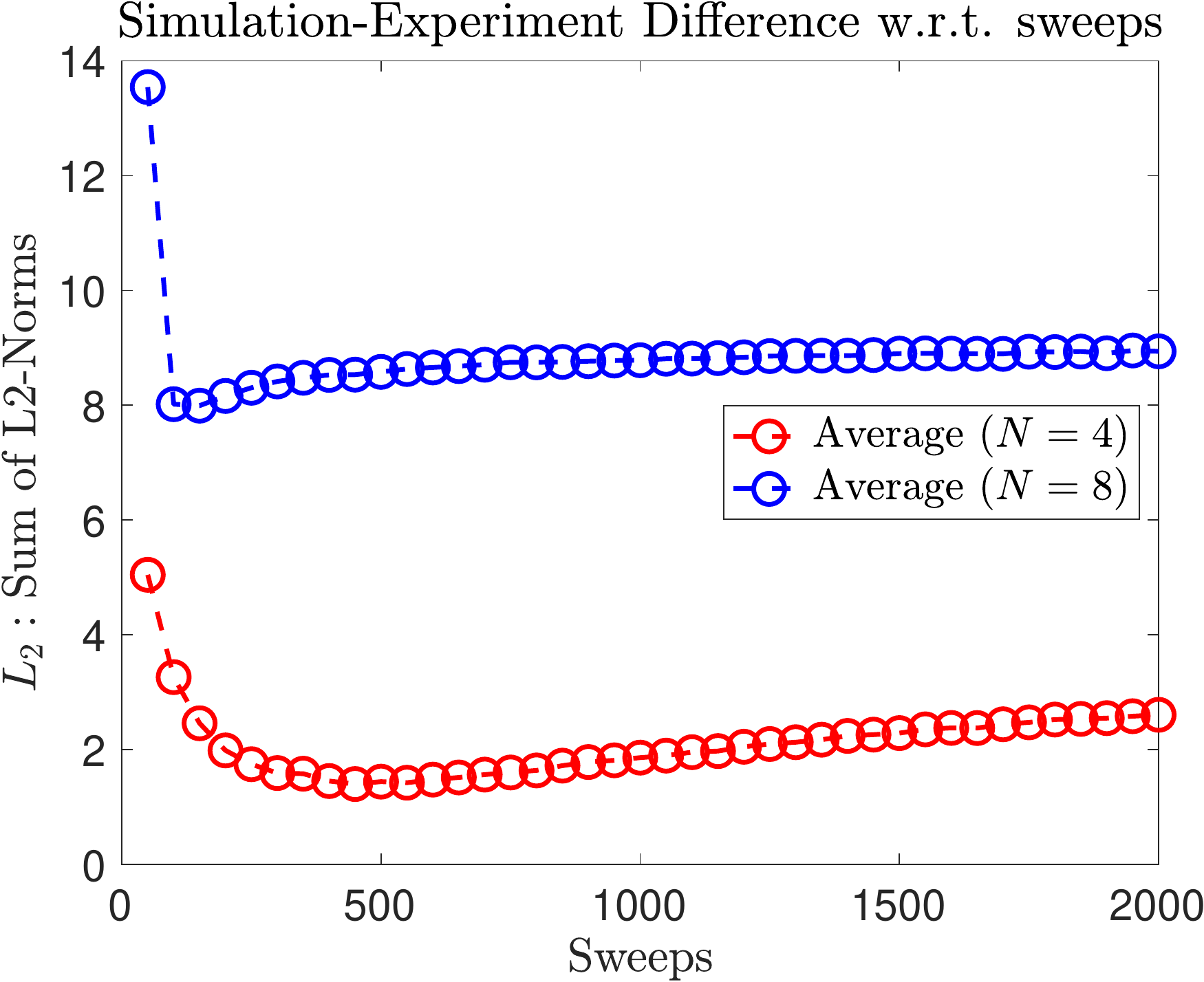}}
\caption{Sum of $\ell_2$-norms (between simulation and experiment data series) over all the eigenstates \textit{vs} the number of sweeps used in SVMC-TF simulations. (a) Subplot legend shows the initial states. (b) For $N=4$, SVMC-TF produces simulation data best matched with the empirical date at $450$ sweeps, while for $N=8$ the optimal number is $150$.}
\label{fig:sweep}
\end{figure}

To rigorously evaluate the differences between the experiment data and simulation data, we consider the sum of such $\ell_2$ norms over all eigenstates of the Hamiltonian, i.e., $\ell_2 = \sum_i \ell^i_2$. 
We plot $\ell_2$ \textit{vs} sweeps in Fig.~\ref{fig:sweep}. We observe that for the initial states shown, with the exception of $\ket{\uparrow\cdots\uparrow\downarrow}$, we can find an optimal sweep number. Considering both initial states at the respective system size, the optimal number of sweeps is $450$ for $N=4$ and $150$ for $N=8$.

\onecolumngrid

\section{Pseudocode for SVMC and SVMC-TF}
\label{append:svmc}

\begin{spacing}{0.8}
\begin{algorithm}[H]
  \tiny
\label{alg:1}
\caption{SVMC and SVMC-TF ($p$-spin reverse annealing)}
\hspace*{\algorithmicindent}
\begin{algorithmic}
\State $\mathcal{H}(s) = -\frac{A(s)}{2} (\sum_{i=1}^{N} \sin \theta_i) - \frac{B(s) N}{2}\left(\frac{1}{N}\sum_{i=1}^{N} \cos \theta_i\right)^{p}$, with a specified reverse annealing dependence $s(t)$.
\Procedure{SVMC}{}
\For{$k=1$ to $K$ (number of samples)}
    \State Initial computational basis state: $\ket{0}_i\rightarrow \theta^{k}_{i,t}: 0$, $\ket{1}_i\rightarrow \theta^{k}_{i,t}: \pi$.
    \For{$t=1$ to $T$ (total number of sweeps)}
        \For{$i=1$ to $n$ (number of qubits)}
            \State Randomly choose a new angle $\theta_{i,t}^{'k} \in [0, \pi]$.
            \State Calculate $\Delta E$.
                \If{$\Delta E \leq 0$}
                    \State $\theta_{i,t}^{k} \rightarrow \theta_{i,t}^{'k}$
                \ElsIf {$p < \exp (-\beta\Delta E)$ where $p\in [0,1]$ is drawn with uniform probability.}
                    \State $\theta_{i,t}^{k} \rightarrow \theta_{i,t}^{'k}$
                \EndIf
    	 \EndFor
	\EndFor
	\State Take the mean of $K$ samples: $\bar{\theta}_{i,t} = (\sum_{k=1}^{K}\theta_{i,t}^{'k})/K$.
\EndFor
\EndProcedure
\State \Return $\bar{\theta}_{i,t}$.
\Procedure{SVMC-TF}{}
\For{$k=1$ to $K$ (number of samples)}
    \State Initial computational basis state: $\ket{0}_i\rightarrow \theta^{k}_{i,t}: 0$, $\ket{1}_i\rightarrow \theta^{k}_{i,t}: \pi$.
    \For{$t=1$ to $T$ (total number of sweeps)}
        \For{$i=1$ to $n$ (number of qubits)}
            \State $\theta_{i,t}^{'k} = \theta_{i,t}^{k} + \epsilon_i(s(t/T))$, 
            \State a random $\epsilon_i(s(t/T) \in [-\min \left(1,\frac{A(s(t/T)}{B(s(t/T)}\right) \pi, \min \left(1,\frac{A(s(t/T)}{B(s(t/T)}\right) \pi]$
            \State Calculate $\Delta E$.
                \If{$\Delta E \leq 0$}
                    \State $\theta_{i,t}^{k} \rightarrow \theta_{i,t}^{'k}$
                \ElsIf {$p < \exp (-\beta\Delta E)$ where $p\in [0,1]$ is drawn with uniform probability.}
                    \State $\theta_{i,t}^{k} \rightarrow \theta_{i,t}^{'k}$
                \EndIf
    	 \EndFor
	\EndFor
	\State Take the mean of $K$ samples: $\bar{\theta}_{i,t} = (\sum_{k=1}^{K}\theta_{i,t}^{'k})/K$.
\EndFor
\EndProcedure
\State \Return $\bar{\theta}_{i,t}$.

\Procedure{Projection onto computational basis}{}
\If{$0\leq\theta_{i,t}^{k}\leq \pi/2$}
    \State $\ket{\psi^k(t)}_i = \ket{0}$
\ElsIf {$\pi/2\leq\theta_{i,t}^{k}\leq \pi$}
    \State $\ket{\psi^k(t)}_i = \ket{1}$
\EndIf
\EndProcedure
\State Remark: $\Delta E$ due to the update of qubit $k$, can be expressed in the following form:
\begin{align*}
\Delta E_k
&= 
\left(-\frac{A(s)}{2} \left(\sum_{\substack{i=1\\i \ne k}}^{N}\sin \theta_i + \sin \theta^{'}_k\right) - \frac{B(s)N}{2}\left(\frac{1}{N}\left(\sum_{\substack{i=1\\i \ne k}}^{N} \cos \theta_i + \cos \theta^{'}_k\right)\right)^{2}  \right)- \mathcal{H}(s)
\,.
\end{align*}
\end{algorithmic}
\end{algorithm}
\end{spacing}

\twocolumngrid
\typeout{}
%

\end{document}